%% file: main.tex
\documentclass[10pt,twocolumn,letterpaper]{article}

\usepackage{iccv}
\usepackage{times}
\usepackage{multirow,nicefrac,xcolor}
\usepackage{pgf}
\usepackage{xspace}

\usepackage{cite}
\usepackage{amsmath,amssymb,amsfonts}
\usepackage{algorithmic}
\usepackage{graphicx}
\usepackage{textcomp}

\usepackage{tikz}
\usepackage{multirow}
\usepackage{subcaption}
\usepackage{array}
\usepackage{pdflscape}
\usepackage{hyperref}
\usepackage{adjustbox}
\usepackage{stmaryrd}

\iccvfinalcopy 

\def\iccvPaperID{12135} 

\ificcvfinal\fi

\begin{document}

\title{How to choose your best allies for a transferable attack?}
\author{Thibault Maho\thanks{Thanks to Rennes Métropole for its funding for international mobility.}\\
Univ. Rennes, Inria, CNRS\\
IRISA, Rennes, France\\
{\tt\small thibault.maho@inria.fr}
\and
Seyed-Mohsen Moosavi-Dezfooli \\
Imperial College London, UK\\
{\tt\small seyed.moosavi@imperial.ac.uk}
\and
Teddy Furon \thanks{Thanks to ANR and AID french agencies for funding Chaire SAIDA.} \\
Univ. Rennes, Inria, CNRS\\
IRISA, Rennes, France\\
{\tt\small teddy.furon@inria.fr}
}

\maketitle
\ificcvfinal\thispagestyle{empty}\fi

\input{sections/notations.tex}

\begin{abstract}

    The transferability of adversarial examples is a key issue in the security of deep neural networks.
    The possibility of an adversarial example crafted for a source model fooling another targeted model makes the threat of adversarial attacks more realistic.
    Measuring transferability is a crucial problem, but the Attack Success Rate alone does not provide a sound evaluation. This paper proposes a new methodology for evaluating transferability by putting distortion in a central position.  This new tool shows that transferable attacks may perform far worse than a black box attack if the attacker randomly picks the source model.
    To address this issue, we propose a new selection mechanism, called \fit, which aims at choosing the best source model with only a few preliminary queries to the target. Our experimental results show that \fit is highly effective at selecting the best source model for multiple scenarios such as single-model attacks, ensemble-model attacks and multiple attacks~\footnote{Code available at \url{https://github.com/t-maho/transferability_measure_fit}}.
    

\end{abstract}

\input{sections/introduction.tex}

\input{sections/related_work.tex}

\input{sections/preliminaries.tex}

\input{sections/transferability.tex}
\input{sections/fit.tex}

\input{sections/results.tex}

\input{sections/conclusion.tex}

{\small
    \bibliographystyle{ieee_fullname.bst}
    \bibliography{references.bib}
}

\clearpage
\onecolumn

\appendix
\input{sections/appendices/setup.tex}
\input{sections/appendices/preliminaries.tex}
\input{sections/appendices/matrix.tex}
\input{sections/appendices/results.tex}

\end{document}

%% file: sections/notations.tex
\newcommand{\set}[1]{\mathcal{#1}}
\def \setA{\set{A}}

\def \T{\textsf{T}}
\def \Prob{\mathbb{P}}
\def \dist{\mathsf{dist}}
\def \PWB{P^{\mathsf{wb}}}
\def \PBB{P^{\mathsf{bb}}}
\def \dBB{d^{\mathsf{bb}}}
\def \dWB{d^{\mathsf{wb}}}

\def \fit{\texttt{FiT}\xspace}
\def \measure{$\hat{T}_{s,t}$\xspace}
\def \measurex{$\hat{T}_{s,x}$\xspace}

\def \di{\texttt{DI}~\cite{xie2019improving}\xspace}
\def \ti{\texttt{TI}~\cite{dong2019evading}\xspace}
\def \admix{\texttt{Admix}~\cite{wang2021admix}\xspace}
\def \pi{\texttt{PI}~\cite{wang2021boosting}\xspace}
\def \taig{\texttt{TAIG}~\cite{huang2022transferable}\xspace}
\def \dwp{\texttt{DWP}~\cite{wang2022enhancing}\xspace}

\def \bp{\texttt{BP}~\cite{boundaryprojection}\xspace}
\def \nmf{\texttt{FMN}~\cite{pintor2021fast}\xspace}
\def \cw{\texttt{CW}~\cite{carlini2017towards}\xspace}
\def \deepfool{\texttt{DeepFool}~\cite{deepfool}\xspace}
\def \pgd{\texttt{PGD}~\cite{madry2017towards}\xspace}
\def \ifgsm{\texttt{I-FGSM}~\cite{ifgsm}\xspace}

\def \rays{\texttt{RayS}~\cite{rays}\xspace}
\def \surfree{\texttt{SurFree}~\cite{surfree}\xspace}
\def \geoda{\texttt{GeoDA}~\cite{geoda}\xspace}

\def \fbi{\texttt{FBI}~\cite{maho2022fbi}\xspace}
\def \ipguard{\texttt{IPGuard}~\cite{ipguard}\xspace}
\def \afa{\texttt{AFA}~\cite{afa}\xspace}

\def \ModSim{\texttt{ModSim}}
\def \ModDist{\texttt{ModDist}}
\def \TransQ{\texttt{TransQ}}
\def \fit{\texttt{FiT}\xspace}
\def \dir{u}

\def \TransQOne{\texttt{\TransQ}$^{(1)}$}
\def \TransQTwo{\texttt{\TransQ}$^{(2)}$}

%% file: sections/introduction.tex
\section{Introduction}

Transferability is one of the most intriguing properties of adversarial examples. A white box attack crafting adversarial examples for an open-source model is likely to fool other models too~\cite{liu2017delving, tramer2017space, naseer2022on, barni2019transferability, yuan2021transferability}. This makes the threat of adversarial examples more realistic.
In practice, the model targeted is usually unknown but accessible as a black box.
This prevents directly applying any white box gradient-based attack~\cite{boundaryprojection, ifgsm, deepfool, madry2017towards}. Black box attacks do exist but they require some thousands of queries to find an adversarial example of low distortion~\cite{geoda, qeba, surfree, rays}. Transferable attacks require no or few queries to fine-tune an adversarial example thanks to the help of a publicly available model similar enough to the target.   

Transferability is usually measured by the Attack Success Rate (ASR), i.e., the probability that the adversarial example crafted for the source model also deludes the target model. 
We argue that this measure leads to an unfair evaluation of transferability.
In the context of adversarial examples, it is not just a matter of discovering data that is not well classified, but rather identifying the perturbation that can fool a classifier with minimal distortion. This principle should also apply to transferable attacks.

\begin{figure}[t]
    \centering
    \resizebox{0.9\linewidth}{!}{\input{./images/introduction/coat_lite_small.pgf}}
    \caption{Evaluation of transferability by comparing the Attack Success Rate vs. distortion trade-off of a white box, transferable, and black box attacks against model $\texttt{CoatLite}_{\texttt{small}}$ (See Sect.~\ref{sec:experimental_setup} for details).
    The blue area is the range of trade-off operated by a transferable attack with random source models.
    A transferable attack may be worse than a black box attack without a good source selection (like \fit).}  
    \label{fig:splash}
\end{figure}
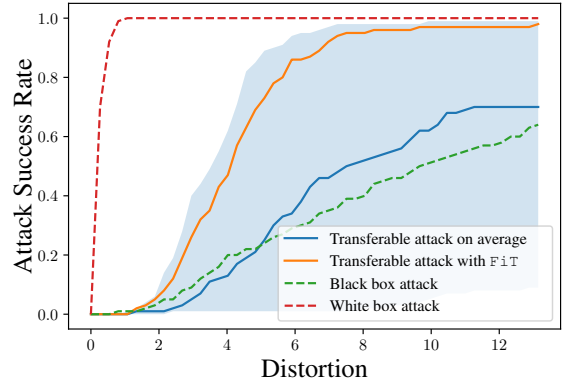
For illustration purposes, let us consider two models, one is robust in the sense that the necessary amount of adversarial perturbation is large, whereas the other model is weak. If the attacker uses the robust model as the source to attack the weak target network, the ASR of the transferable attack will certainly be big. It does not mean that this is the right choice. The ASR is high because the robust source model needs large perturbation to be deluded, which will fool any weaker model. The ASR alone does not reflect the overshooting in distortion. The converse, using the weak to attack the robust, would yield a low ASR. To summarize, the ASR alone fails to capture how relevant the \emph{direction of the perturbation} given by the source is for attacking the target.

The first contribution of this paper is to put distortion back into the picture. Section~\ref{sec:Methodology} evaluates transferability by comparing the distortion of a transferable attack to the ones of two reference attacks: On one hand, the strongest attack, i.e., the white box attack directly applied on the target model; on the other hand, the weakest attack, i.e., the black box attack.

The second contribution shows the great variability of the performance of transferable attacks. \autoref{fig:splash} summarizes this observation by plotting the ASR as a function of the distortion (the experimental protocol is explained in Sect.~\ref{sec:experimental_setup}). Naturally, the black box attack needs much more distortion than the white box attack.  For instance, the white box attack yields an ASR of 50\% with a distortion of 0.19, whereas the black box attack needs a distortion of 9.7. The surprise is that if the attacker resorts to a transferable attack and picks a source model at random, there is almost a 50\% chance that the attack performs even worse than the black box attack. Section~\ref{sec:Transferability} outlines a triad of factors: input, source model, and attack.

This observation challenges the prevalent notion that adversarial examples transfer easily between models, and highlights the need to carefully choose the source model to attack a target. Under the assumption that the attacker has indeed several candidate models, our third contribution, named \fit in Sect.~\ref{sec:HowTo}, provides an affordable measure for model selection, allowing the attacker to choose a good source model with only a few queries to the target.

%% file: images/introduction/coat_lite_small.pgf
\begingroup%
\makeatletter%
\begin{pgfpicture}%
\pgfpathrectangle{\pgfpointorigin}{\pgfqpoint{5.332292in}{3.703148in}}%
\pgfusepath{use as bounding box}%
\begin{pgfscope}%
\pgfsetbuttcap%
\pgfsetmiterjoin%
\definecolor{currentfill}{rgb}{1.000000,1.000000,1.000000}%
\pgfsetfillcolor{currentfill}%
\pgfsetlinewidth{0.000000pt}%
\definecolor{currentstroke}{rgb}{1.000000,1.000000,1.000000}%
\pgfsetstrokecolor{currentstroke}%
\pgfsetdash{}{0pt}%
\pgfpathmoveto{\pgfqpoint{0.000000in}{0.000000in}}%
\pgfpathlineto{\pgfqpoint{5.332292in}{0.000000in}}%
\pgfpathlineto{\pgfqpoint{5.332292in}{3.703148in}}%
\pgfpathlineto{\pgfqpoint{0.000000in}{3.703148in}}%
\pgfpathlineto{\pgfqpoint{0.000000in}{0.000000in}}%
\pgfpathclose%
\pgfusepath{fill}%
\end{pgfscope}%
\begin{pgfscope}%
\pgfsetbuttcap%
\pgfsetmiterjoin%
\definecolor{currentfill}{rgb}{1.000000,1.000000,1.000000}%
\pgfsetfillcolor{currentfill}%
\pgfsetlinewidth{0.000000pt}%
\definecolor{currentstroke}{rgb}{0.000000,0.000000,0.000000}%
\pgfsetstrokecolor{currentstroke}%
\pgfsetstrokeopacity{0.000000}%
\pgfsetdash{}{0pt}%
\pgfpathmoveto{\pgfqpoint{0.582292in}{0.523148in}}%
\pgfpathlineto{\pgfqpoint{5.232292in}{0.523148in}}%
\pgfpathlineto{\pgfqpoint{5.232292in}{3.603148in}}%
\pgfpathlineto{\pgfqpoint{0.582292in}{3.603148in}}%
\pgfpathlineto{\pgfqpoint{0.582292in}{0.523148in}}%
\pgfpathclose%
\pgfusepath{fill}%
\end{pgfscope}%
\begin{pgfscope}%
\pgfpathrectangle{\pgfqpoint{0.582292in}{0.523148in}}{\pgfqpoint{4.650000in}{3.080000in}}%
\pgfusepath{clip}%
\pgfsetbuttcap%
\pgfsetroundjoin%
\definecolor{currentfill}{rgb}{0.121569,0.466667,0.705882}%
\pgfsetfillcolor{currentfill}%
\pgfsetfillopacity{0.200000}%
\pgfsetlinewidth{0.000000pt}%
\definecolor{currentstroke}{rgb}{0.000000,0.000000,0.000000}%
\pgfsetstrokecolor{currentstroke}%
\pgfsetdash{}{0pt}%
\pgfpathmoveto{\pgfqpoint{0.793655in}{0.663148in}}%
\pgfpathlineto{\pgfqpoint{0.793655in}{0.663148in}}%
\pgfpathlineto{\pgfqpoint{0.879926in}{0.663148in}}%
\pgfpathlineto{\pgfqpoint{0.966197in}{0.663148in}}%
\pgfpathlineto{\pgfqpoint{1.052468in}{0.663148in}}%
\pgfpathlineto{\pgfqpoint{1.138739in}{0.663148in}}%
\pgfpathlineto{\pgfqpoint{1.225010in}{0.663148in}}%
\pgfpathlineto{\pgfqpoint{1.311281in}{0.663148in}}%
\pgfpathlineto{\pgfqpoint{1.397552in}{0.663148in}}%
\pgfpathlineto{\pgfqpoint{1.483822in}{0.663148in}}%
\pgfpathlineto{\pgfqpoint{1.570093in}{0.691148in}}%
\pgfpathlineto{\pgfqpoint{1.656364in}{0.691148in}}%
\pgfpathlineto{\pgfqpoint{1.742635in}{0.691148in}}%
\pgfpathlineto{\pgfqpoint{1.828906in}{0.691148in}}%
\pgfpathlineto{\pgfqpoint{1.915177in}{0.691148in}}%
\pgfpathlineto{\pgfqpoint{2.001448in}{0.691148in}}%
\pgfpathlineto{\pgfqpoint{2.087719in}{0.691148in}}%
\pgfpathlineto{\pgfqpoint{2.173989in}{0.691148in}}%
\pgfpathlineto{\pgfqpoint{2.260260in}{0.691148in}}%
\pgfpathlineto{\pgfqpoint{2.346531in}{0.691148in}}%
\pgfpathlineto{\pgfqpoint{2.432802in}{0.691148in}}%
\pgfpathlineto{\pgfqpoint{2.519073in}{0.691148in}}%
\pgfpathlineto{\pgfqpoint{2.605344in}{0.691148in}}%
\pgfpathlineto{\pgfqpoint{2.691615in}{0.691148in}}%
\pgfpathlineto{\pgfqpoint{2.777885in}{0.691148in}}%
\pgfpathlineto{\pgfqpoint{2.864156in}{0.691148in}}%
\pgfpathlineto{\pgfqpoint{2.950427in}{0.691148in}}%
\pgfpathlineto{\pgfqpoint{3.036698in}{0.691148in}}%
\pgfpathlineto{\pgfqpoint{3.122969in}{0.691148in}}%
\pgfpathlineto{\pgfqpoint{3.209240in}{0.691148in}}%
\pgfpathlineto{\pgfqpoint{3.295511in}{0.691148in}}%
\pgfpathlineto{\pgfqpoint{3.381782in}{0.719148in}}%
\pgfpathlineto{\pgfqpoint{3.468052in}{0.719148in}}%
\pgfpathlineto{\pgfqpoint{3.554323in}{0.719148in}}%
\pgfpathlineto{\pgfqpoint{3.640594in}{0.747148in}}%
\pgfpathlineto{\pgfqpoint{3.726865in}{0.747148in}}%
\pgfpathlineto{\pgfqpoint{3.813136in}{0.775148in}}%
\pgfpathlineto{\pgfqpoint{3.899407in}{0.775148in}}%
\pgfpathlineto{\pgfqpoint{3.985678in}{0.803148in}}%
\pgfpathlineto{\pgfqpoint{4.071949in}{0.831148in}}%
\pgfpathlineto{\pgfqpoint{4.158219in}{0.859148in}}%
\pgfpathlineto{\pgfqpoint{4.244490in}{0.859148in}}%
\pgfpathlineto{\pgfqpoint{4.330761in}{0.859148in}}%
\pgfpathlineto{\pgfqpoint{4.417032in}{0.887148in}}%
\pgfpathlineto{\pgfqpoint{4.503303in}{0.887148in}}%
\pgfpathlineto{\pgfqpoint{4.589574in}{0.887148in}}%
\pgfpathlineto{\pgfqpoint{4.675845in}{0.887148in}}%
\pgfpathlineto{\pgfqpoint{4.762116in}{0.887148in}}%
\pgfpathlineto{\pgfqpoint{4.848386in}{0.887148in}}%
\pgfpathlineto{\pgfqpoint{4.934657in}{0.915148in}}%
\pgfpathlineto{\pgfqpoint{5.020928in}{0.915148in}}%
\pgfpathlineto{\pgfqpoint{5.020928in}{3.435148in}}%
\pgfpathlineto{\pgfqpoint{5.020928in}{3.435148in}}%
\pgfpathlineto{\pgfqpoint{4.934657in}{3.435148in}}%
\pgfpathlineto{\pgfqpoint{4.848386in}{3.435148in}}%
\pgfpathlineto{\pgfqpoint{4.762116in}{3.435148in}}%
\pgfpathlineto{\pgfqpoint{4.675845in}{3.435148in}}%
\pgfpathlineto{\pgfqpoint{4.589574in}{3.435148in}}%
\pgfpathlineto{\pgfqpoint{4.503303in}{3.435148in}}%
\pgfpathlineto{\pgfqpoint{4.417032in}{3.435148in}}%
\pgfpathlineto{\pgfqpoint{4.330761in}{3.435148in}}%
\pgfpathlineto{\pgfqpoint{4.244490in}{3.435148in}}%
\pgfpathlineto{\pgfqpoint{4.158219in}{3.435148in}}%
\pgfpathlineto{\pgfqpoint{4.071949in}{3.435148in}}%
\pgfpathlineto{\pgfqpoint{3.985678in}{3.435148in}}%
\pgfpathlineto{\pgfqpoint{3.899407in}{3.407148in}}%
\pgfpathlineto{\pgfqpoint{3.813136in}{3.407148in}}%
\pgfpathlineto{\pgfqpoint{3.726865in}{3.407148in}}%
\pgfpathlineto{\pgfqpoint{3.640594in}{3.407148in}}%
\pgfpathlineto{\pgfqpoint{3.554323in}{3.407148in}}%
\pgfpathlineto{\pgfqpoint{3.468052in}{3.407148in}}%
\pgfpathlineto{\pgfqpoint{3.381782in}{3.407148in}}%
\pgfpathlineto{\pgfqpoint{3.295511in}{3.407148in}}%
\pgfpathlineto{\pgfqpoint{3.209240in}{3.407148in}}%
\pgfpathlineto{\pgfqpoint{3.122969in}{3.407148in}}%
\pgfpathlineto{\pgfqpoint{3.036698in}{3.379148in}}%
\pgfpathlineto{\pgfqpoint{2.950427in}{3.351148in}}%
\pgfpathlineto{\pgfqpoint{2.864156in}{3.323148in}}%
\pgfpathlineto{\pgfqpoint{2.777885in}{3.323148in}}%
\pgfpathlineto{\pgfqpoint{2.691615in}{3.295148in}}%
\pgfpathlineto{\pgfqpoint{2.605344in}{3.211148in}}%
\pgfpathlineto{\pgfqpoint{2.519073in}{3.183148in}}%
\pgfpathlineto{\pgfqpoint{2.432802in}{3.155148in}}%
\pgfpathlineto{\pgfqpoint{2.346531in}{3.043148in}}%
\pgfpathlineto{\pgfqpoint{2.260260in}{2.959148in}}%
\pgfpathlineto{\pgfqpoint{2.173989in}{2.651148in}}%
\pgfpathlineto{\pgfqpoint{2.087719in}{2.399148in}}%
\pgfpathlineto{\pgfqpoint{2.001448in}{2.203148in}}%
\pgfpathlineto{\pgfqpoint{1.915177in}{2.035148in}}%
\pgfpathlineto{\pgfqpoint{1.828906in}{1.895148in}}%
\pgfpathlineto{\pgfqpoint{1.742635in}{1.783148in}}%
\pgfpathlineto{\pgfqpoint{1.656364in}{1.447148in}}%
\pgfpathlineto{\pgfqpoint{1.570093in}{1.195148in}}%
\pgfpathlineto{\pgfqpoint{1.483822in}{1.055148in}}%
\pgfpathlineto{\pgfqpoint{1.397552in}{0.831148in}}%
\pgfpathlineto{\pgfqpoint{1.311281in}{0.775148in}}%
\pgfpathlineto{\pgfqpoint{1.225010in}{0.719148in}}%
\pgfpathlineto{\pgfqpoint{1.138739in}{0.691148in}}%
\pgfpathlineto{\pgfqpoint{1.052468in}{0.663148in}}%
\pgfpathlineto{\pgfqpoint{0.966197in}{0.663148in}}%
\pgfpathlineto{\pgfqpoint{0.879926in}{0.663148in}}%
\pgfpathlineto{\pgfqpoint{0.793655in}{0.663148in}}%
\pgfpathlineto{\pgfqpoint{0.793655in}{0.663148in}}%
\pgfpathclose%
\pgfusepath{fill}%
\end{pgfscope}%
\begin{pgfscope}%
\pgfsetbuttcap%
\pgfsetroundjoin%
\definecolor{currentfill}{rgb}{0.000000,0.000000,0.000000}%
\pgfsetfillcolor{currentfill}%
\pgfsetlinewidth{0.803000pt}%
\definecolor{currentstroke}{rgb}{0.000000,0.000000,0.000000}%
\pgfsetstrokecolor{currentstroke}%
\pgfsetdash{}{0pt}%
\pgfsys@defobject{currentmarker}{\pgfqpoint{0.000000in}{-0.048611in}}{\pgfqpoint{0.000000in}{0.000000in}}{%
\pgfpathmoveto{\pgfqpoint{0.000000in}{0.000000in}}%
\pgfpathlineto{\pgfqpoint{0.000000in}{-0.048611in}}%
\pgfusepath{stroke,fill}%
}%
\begin{pgfscope}%
\pgfsys@transformshift{0.793655in}{0.523148in}%
\pgfsys@useobject{currentmarker}{}%
\end{pgfscope}%
\end{pgfscope}%
\begin{pgfscope}%
\definecolor{textcolor}{rgb}{0.000000,0.000000,0.000000}%
\pgfsetstrokecolor{textcolor}%
\pgfsetfillcolor{textcolor}%
\pgftext[x=0.793655in,y=0.425926in,,top]{\color{textcolor}\rmfamily\fontsize{11.000000}{13.200000}\selectfont \(\displaystyle {0}\)}%
\end{pgfscope}%
\begin{pgfscope}%
\pgfsetbuttcap%
\pgfsetroundjoin%
\definecolor{currentfill}{rgb}{0.000000,0.000000,0.000000}%
\pgfsetfillcolor{currentfill}%
\pgfsetlinewidth{0.803000pt}%
\definecolor{currentstroke}{rgb}{0.000000,0.000000,0.000000}%
\pgfsetstrokecolor{currentstroke}%
\pgfsetdash{}{0pt}%
\pgfsys@defobject{currentmarker}{\pgfqpoint{0.000000in}{-0.048611in}}{\pgfqpoint{0.000000in}{0.000000in}}{%
\pgfpathmoveto{\pgfqpoint{0.000000in}{0.000000in}}%
\pgfpathlineto{\pgfqpoint{0.000000in}{-0.048611in}}%
\pgfusepath{stroke,fill}%
}%
\begin{pgfscope}%
\pgfsys@transformshift{1.436830in}{0.523148in}%
\pgfsys@useobject{currentmarker}{}%
\end{pgfscope}%
\end{pgfscope}%
\begin{pgfscope}%
\definecolor{textcolor}{rgb}{0.000000,0.000000,0.000000}%
\pgfsetstrokecolor{textcolor}%
\pgfsetfillcolor{textcolor}%
\pgftext[x=1.436830in,y=0.425926in,,top]{\color{textcolor}\rmfamily\fontsize{11.000000}{13.200000}\selectfont \(\displaystyle {2}\)}%
\end{pgfscope}%
\begin{pgfscope}%
\pgfsetbuttcap%
\pgfsetroundjoin%
\definecolor{currentfill}{rgb}{0.000000,0.000000,0.000000}%
\pgfsetfillcolor{currentfill}%
\pgfsetlinewidth{0.803000pt}%
\definecolor{currentstroke}{rgb}{0.000000,0.000000,0.000000}%
\pgfsetstrokecolor{currentstroke}%
\pgfsetdash{}{0pt}%
\pgfsys@defobject{currentmarker}{\pgfqpoint{0.000000in}{-0.048611in}}{\pgfqpoint{0.000000in}{0.000000in}}{%
\pgfpathmoveto{\pgfqpoint{0.000000in}{0.000000in}}%
\pgfpathlineto{\pgfqpoint{0.000000in}{-0.048611in}}%
\pgfusepath{stroke,fill}%
}%
\begin{pgfscope}%
\pgfsys@transformshift{2.080004in}{0.523148in}%
\pgfsys@useobject{currentmarker}{}%
\end{pgfscope}%
\end{pgfscope}%
\begin{pgfscope}%
\definecolor{textcolor}{rgb}{0.000000,0.000000,0.000000}%
\pgfsetstrokecolor{textcolor}%
\pgfsetfillcolor{textcolor}%
\pgftext[x=2.080004in,y=0.425926in,,top]{\color{textcolor}\rmfamily\fontsize{11.000000}{13.200000}\selectfont \(\displaystyle {4}\)}%
\end{pgfscope}%
\begin{pgfscope}%
\pgfsetbuttcap%
\pgfsetroundjoin%
\definecolor{currentfill}{rgb}{0.000000,0.000000,0.000000}%
\pgfsetfillcolor{currentfill}%
\pgfsetlinewidth{0.803000pt}%
\definecolor{currentstroke}{rgb}{0.000000,0.000000,0.000000}%
\pgfsetstrokecolor{currentstroke}%
\pgfsetdash{}{0pt}%
\pgfsys@defobject{currentmarker}{\pgfqpoint{0.000000in}{-0.048611in}}{\pgfqpoint{0.000000in}{0.000000in}}{%
\pgfpathmoveto{\pgfqpoint{0.000000in}{0.000000in}}%
\pgfpathlineto{\pgfqpoint{0.000000in}{-0.048611in}}%
\pgfusepath{stroke,fill}%
}%
\begin{pgfscope}%
\pgfsys@transformshift{2.723179in}{0.523148in}%
\pgfsys@useobject{currentmarker}{}%
\end{pgfscope}%
\end{pgfscope}%
\begin{pgfscope}%
\definecolor{textcolor}{rgb}{0.000000,0.000000,0.000000}%
\pgfsetstrokecolor{textcolor}%
\pgfsetfillcolor{textcolor}%
\pgftext[x=2.723179in,y=0.425926in,,top]{\color{textcolor}\rmfamily\fontsize{11.000000}{13.200000}\selectfont \(\displaystyle {6}\)}%
\end{pgfscope}%
\begin{pgfscope}%
\pgfsetbuttcap%
\pgfsetroundjoin%
\definecolor{currentfill}{rgb}{0.000000,0.000000,0.000000}%
\pgfsetfillcolor{currentfill}%
\pgfsetlinewidth{0.803000pt}%
\definecolor{currentstroke}{rgb}{0.000000,0.000000,0.000000}%
\pgfsetstrokecolor{currentstroke}%
\pgfsetdash{}{0pt}%
\pgfsys@defobject{currentmarker}{\pgfqpoint{0.000000in}{-0.048611in}}{\pgfqpoint{0.000000in}{0.000000in}}{%
\pgfpathmoveto{\pgfqpoint{0.000000in}{0.000000in}}%
\pgfpathlineto{\pgfqpoint{0.000000in}{-0.048611in}}%
\pgfusepath{stroke,fill}%
}%
\begin{pgfscope}%
\pgfsys@transformshift{3.366353in}{0.523148in}%
\pgfsys@useobject{currentmarker}{}%
\end{pgfscope}%
\end{pgfscope}%
\begin{pgfscope}%
\definecolor{textcolor}{rgb}{0.000000,0.000000,0.000000}%
\pgfsetstrokecolor{textcolor}%
\pgfsetfillcolor{textcolor}%
\pgftext[x=3.366353in,y=0.425926in,,top]{\color{textcolor}\rmfamily\fontsize{11.000000}{13.200000}\selectfont \(\displaystyle {8}\)}%
\end{pgfscope}%
\begin{pgfscope}%
\pgfsetbuttcap%
\pgfsetroundjoin%
\definecolor{currentfill}{rgb}{0.000000,0.000000,0.000000}%
\pgfsetfillcolor{currentfill}%
\pgfsetlinewidth{0.803000pt}%
\definecolor{currentstroke}{rgb}{0.000000,0.000000,0.000000}%
\pgfsetstrokecolor{currentstroke}%
\pgfsetdash{}{0pt}%
\pgfsys@defobject{currentmarker}{\pgfqpoint{0.000000in}{-0.048611in}}{\pgfqpoint{0.000000in}{0.000000in}}{%
\pgfpathmoveto{\pgfqpoint{0.000000in}{0.000000in}}%
\pgfpathlineto{\pgfqpoint{0.000000in}{-0.048611in}}%
\pgfusepath{stroke,fill}%
}%
\begin{pgfscope}%
\pgfsys@transformshift{4.009527in}{0.523148in}%
\pgfsys@useobject{currentmarker}{}%
\end{pgfscope}%
\end{pgfscope}%
\begin{pgfscope}%
\definecolor{textcolor}{rgb}{0.000000,0.000000,0.000000}%
\pgfsetstrokecolor{textcolor}%
\pgfsetfillcolor{textcolor}%
\pgftext[x=4.009527in,y=0.425926in,,top]{\color{textcolor}\rmfamily\fontsize{11.000000}{13.200000}\selectfont \(\displaystyle {10}\)}%
\end{pgfscope}%
\begin{pgfscope}%
\pgfsetbuttcap%
\pgfsetroundjoin%
\definecolor{currentfill}{rgb}{0.000000,0.000000,0.000000}%
\pgfsetfillcolor{currentfill}%
\pgfsetlinewidth{0.803000pt}%
\definecolor{currentstroke}{rgb}{0.000000,0.000000,0.000000}%
\pgfsetstrokecolor{currentstroke}%
\pgfsetdash{}{0pt}%
\pgfsys@defobject{currentmarker}{\pgfqpoint{0.000000in}{-0.048611in}}{\pgfqpoint{0.000000in}{0.000000in}}{%
\pgfpathmoveto{\pgfqpoint{0.000000in}{0.000000in}}%
\pgfpathlineto{\pgfqpoint{0.000000in}{-0.048611in}}%
\pgfusepath{stroke,fill}%
}%
\begin{pgfscope}%
\pgfsys@transformshift{4.652702in}{0.523148in}%
\pgfsys@useobject{currentmarker}{}%
\end{pgfscope}%
\end{pgfscope}%
\begin{pgfscope}%
\definecolor{textcolor}{rgb}{0.000000,0.000000,0.000000}%
\pgfsetstrokecolor{textcolor}%
\pgfsetfillcolor{textcolor}%
\pgftext[x=4.652702in,y=0.425926in,,top]{\color{textcolor}\rmfamily\fontsize{11.000000}{13.200000}\selectfont \(\displaystyle {12}\)}%
\end{pgfscope}%
\begin{pgfscope}%
\definecolor{textcolor}{rgb}{0.000000,0.000000,0.000000}%
\pgfsetstrokecolor{textcolor}%
\pgfsetfillcolor{textcolor}%
\pgftext[x=2.907292in,y=0.235185in,,top]{\color{textcolor}\rmfamily\fontsize{18.000000}{13.200000}\selectfont Distortion}%
\end{pgfscope}%
\begin{pgfscope}%
\pgfsetbuttcap%
\pgfsetroundjoin%
\definecolor{currentfill}{rgb}{0.000000,0.000000,0.000000}%
\pgfsetfillcolor{currentfill}%
\pgfsetlinewidth{0.803000pt}%
\definecolor{currentstroke}{rgb}{0.000000,0.000000,0.000000}%
\pgfsetstrokecolor{currentstroke}%
\pgfsetdash{}{0pt}%
\pgfsys@defobject{currentmarker}{\pgfqpoint{-0.048611in}{0.000000in}}{\pgfqpoint{-0.000000in}{0.000000in}}{%
\pgfpathmoveto{\pgfqpoint{-0.000000in}{0.000000in}}%
\pgfpathlineto{\pgfqpoint{-0.048611in}{0.000000in}}%
\pgfusepath{stroke,fill}%
}%
\begin{pgfscope}%
\pgfsys@transformshift{0.582292in}{0.663148in}%
\pgfsys@useobject{currentmarker}{}%
\end{pgfscope}%
\end{pgfscope}%
\begin{pgfscope}%
\definecolor{textcolor}{rgb}{0.000000,0.000000,0.000000}%
\pgfsetstrokecolor{textcolor}%
\pgfsetfillcolor{textcolor}%
\pgftext[x=0.290741in, y=0.610341in, left, base]{\color{textcolor}\rmfamily\fontsize{11.000000}{13.200000}\selectfont \(\displaystyle {0.0}\)}%
\end{pgfscope}%
\begin{pgfscope}%
\pgfsetbuttcap%
\pgfsetroundjoin%
\definecolor{currentfill}{rgb}{0.000000,0.000000,0.000000}%
\pgfsetfillcolor{currentfill}%
\pgfsetlinewidth{0.803000pt}%
\definecolor{currentstroke}{rgb}{0.000000,0.000000,0.000000}%
\pgfsetstrokecolor{currentstroke}%
\pgfsetdash{}{0pt}%
\pgfsys@defobject{currentmarker}{\pgfqpoint{-0.048611in}{0.000000in}}{\pgfqpoint{-0.000000in}{0.000000in}}{%
\pgfpathmoveto{\pgfqpoint{-0.000000in}{0.000000in}}%
\pgfpathlineto{\pgfqpoint{-0.048611in}{0.000000in}}%
\pgfusepath{stroke,fill}%
}%
\begin{pgfscope}%
\pgfsys@transformshift{0.582292in}{1.223148in}%
\pgfsys@useobject{currentmarker}{}%
\end{pgfscope}%
\end{pgfscope}%
\begin{pgfscope}%
\definecolor{textcolor}{rgb}{0.000000,0.000000,0.000000}%
\pgfsetstrokecolor{textcolor}%
\pgfsetfillcolor{textcolor}%
\pgftext[x=0.290741in, y=1.170341in, left, base]{\color{textcolor}\rmfamily\fontsize{11.000000}{13.200000}\selectfont \(\displaystyle {0.2}\)}%
\end{pgfscope}%
\begin{pgfscope}%
\pgfsetbuttcap%
\pgfsetroundjoin%
\definecolor{currentfill}{rgb}{0.000000,0.000000,0.000000}%
\pgfsetfillcolor{currentfill}%
\pgfsetlinewidth{0.803000pt}%
\definecolor{currentstroke}{rgb}{0.000000,0.000000,0.000000}%
\pgfsetstrokecolor{currentstroke}%
\pgfsetdash{}{0pt}%
\pgfsys@defobject{currentmarker}{\pgfqpoint{-0.048611in}{0.000000in}}{\pgfqpoint{-0.000000in}{0.000000in}}{%
\pgfpathmoveto{\pgfqpoint{-0.000000in}{0.000000in}}%
\pgfpathlineto{\pgfqpoint{-0.048611in}{0.000000in}}%
\pgfusepath{stroke,fill}%
}%
\begin{pgfscope}%
\pgfsys@transformshift{0.582292in}{1.783148in}%
\pgfsys@useobject{currentmarker}{}%
\end{pgfscope}%
\end{pgfscope}%
\begin{pgfscope}%
\definecolor{textcolor}{rgb}{0.000000,0.000000,0.000000}%
\pgfsetstrokecolor{textcolor}%
\pgfsetfillcolor{textcolor}%
\pgftext[x=0.290741in, y=1.730341in, left, base]{\color{textcolor}\rmfamily\fontsize{11.000000}{13.200000}\selectfont \(\displaystyle {0.4}\)}%
\end{pgfscope}%
\begin{pgfscope}%
\pgfsetbuttcap%
\pgfsetroundjoin%
\definecolor{currentfill}{rgb}{0.000000,0.000000,0.000000}%
\pgfsetfillcolor{currentfill}%
\pgfsetlinewidth{0.803000pt}%
\definecolor{currentstroke}{rgb}{0.000000,0.000000,0.000000}%
\pgfsetstrokecolor{currentstroke}%
\pgfsetdash{}{0pt}%
\pgfsys@defobject{currentmarker}{\pgfqpoint{-0.048611in}{0.000000in}}{\pgfqpoint{-0.000000in}{0.000000in}}{%
\pgfpathmoveto{\pgfqpoint{-0.000000in}{0.000000in}}%
\pgfpathlineto{\pgfqpoint{-0.048611in}{0.000000in}}%
\pgfusepath{stroke,fill}%
}%
\begin{pgfscope}%
\pgfsys@transformshift{0.582292in}{2.343148in}%
\pgfsys@useobject{currentmarker}{}%
\end{pgfscope}%
\end{pgfscope}%
\begin{pgfscope}%
\definecolor{textcolor}{rgb}{0.000000,0.000000,0.000000}%
\pgfsetstrokecolor{textcolor}%
\pgfsetfillcolor{textcolor}%
\pgftext[x=0.290741in, y=2.290341in, left, base]{\color{textcolor}\rmfamily\fontsize{11.000000}{13.200000}\selectfont \(\displaystyle {0.6}\)}%
\end{pgfscope}%
\begin{pgfscope}%
\pgfsetbuttcap%
\pgfsetroundjoin%
\definecolor{currentfill}{rgb}{0.000000,0.000000,0.000000}%
\pgfsetfillcolor{currentfill}%
\pgfsetlinewidth{0.803000pt}%
\definecolor{currentstroke}{rgb}{0.000000,0.000000,0.000000}%
\pgfsetstrokecolor{currentstroke}%
\pgfsetdash{}{0pt}%
\pgfsys@defobject{currentmarker}{\pgfqpoint{-0.048611in}{0.000000in}}{\pgfqpoint{-0.000000in}{0.000000in}}{%
\pgfpathmoveto{\pgfqpoint{-0.000000in}{0.000000in}}%
\pgfpathlineto{\pgfqpoint{-0.048611in}{0.000000in}}%
\pgfusepath{stroke,fill}%
}%
\begin{pgfscope}%
\pgfsys@transformshift{0.582292in}{2.903148in}%
\pgfsys@useobject{currentmarker}{}%
\end{pgfscope}%
\end{pgfscope}%
\begin{pgfscope}%
\definecolor{textcolor}{rgb}{0.000000,0.000000,0.000000}%
\pgfsetstrokecolor{textcolor}%
\pgfsetfillcolor{textcolor}%
\pgftext[x=0.290741in, y=2.850341in, left, base]{\color{textcolor}\rmfamily\fontsize{11.000000}{13.200000}\selectfont \(\displaystyle {0.8}\)}%
\end{pgfscope}%
\begin{pgfscope}%
\pgfsetbuttcap%
\pgfsetroundjoin%
\definecolor{currentfill}{rgb}{0.000000,0.000000,0.000000}%
\pgfsetfillcolor{currentfill}%
\pgfsetlinewidth{0.803000pt}%
\definecolor{currentstroke}{rgb}{0.000000,0.000000,0.000000}%
\pgfsetstrokecolor{currentstroke}%
\pgfsetdash{}{0pt}%
\pgfsys@defobject{currentmarker}{\pgfqpoint{-0.048611in}{0.000000in}}{\pgfqpoint{-0.000000in}{0.000000in}}{%
\pgfpathmoveto{\pgfqpoint{-0.000000in}{0.000000in}}%
\pgfpathlineto{\pgfqpoint{-0.048611in}{0.000000in}}%
\pgfusepath{stroke,fill}%
}%
\begin{pgfscope}%
\pgfsys@transformshift{0.582292in}{3.463148in}%
\pgfsys@useobject{currentmarker}{}%
\end{pgfscope}%
\end{pgfscope}%
\begin{pgfscope}%
\definecolor{textcolor}{rgb}{0.000000,0.000000,0.000000}%
\pgfsetstrokecolor{textcolor}%
\pgfsetfillcolor{textcolor}%
\pgftext[x=0.290741in, y=3.410341in, left, base]{\color{textcolor}\rmfamily\fontsize{11.000000}{13.200000}\selectfont \(\displaystyle {1.0}\)}%
\end{pgfscope}%
\begin{pgfscope}%
\definecolor{textcolor}{rgb}{0.000000,0.000000,0.000000}%
\pgfsetstrokecolor{textcolor}%
\pgfsetfillcolor{textcolor}%
\pgftext[x=0.235185in,y=2.063148in,,bottom,rotate=90.000000]{\color{textcolor}\rmfamily\fontsize{18.000000}{13.200000}\selectfont Attack Success Rate}%
\end{pgfscope}%
\begin{pgfscope}%
\pgfpathrectangle{\pgfqpoint{0.582292in}{0.523148in}}{\pgfqpoint{4.650000in}{3.080000in}}%
\pgfusepath{clip}%
\pgfsetrectcap%
\pgfsetroundjoin%
\pgfsetlinewidth{1.505625pt}%
\definecolor{currentstroke}{rgb}{0.121569,0.466667,0.705882}%
\pgfsetstrokecolor{currentstroke}%
\pgfsetdash{}{0pt}%
\pgfpathmoveto{\pgfqpoint{0.793655in}{0.663148in}}%
\pgfpathlineto{\pgfqpoint{0.879926in}{0.663148in}}%
\pgfpathlineto{\pgfqpoint{0.966197in}{0.663148in}}%
\pgfpathlineto{\pgfqpoint{1.052468in}{0.663148in}}%
\pgfpathlineto{\pgfqpoint{1.138739in}{0.663148in}}%
\pgfpathlineto{\pgfqpoint{1.225010in}{0.691148in}}%
\pgfpathlineto{\pgfqpoint{1.311281in}{0.691148in}}%
\pgfpathlineto{\pgfqpoint{1.397552in}{0.691148in}}%
\pgfpathlineto{\pgfqpoint{1.483822in}{0.691148in}}%
\pgfpathlineto{\pgfqpoint{1.570093in}{0.719148in}}%
\pgfpathlineto{\pgfqpoint{1.656364in}{0.747148in}}%
\pgfpathlineto{\pgfqpoint{1.742635in}{0.803148in}}%
\pgfpathlineto{\pgfqpoint{1.828906in}{0.859148in}}%
\pgfpathlineto{\pgfqpoint{1.915177in}{0.971148in}}%
\pgfpathlineto{\pgfqpoint{2.001448in}{0.999148in}}%
\pgfpathlineto{\pgfqpoint{2.087719in}{1.027148in}}%
\pgfpathlineto{\pgfqpoint{2.173989in}{1.139148in}}%
\pgfpathlineto{\pgfqpoint{2.260260in}{1.195148in}}%
\pgfpathlineto{\pgfqpoint{2.346531in}{1.251148in}}%
\pgfpathlineto{\pgfqpoint{2.432802in}{1.363148in}}%
\pgfpathlineto{\pgfqpoint{2.519073in}{1.503148in}}%
\pgfpathlineto{\pgfqpoint{2.605344in}{1.587148in}}%
\pgfpathlineto{\pgfqpoint{2.691615in}{1.615148in}}%
\pgfpathlineto{\pgfqpoint{2.777885in}{1.727148in}}%
\pgfpathlineto{\pgfqpoint{2.864156in}{1.867148in}}%
\pgfpathlineto{\pgfqpoint{2.950427in}{1.951148in}}%
\pgfpathlineto{\pgfqpoint{3.036698in}{1.951148in}}%
\pgfpathlineto{\pgfqpoint{3.122969in}{2.007148in}}%
\pgfpathlineto{\pgfqpoint{3.209240in}{2.063148in}}%
\pgfpathlineto{\pgfqpoint{3.295511in}{2.091148in}}%
\pgfpathlineto{\pgfqpoint{3.381782in}{2.119148in}}%
\pgfpathlineto{\pgfqpoint{3.468052in}{2.147148in}}%
\pgfpathlineto{\pgfqpoint{3.554323in}{2.175148in}}%
\pgfpathlineto{\pgfqpoint{3.640594in}{2.203148in}}%
\pgfpathlineto{\pgfqpoint{3.726865in}{2.231148in}}%
\pgfpathlineto{\pgfqpoint{3.813136in}{2.315148in}}%
\pgfpathlineto{\pgfqpoint{3.899407in}{2.399148in}}%
\pgfpathlineto{\pgfqpoint{3.985678in}{2.399148in}}%
\pgfpathlineto{\pgfqpoint{4.071949in}{2.455148in}}%
\pgfpathlineto{\pgfqpoint{4.158219in}{2.567148in}}%
\pgfpathlineto{\pgfqpoint{4.244490in}{2.567148in}}%
\pgfpathlineto{\pgfqpoint{4.330761in}{2.595148in}}%
\pgfpathlineto{\pgfqpoint{4.417032in}{2.623148in}}%
\pgfpathlineto{\pgfqpoint{4.503303in}{2.623148in}}%
\pgfpathlineto{\pgfqpoint{4.589574in}{2.623148in}}%
\pgfpathlineto{\pgfqpoint{4.675845in}{2.623148in}}%
\pgfpathlineto{\pgfqpoint{4.762116in}{2.623148in}}%
\pgfpathlineto{\pgfqpoint{4.848386in}{2.623148in}}%
\pgfpathlineto{\pgfqpoint{4.934657in}{2.623148in}}%
\pgfpathlineto{\pgfqpoint{5.020928in}{2.623148in}}%
\pgfusepath{stroke}%
\end{pgfscope}%
\begin{pgfscope}%
\pgfpathrectangle{\pgfqpoint{0.582292in}{0.523148in}}{\pgfqpoint{4.650000in}{3.080000in}}%
\pgfusepath{clip}%
\pgfsetrectcap%
\pgfsetroundjoin%
\pgfsetlinewidth{1.505625pt}%
\definecolor{currentstroke}{rgb}{1.000000,0.498039,0.054902}%
\pgfsetstrokecolor{currentstroke}%
\pgfsetdash{}{0pt}%
\pgfpathmoveto{\pgfqpoint{0.793655in}{0.663148in}}%
\pgfpathlineto{\pgfqpoint{0.879926in}{0.663148in}}%
\pgfpathlineto{\pgfqpoint{0.966197in}{0.663148in}}%
\pgfpathlineto{\pgfqpoint{1.052468in}{0.663148in}}%
\pgfpathlineto{\pgfqpoint{1.138739in}{0.663148in}}%
\pgfpathlineto{\pgfqpoint{1.225010in}{0.719148in}}%
\pgfpathlineto{\pgfqpoint{1.311281in}{0.747148in}}%
\pgfpathlineto{\pgfqpoint{1.397552in}{0.803148in}}%
\pgfpathlineto{\pgfqpoint{1.483822in}{0.887148in}}%
\pgfpathlineto{\pgfqpoint{1.570093in}{0.999148in}}%
\pgfpathlineto{\pgfqpoint{1.656364in}{1.195148in}}%
\pgfpathlineto{\pgfqpoint{1.742635in}{1.391148in}}%
\pgfpathlineto{\pgfqpoint{1.828906in}{1.559148in}}%
\pgfpathlineto{\pgfqpoint{1.915177in}{1.643148in}}%
\pgfpathlineto{\pgfqpoint{2.001448in}{1.867148in}}%
\pgfpathlineto{\pgfqpoint{2.087719in}{1.979148in}}%
\pgfpathlineto{\pgfqpoint{2.173989in}{2.259148in}}%
\pgfpathlineto{\pgfqpoint{2.260260in}{2.427148in}}%
\pgfpathlineto{\pgfqpoint{2.346531in}{2.595148in}}%
\pgfpathlineto{\pgfqpoint{2.432802in}{2.707148in}}%
\pgfpathlineto{\pgfqpoint{2.519073in}{2.847148in}}%
\pgfpathlineto{\pgfqpoint{2.605344in}{2.903148in}}%
\pgfpathlineto{\pgfqpoint{2.691615in}{3.071148in}}%
\pgfpathlineto{\pgfqpoint{2.777885in}{3.071148in}}%
\pgfpathlineto{\pgfqpoint{2.864156in}{3.099148in}}%
\pgfpathlineto{\pgfqpoint{2.950427in}{3.155148in}}%
\pgfpathlineto{\pgfqpoint{3.036698in}{3.239148in}}%
\pgfpathlineto{\pgfqpoint{3.122969in}{3.295148in}}%
\pgfpathlineto{\pgfqpoint{3.209240in}{3.323148in}}%
\pgfpathlineto{\pgfqpoint{3.295511in}{3.323148in}}%
\pgfpathlineto{\pgfqpoint{3.381782in}{3.323148in}}%
\pgfpathlineto{\pgfqpoint{3.468052in}{3.351148in}}%
\pgfpathlineto{\pgfqpoint{3.554323in}{3.351148in}}%
\pgfpathlineto{\pgfqpoint{3.640594in}{3.351148in}}%
\pgfpathlineto{\pgfqpoint{3.726865in}{3.351148in}}%
\pgfpathlineto{\pgfqpoint{3.813136in}{3.351148in}}%
\pgfpathlineto{\pgfqpoint{3.899407in}{3.379148in}}%
\pgfpathlineto{\pgfqpoint{3.985678in}{3.379148in}}%
\pgfpathlineto{\pgfqpoint{4.071949in}{3.379148in}}%
\pgfpathlineto{\pgfqpoint{4.158219in}{3.379148in}}%
\pgfpathlineto{\pgfqpoint{4.244490in}{3.379148in}}%
\pgfpathlineto{\pgfqpoint{4.330761in}{3.379148in}}%
\pgfpathlineto{\pgfqpoint{4.417032in}{3.379148in}}%
\pgfpathlineto{\pgfqpoint{4.503303in}{3.379148in}}%
\pgfpathlineto{\pgfqpoint{4.589574in}{3.379148in}}%
\pgfpathlineto{\pgfqpoint{4.675845in}{3.379148in}}%
\pgfpathlineto{\pgfqpoint{4.762116in}{3.379148in}}%
\pgfpathlineto{\pgfqpoint{4.848386in}{3.379148in}}%
\pgfpathlineto{\pgfqpoint{4.934657in}{3.379148in}}%
\pgfpathlineto{\pgfqpoint{5.020928in}{3.407148in}}%
\pgfusepath{stroke}%
\end{pgfscope}%
\begin{pgfscope}%
\pgfpathrectangle{\pgfqpoint{0.582292in}{0.523148in}}{\pgfqpoint{4.650000in}{3.080000in}}%
\pgfusepath{clip}%
\pgfsetbuttcap%
\pgfsetroundjoin%
\pgfsetlinewidth{1.505625pt}%
\definecolor{currentstroke}{rgb}{0.172549,0.627451,0.172549}%
\pgfsetstrokecolor{currentstroke}%
\pgfsetdash{{5.550000pt}{2.400000pt}}{0.000000pt}%
\pgfpathmoveto{\pgfqpoint{0.793655in}{0.663148in}}%
\pgfpathlineto{\pgfqpoint{0.879926in}{0.663148in}}%
\pgfpathlineto{\pgfqpoint{0.966197in}{0.663148in}}%
\pgfpathlineto{\pgfqpoint{1.052468in}{0.691148in}}%
\pgfpathlineto{\pgfqpoint{1.138739in}{0.691148in}}%
\pgfpathlineto{\pgfqpoint{1.225010in}{0.691148in}}%
\pgfpathlineto{\pgfqpoint{1.311281in}{0.719148in}}%
\pgfpathlineto{\pgfqpoint{1.397552in}{0.747148in}}%
\pgfpathlineto{\pgfqpoint{1.483822in}{0.803148in}}%
\pgfpathlineto{\pgfqpoint{1.570093in}{0.803148in}}%
\pgfpathlineto{\pgfqpoint{1.656364in}{0.887148in}}%
\pgfpathlineto{\pgfqpoint{1.742635in}{0.915148in}}%
\pgfpathlineto{\pgfqpoint{1.828906in}{0.999148in}}%
\pgfpathlineto{\pgfqpoint{1.915177in}{1.055148in}}%
\pgfpathlineto{\pgfqpoint{2.001448in}{1.111148in}}%
\pgfpathlineto{\pgfqpoint{2.087719in}{1.223148in}}%
\pgfpathlineto{\pgfqpoint{2.173989in}{1.223148in}}%
\pgfpathlineto{\pgfqpoint{2.260260in}{1.279148in}}%
\pgfpathlineto{\pgfqpoint{2.346531in}{1.279148in}}%
\pgfpathlineto{\pgfqpoint{2.432802in}{1.335148in}}%
\pgfpathlineto{\pgfqpoint{2.519073in}{1.391148in}}%
\pgfpathlineto{\pgfqpoint{2.605344in}{1.419148in}}%
\pgfpathlineto{\pgfqpoint{2.691615in}{1.475148in}}%
\pgfpathlineto{\pgfqpoint{2.777885in}{1.503148in}}%
\pgfpathlineto{\pgfqpoint{2.864156in}{1.531148in}}%
\pgfpathlineto{\pgfqpoint{2.950427in}{1.615148in}}%
\pgfpathlineto{\pgfqpoint{3.036698in}{1.643148in}}%
\pgfpathlineto{\pgfqpoint{3.122969in}{1.671148in}}%
\pgfpathlineto{\pgfqpoint{3.209240in}{1.755148in}}%
\pgfpathlineto{\pgfqpoint{3.295511in}{1.755148in}}%
\pgfpathlineto{\pgfqpoint{3.381782in}{1.783148in}}%
\pgfpathlineto{\pgfqpoint{3.468052in}{1.895148in}}%
\pgfpathlineto{\pgfqpoint{3.554323in}{1.923148in}}%
\pgfpathlineto{\pgfqpoint{3.640594in}{1.951148in}}%
\pgfpathlineto{\pgfqpoint{3.726865in}{1.951148in}}%
\pgfpathlineto{\pgfqpoint{3.813136in}{2.007148in}}%
\pgfpathlineto{\pgfqpoint{3.899407in}{2.063148in}}%
\pgfpathlineto{\pgfqpoint{3.985678in}{2.091148in}}%
\pgfpathlineto{\pgfqpoint{4.071949in}{2.119148in}}%
\pgfpathlineto{\pgfqpoint{4.158219in}{2.147148in}}%
\pgfpathlineto{\pgfqpoint{4.244490in}{2.175148in}}%
\pgfpathlineto{\pgfqpoint{4.330761in}{2.203148in}}%
\pgfpathlineto{\pgfqpoint{4.417032in}{2.231148in}}%
\pgfpathlineto{\pgfqpoint{4.503303in}{2.259148in}}%
\pgfpathlineto{\pgfqpoint{4.589574in}{2.259148in}}%
\pgfpathlineto{\pgfqpoint{4.675845in}{2.287148in}}%
\pgfpathlineto{\pgfqpoint{4.762116in}{2.343148in}}%
\pgfpathlineto{\pgfqpoint{4.848386in}{2.343148in}}%
\pgfpathlineto{\pgfqpoint{4.934657in}{2.427148in}}%
\pgfpathlineto{\pgfqpoint{5.020928in}{2.455148in}}%
\pgfusepath{stroke}%
\end{pgfscope}%
\begin{pgfscope}%
\pgfpathrectangle{\pgfqpoint{0.582292in}{0.523148in}}{\pgfqpoint{4.650000in}{3.080000in}}%
\pgfusepath{clip}%
\pgfsetbuttcap%
\pgfsetroundjoin%
\pgfsetlinewidth{1.505625pt}%
\definecolor{currentstroke}{rgb}{0.839216,0.152941,0.156863}%
\pgfsetstrokecolor{currentstroke}%
\pgfsetdash{{5.550000pt}{2.400000pt}}{0.000000pt}%
\pgfpathmoveto{\pgfqpoint{0.793655in}{0.663148in}}%
\pgfpathlineto{\pgfqpoint{0.879926in}{2.623148in}}%
\pgfpathlineto{\pgfqpoint{0.966197in}{3.239148in}}%
\pgfpathlineto{\pgfqpoint{1.052468in}{3.435148in}}%
\pgfpathlineto{\pgfqpoint{1.138739in}{3.463148in}}%
\pgfpathlineto{\pgfqpoint{1.225010in}{3.463148in}}%
\pgfpathlineto{\pgfqpoint{1.311281in}{3.463148in}}%
\pgfpathlineto{\pgfqpoint{1.397552in}{3.463148in}}%
\pgfpathlineto{\pgfqpoint{1.483822in}{3.463148in}}%
\pgfpathlineto{\pgfqpoint{1.570093in}{3.463148in}}%
\pgfpathlineto{\pgfqpoint{1.656364in}{3.463148in}}%
\pgfpathlineto{\pgfqpoint{1.742635in}{3.463148in}}%
\pgfpathlineto{\pgfqpoint{1.828906in}{3.463148in}}%
\pgfpathlineto{\pgfqpoint{1.915177in}{3.463148in}}%
\pgfpathlineto{\pgfqpoint{2.001448in}{3.463148in}}%
\pgfpathlineto{\pgfqpoint{2.087719in}{3.463148in}}%
\pgfpathlineto{\pgfqpoint{2.173989in}{3.463148in}}%
\pgfpathlineto{\pgfqpoint{2.260260in}{3.463148in}}%
\pgfpathlineto{\pgfqpoint{2.346531in}{3.463148in}}%
\pgfpathlineto{\pgfqpoint{2.432802in}{3.463148in}}%
\pgfpathlineto{\pgfqpoint{2.519073in}{3.463148in}}%
\pgfpathlineto{\pgfqpoint{2.605344in}{3.463148in}}%
\pgfpathlineto{\pgfqpoint{2.691615in}{3.463148in}}%
\pgfpathlineto{\pgfqpoint{2.777885in}{3.463148in}}%
\pgfpathlineto{\pgfqpoint{2.864156in}{3.463148in}}%
\pgfpathlineto{\pgfqpoint{2.950427in}{3.463148in}}%
\pgfpathlineto{\pgfqpoint{3.036698in}{3.463148in}}%
\pgfpathlineto{\pgfqpoint{3.122969in}{3.463148in}}%
\pgfpathlineto{\pgfqpoint{3.209240in}{3.463148in}}%
\pgfpathlineto{\pgfqpoint{3.295511in}{3.463148in}}%
\pgfpathlineto{\pgfqpoint{3.381782in}{3.463148in}}%
\pgfpathlineto{\pgfqpoint{3.468052in}{3.463148in}}%
\pgfpathlineto{\pgfqpoint{3.554323in}{3.463148in}}%
\pgfpathlineto{\pgfqpoint{3.640594in}{3.463148in}}%
\pgfpathlineto{\pgfqpoint{3.726865in}{3.463148in}}%
\pgfpathlineto{\pgfqpoint{3.813136in}{3.463148in}}%
\pgfpathlineto{\pgfqpoint{3.899407in}{3.463148in}}%
\pgfpathlineto{\pgfqpoint{3.985678in}{3.463148in}}%
\pgfpathlineto{\pgfqpoint{4.071949in}{3.463148in}}%
\pgfpathlineto{\pgfqpoint{4.158219in}{3.463148in}}%
\pgfpathlineto{\pgfqpoint{4.244490in}{3.463148in}}%
\pgfpathlineto{\pgfqpoint{4.330761in}{3.463148in}}%
\pgfpathlineto{\pgfqpoint{4.417032in}{3.463148in}}%
\pgfpathlineto{\pgfqpoint{4.503303in}{3.463148in}}%
\pgfpathlineto{\pgfqpoint{4.589574in}{3.463148in}}%
\pgfpathlineto{\pgfqpoint{4.675845in}{3.463148in}}%
\pgfpathlineto{\pgfqpoint{4.762116in}{3.463148in}}%
\pgfpathlineto{\pgfqpoint{4.848386in}{3.463148in}}%
\pgfpathlineto{\pgfqpoint{4.934657in}{3.463148in}}%
\pgfpathlineto{\pgfqpoint{5.020928in}{3.463148in}}%
\pgfusepath{stroke}%
\end{pgfscope}%
\begin{pgfscope}%
\pgfsetrectcap%
\pgfsetmiterjoin%
\pgfsetlinewidth{0.803000pt}%
\definecolor{currentstroke}{rgb}{0.000000,0.000000,0.000000}%
\pgfsetstrokecolor{currentstroke}%
\pgfsetdash{}{0pt}%
\pgfpathmoveto{\pgfqpoint{0.582292in}{0.523148in}}%
\pgfpathlineto{\pgfqpoint{0.582292in}{3.603148in}}%
\pgfusepath{stroke}%
\end{pgfscope}%
\begin{pgfscope}%
\pgfsetrectcap%
\pgfsetmiterjoin%
\pgfsetlinewidth{0.803000pt}%
\definecolor{currentstroke}{rgb}{0.000000,0.000000,0.000000}%
\pgfsetstrokecolor{currentstroke}%
\pgfsetdash{}{0pt}%
\pgfpathmoveto{\pgfqpoint{5.232292in}{0.523148in}}%
\pgfpathlineto{\pgfqpoint{5.232292in}{3.603148in}}%
\pgfusepath{stroke}%
\end{pgfscope}%
\begin{pgfscope}%
\pgfsetrectcap%
\pgfsetmiterjoin%
\pgfsetlinewidth{0.803000pt}%
\definecolor{currentstroke}{rgb}{0.000000,0.000000,0.000000}%
\pgfsetstrokecolor{currentstroke}%
\pgfsetdash{}{0pt}%
\pgfpathmoveto{\pgfqpoint{0.582292in}{0.523148in}}%
\pgfpathlineto{\pgfqpoint{5.232292in}{0.523148in}}%
\pgfusepath{stroke}%
\end{pgfscope}%
\begin{pgfscope}%
\pgfsetrectcap%
\pgfsetmiterjoin%
\pgfsetlinewidth{0.803000pt}%
\definecolor{currentstroke}{rgb}{0.000000,0.000000,0.000000}%
\pgfsetstrokecolor{currentstroke}%
\pgfsetdash{}{0pt}%
\pgfpathmoveto{\pgfqpoint{0.582292in}{3.603148in}}%
\pgfpathlineto{\pgfqpoint{5.232292in}{3.603148in}}%
\pgfusepath{stroke}%
\end{pgfscope}%
\begin{pgfscope}%
\pgfsetbuttcap%
\pgfsetmiterjoin%
\definecolor{currentfill}{rgb}{1.000000,1.000000,1.000000}%
\pgfsetfillcolor{currentfill}%
\pgfsetfillopacity{0.800000}%
\pgfsetlinewidth{1.003750pt}%
\definecolor{currentstroke}{rgb}{0.800000,0.800000,0.800000}%
\pgfsetstrokecolor{currentstroke}%
\pgfsetstrokeopacity{0.800000}%
\pgfsetdash{}{0pt}%
\pgfpathmoveto{\pgfqpoint{2.593893in}{0.599537in}}%
\pgfpathlineto{\pgfqpoint{5.125347in}{0.599537in}}%
\pgfpathquadraticcurveto{\pgfqpoint{5.155903in}{0.599537in}}{\pgfqpoint{5.155903in}{0.630092in}}%
\pgfpathlineto{\pgfqpoint{5.155903in}{1.466435in}}%
\pgfpathquadraticcurveto{\pgfqpoint{5.155903in}{1.496991in}}{\pgfqpoint{5.125347in}{1.496991in}}%
\pgfpathlineto{\pgfqpoint{2.593893in}{1.496991in}}%
\pgfpathquadraticcurveto{\pgfqpoint{2.563338in}{1.496991in}}{\pgfqpoint{2.563338in}{1.466435in}}%
\pgfpathlineto{\pgfqpoint{2.563338in}{0.630092in}}%
\pgfpathquadraticcurveto{\pgfqpoint{2.563338in}{0.599537in}}{\pgfqpoint{2.593893in}{0.599537in}}%
\pgfpathlineto{\pgfqpoint{2.593893in}{0.599537in}}%
\pgfpathclose%
\pgfusepath{stroke,fill}%
\end{pgfscope}%
\begin{pgfscope}%
\pgfsetrectcap%
\pgfsetroundjoin%
\pgfsetlinewidth{1.505625pt}%
\definecolor{currentstroke}{rgb}{0.121569,0.466667,0.705882}%
\pgfsetstrokecolor{currentstroke}%
\pgfsetdash{}{0pt}%
\pgfpathmoveto{\pgfqpoint{2.624449in}{1.382408in}}%
\pgfpathlineto{\pgfqpoint{2.777227in}{1.382408in}}%
\pgfpathlineto{\pgfqpoint{2.930004in}{1.382408in}}%
\pgfusepath{stroke}%
\end{pgfscope}%
\begin{pgfscope}%
\definecolor{textcolor}{rgb}{0.000000,0.000000,0.000000}%
\pgfsetstrokecolor{textcolor}%
\pgfsetfillcolor{textcolor}%
\pgftext[x=3.052227in,y=1.328935in,left,base]{\color{textcolor}\rmfamily\fontsize{11.000000}{13.200000}\selectfont Transferable attack on average}%
\end{pgfscope}%
\begin{pgfscope}%
\pgfsetrectcap%
\pgfsetroundjoin%
\pgfsetlinewidth{1.505625pt}%
\definecolor{currentstroke}{rgb}{1.000000,0.498039,0.054902}%
\pgfsetstrokecolor{currentstroke}%
\pgfsetdash{}{0pt}%
\pgfpathmoveto{\pgfqpoint{2.624449in}{1.169502in}}%
\pgfpathlineto{\pgfqpoint{2.777227in}{1.169502in}}%
\pgfpathlineto{\pgfqpoint{2.930004in}{1.169502in}}%
\pgfusepath{stroke}%
\end{pgfscope}%
\begin{pgfscope}%
\definecolor{textcolor}{rgb}{0.000000,0.000000,0.000000}%
\pgfsetstrokecolor{textcolor}%
\pgfsetfillcolor{textcolor}%
\pgftext[x=3.052227in,y=1.116030in,left,base]{\color{textcolor}\rmfamily\fontsize{11.000000}{13.200000}\selectfont Transferable attack with \fit}%
\end{pgfscope}%
\begin{pgfscope}%
\pgfsetbuttcap%
\pgfsetroundjoin%
\pgfsetlinewidth{1.505625pt}%
\definecolor{currentstroke}{rgb}{0.172549,0.627451,0.172549}%
\pgfsetstrokecolor{currentstroke}%
\pgfsetdash{{5.550000pt}{2.400000pt}}{0.000000pt}%
\pgfpathmoveto{\pgfqpoint{2.624449in}{0.956597in}}%
\pgfpathlineto{\pgfqpoint{2.777227in}{0.956597in}}%
\pgfpathlineto{\pgfqpoint{2.930004in}{0.956597in}}%
\pgfusepath{stroke}%
\end{pgfscope}%
\begin{pgfscope}%
\definecolor{textcolor}{rgb}{0.000000,0.000000,0.000000}%
\pgfsetstrokecolor{textcolor}%
\pgfsetfillcolor{textcolor}%
\pgftext[x=3.052227in,y=0.903125in,left,base]{\color{textcolor}\rmfamily\fontsize{11.000000}{13.200000}\selectfont Black box attack}%
\end{pgfscope}%
\begin{pgfscope}%
\pgfsetbuttcap%
\pgfsetroundjoin%
\pgfsetlinewidth{1.505625pt}%
\definecolor{currentstroke}{rgb}{0.839216,0.152941,0.156863}%
\pgfsetstrokecolor{currentstroke}%
\pgfsetdash{{5.550000pt}{2.400000pt}}{0.000000pt}%
\pgfpathmoveto{\pgfqpoint{2.624449in}{0.743692in}}%
\pgfpathlineto{\pgfqpoint{2.777227in}{0.743692in}}%
\pgfpathlineto{\pgfqpoint{2.930004in}{0.743692in}}%
\pgfusepath{stroke}%
\end{pgfscope}%
\begin{pgfscope}%
\definecolor{textcolor}{rgb}{0.000000,0.000000,0.000000}%
\pgfsetstrokecolor{textcolor}%
\pgfsetfillcolor{textcolor}%
\pgftext[x=3.052227in,y=0.690220in,left,base]{\color{textcolor}\rmfamily\fontsize{11.000000}{13.200000}\selectfont White box attack}%
\end{pgfscope}%
\end{pgfpicture}%
\makeatother%
\endgroup%

%% file: sections/related_work.tex
\section{Related work}

\subsection{White box, black box, and transferable attacks}
\label{sec:RelatedAttacks}
The threat analysis related to adversarial examples usually considers two scenarios: the white box assumption where the attacker knows the internals of the target model, and the black box assumption where he does not but has limited access to the target.

White box attacks are now performing very well in terms of Attack Success Rate and distortion. We distinguish two kinds of attacks: 1) attacks constrained by a distortion budget like \pgd whose performance is measured by the ASR, and 2) attacks yielding almost surely an adversarial example like \deepfool and \cw whose performance is measured by the distortion.
The trend is to make them speed efficient as well with fewer computations of the gradient of the neural network, like~\nmf and~\bp.

Our paper also considers so-called decision-based black box attacks like \geoda, \surfree, and \rays, where the attacker only sees the class predicted by the model, but not the confidence score, neither the probit nor the logit. A decision-based black box attack typically needs thousands of queries to find an adversarial example with a relatively low distortion, yet, much higher than with a white box attack.
This is the price the attacker has to pay when he does not know the target model.

The transferable attacks, which pertain to the black box scenario, disrupt this last statement.
The assumption is that attacker knows another model, called source, trained for the same classification problem as the target. Yet, leading a white box attack on the source model with the hope that this adversarial example will also fool the target yields poor results because the adversarial perturbation is too specific to the source. 
Modern transferable attacks increase the ASR by avoiding the overfitting of examples on the source. 
Input transformations have been purposed by \di, \ti and \admix.
Papers~\cite{wang2021boosting, linnesterov, huang2022direction} stabilized the gradient while~\cite{Wang_2021_ICCV, huang2019enhancing, zhang2022improving, zhou2018transferable} focus on the importance of intermediate features.
The combination of several methods has also been investigated~\cite{zou2020improving, huang2022transferable}.
For instance, \taig uses input transformation, integrated gradients, and attention reduction.

Ensemble-model attacks make the assumption that the attacker leverages not a single source but several source models.
Paper~\cite{wang2021boosting} averages the logits of several sources and leads a white box attack on this aggregation of models, whereas~\cite{xiong2022stochastic} reduces model variance.
The main obstacle to these methods is the computational complexity.
To counteract this drawback, some methods design a specific ensemble of sources like ghost networks~\cite{li2020learning} or pruned network~\cite{wang2022enhancing}. This amounts to creating several sources without the need to compute the gradient of each model.

All this literature uses the ASR as the figure of merit for a given distortion while our paper draws the full operating characteristic of ASR vs. distortion. Moreover, the performance of the transferability is compared neither to the white box nor the  decision-based black box attack.

\subsection{Fingerprinting}
\label{sec:Fingerprinting}
Models are valuable assets because of their training costs, be it the expertise, the annotated data, or the computational power. 
Fingerprinting methods have been proposed to protect this intellectual property. 
They provide a similarity score between the two models by looking at similar decisions in submitted queries.
They often use adversarial examples specifically designed to be unique for a model. The models are considered similar if they share the same decision for these inputs.
\ipguard creates adversarial examples close to the decision boundary to frame it. 
Universal adversarial perturbations are used to characterize the boundary in \cite{https://doi.org/10.48550/arxiv.2202.08602}.
The paper~\cite{lukasdeep} built adversarial examples with a transferable property called conferrable adversarial examples.
\afa crafts adversarial examples model-specific using dropout.
\fbi is the only method using unmodified images. 
It estimates the mutual information to measure the statistical dependence of two models' predictions.  

Our paper investigates the idea that a model similar to the target may be a good source for transferable attacks.

%% file: sections/preliminaries.tex
\section{Methodology}
\label{sec:Methodology}
This paper introduces a new way to gauge transferability.
The main motivation is that transferability should indicate whether a source $s$ is useful for attacking a target $t$, independently from the inherent robustness of $t$.

\subsection{Notations}
Let $f: [0,255]^N \rightarrow \mathcal{R}^K$ denote a classifier computing the predicted probabilities $f(x)$ of $K$ classes for input $x$.
For a given input $x$ of class $y$, the adversarial example $x_a$ is the result of an attack
(be it white box, black box, or by transferability) such that
\begin{equation}
    \arg \max_{1\leq k\leq K} f_k(x_a) \neq \arg \max_{1\leq k\leq K} f_k(x) = y.
\end{equation}


In the case of transferable attacks, the attacker uses a source model $f_s$ to build a transferable adversarial example against the target model $f_t$.
This paper considers a collection of $m$ source models denoted by $\mathcal{F}_s = \{f_s^1, ..., f_s^m\}$, and a set of $n$ inputs $\mathcal{X}$.


\subsection{Measurement}
\label{sec:measure}

\textbf{Distortion.} 
We measure the distortion of an adversarial perturbation using the Euclidean norm which is a common choice as we deal with natural images:
\begin{equation}
    \dist(x_a,x) := \|x_a - x\|_2/\sqrt{N},
\end{equation}
with $N = 3\times H\times L$ pixels. This distortion can be seen as the average amplitude of the perturbation per pixel.

\textbf{Operating characteristic.}
Given an attack (be it white box, black box, or by transferability) and a target model $f_t$, we call operating characteristic the following function:
\begin{equation}
    P(D) := \Prob(\dist(x_a,x) < D).
    \label{eq:OpCharac}
\end{equation}
It is thus the Attack Success Rate as a function of the distortion. Contrary to the literature which measures the ASR for a few distortion levels, we consider the full range of $D$ values.

\subsection{Transferability}
Given a white box and a black box attacks, The proposed methodology first computes the operating characteristic  $\PWB_t(D)$ of a state-of-the-art white box attack directly applied to the target model, and the operating characteristic  $\PBB_t(D)$ of a state-of-the-art black box attack. 
It then measures where the operating characteristic of the transferable attack from the source $s$ to the target $t$ lies in between the two characteristics as follows: 
\begin{equation}
    \label{eq:transferable_gain}
    T_{s, t} := \frac{\int_{0}^{\infty} P_{s, t}(u) - \PBB_t(u) du}{\int_{0}^{\infty} \PWB_t(u) - \PBB_t(u) du}.
\end{equation}

The proposed score is calculated as the ratio of the areas between the different operational characteristics, which are defined as Cumulative Distribution Functions (CDFs). The numerator is therefore close to the 1-Wasserstein distance~\cite{wassersteinarea} between the ASR function of the distortions of the transferable attack and that of the black box attack, where the absolute value is removed to obtain a signed score.
If the transferable attack performs as well as the white box attack (resp. as bad as the black box attack), then $T_{s,t} =1$ (resp. $T_{s,t} =0$). \autoref{fig:splash} shows that the numerator of~\eqref{eq:transferable_gain} can indeed be negative as transferability can be even worse than the black box if the source $s$ is not well chosen. Therefore, zero is not a lower bound for $T_{s,t}$. In the denominator, the 1-Wasserstein distance is obtained exactly between the white box and the black box because the white box consistently generates adversarial examples with lower distortion.

\subsection{Practical implementation}
\textbf{Attacks.} We only consider state-of-the-art attacks which almost surely deliver an adversarial example. As said in Sect.~\ref{sec:RelatedAttacks}, this is usually the case for black box attacks, but we are limited to using white box attacks that are \emph{not} subject to a distortion budget. 
As for transferability, the attack uses the publicly available model $s$ and input $x$ to craft an adversarial direction $\dir_{x,s}$ of Euclidean norm $\sqrt{N}$. The use of $\sqrt{N}$ for the norm is for the sake of simplicity in notation. We assume that there is an oracle giving the minimal distortion along this direction to fool the target $t$. In other words, $x_a = x + d\;\dir_{x,s}$, with
\begin{equation}
\label{eq:MinD}
    d = \min\{\delta: \arg \max_k f_{t,k}(x+\delta\;\dir_{x,s})\neq y\}.
\end{equation}
Note that $\dist(x_a,x) = d$.
This definition favours transferability as its best. 
In practice, such an oracle does not exist but the attacker finds a good estimate of~\eqref{eq:MinD} thanks to a line search within a few queries to the target. 

\textbf{Transferability.} We run a given attack over a collection $\mathcal{X}$ of $n$ inputs correctly classified by the target model.
We compute the distortions $d(j) = \dist(x_{a,j},x_j)$ with $x_j\in\mathcal{X}$ and sort them so that $d(1)\leq d(2)\leq \ldots \leq d(n)$. We set $d(0)=0$ and $d(n+1) = \infty$ to properly define the empirical operating characteristic by the following step function:
\begin{equation}
    \hat{P}(D) := j/n \quad \forall D\in [d(j),d(j+1)).
    \label{eq:EmpiricalOpCar}
\end{equation}
It is then easier to estimate the integrals appearing in~\eqref{eq:transferable_gain}, which are indeed areas under two curves, by a Lebesgue sum rather than a Riemann sum. This gives:
\begin{equation}
    \hat{T}_{s, t} := \frac{\sum_{j=1}^n d_{s, t}(j) - \dBB_t(j)}{\sum_{j=1}^n \dWB_t(j) - \dBB_t(j)},
    \label{eq:TScoreEmpirical}
\end{equation}
where the distortions $(\dBB_t(j))_j$ (resp. $(\dWB_t(j))_j$) resulting from the black box (resp. white box) attack against model $t$ are also sorted in increasing order.

%% file: sections/transferability.tex
\section{Triad of transferability: data, model, attack}
This section is an experimental investigation of the factors affecting transferability.

\label{sec:Transferability}
\subsection{Experimental setup}
\label{sec:experimental_setup}

The study assesses transferable attacks on a total of 48 models, with 47 of them sourced from the Timm library~\cite{rw2019timm} and one very robust, namely \texttt{ResNet50}$_\texttt{AdvTrain}$, obtained from the GitHub repository~\footnote{\url{https://github.com/MadryLab/robustness}}.
Appendix~\ref{app:setup} lists all the models. The experiments utilize 100 images from the validation set of ILSVRC'12, all of which are correctly classified by all models under consideration.

The study considers three transferable attacks - \di, \taig, and \dwp - each using a different approach to improve transferability (see Sect.~\ref{sec:RelatedAttacks}). These attacks are selected as the best in their categories in~\cite{hanweitransferability}.
They all share an $\epsilon$ parameter to control the maximum perturbation added per pixel. The effect of the parameter $\epsilon$ indeed happens to be negligible in our protocol, as demonstrated in Appendix~\ref{app:epsilon_parameter}. We choose $\epsilon=8$.

To measure transferability~\eqref{eq:TScoreEmpirical}, we need state-of-the-art black box and white box attacks. Certain methods may exhibit a preference towards one model over the other, necessitating the use of multiple attacks. Our study employs four white box attacks (\bp, \deepfool, \ifgsm, and \pgd) and three black box attacks (\surfree, \rays, and \geoda).
All black box attacks are run with 2,000 queries, which has been determined to be sufficient for achieving convergence. 
For distortion-constrained white box attack, the $\epsilon$ parameter is set to 4.
We record the smallest perturbation distortion over the black box (white box) for each image and draw the operating characteristic~\eqref{eq:EmpiricalOpCar} as appearing in green (resp. red) dashed line in Fig.~\ref{fig:splash}.

As for the transferable attack, for a given target model, the attacker has access to a subset of all the other models whose architecture differs from the target. This amounts to an average of 45 models out of 48.
For instance, in Fig.~\ref{fig:splash}, $\texttt{CoatLite}_{\texttt{small}}$ being the target, we exclude all other $\texttt{CoatLite}$ models from being a source. The light blue area delimits the operating characteristics of transferable attack~\di using as the source one of the 45 remaining models.



\subsection{Model dependence}
\label{sec:model_dependence}



\begin{figure}[ht]
        \centering
        \resizebox{\linewidth}{!}{\input{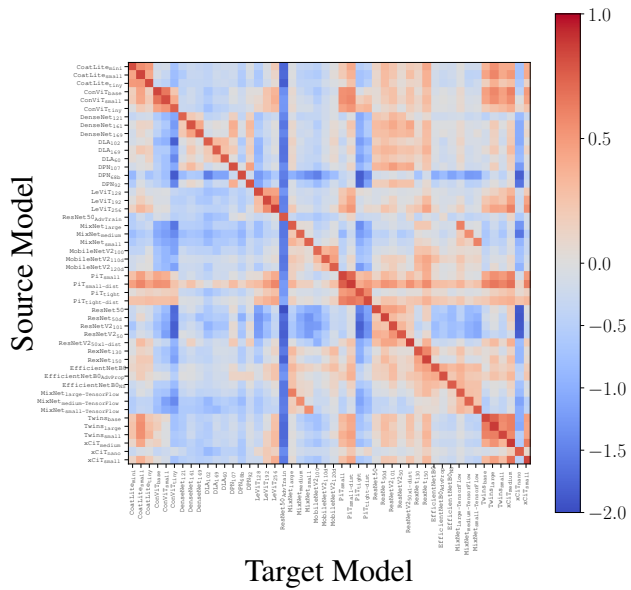}}
    \caption{Transferability score \measure matrix of 48 sources and 48 targets listed in App.~\ref{app:setup} with attack \di.}
    \label{fig:gain_score_matrix_di}
\end{figure}

\def\pp{0.8}
\begin{figure}[bt]
    \centering
    \resizebox{\linewidth}{!}{\input{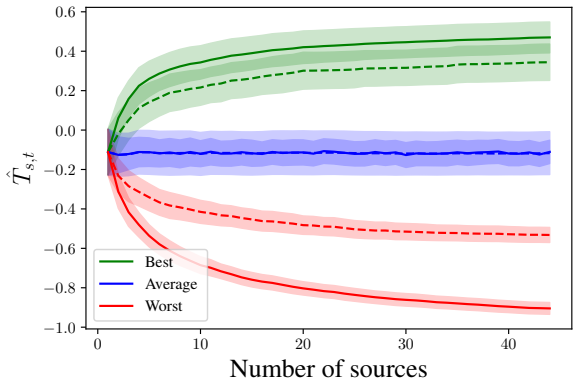}}
    \caption{\measure function of the number of available sources and attack~\di. The best or worst model selected per image (solid line) or on average (dashed line).}
    \label{fig:model_dependence_n_sources}
\end{figure}

\textbf{Large transferability variation.}
\autoref{fig:gain_score_matrix_di} shows the matrix \measure, where $s$ and $t$ are any of the $48^2$ pairs, for attack~\di.
Not all models possess the same transferability capabilities.
At first look, the figure contains more blue than red cells which means that transferability takes negative value more often.  

Some rare models, like \texttt{PiT}$_\texttt{small-dist}$, exhibit good transferability towards any target. On the contrary, \texttt{DPN68b} is always a bad source.
On the other hand, \texttt{ResNet50}$_\texttt{AdvTrain}$ is a very difficult target (note however that its accuracy is low), followed by \texttt{PiT}$_\texttt{light}$ and \texttt{ConViT}$_\texttt{tiny}$, whereas the family of \texttt{ResNet} are fairly easy target.

Prior works have shown that models with architectures similar to the target transfer better~\cite{wu2020towards, petrov2019measuring}. The red squares appearing in the matrix confirm this (models are ordered according to architecture family).
For instance, \texttt{MobileNetV2} and \texttt{RexNet} architectures exhibit transferability close to 0.7. However, this is not an absolute truth. 
For example, \texttt{EfficientNet-B0} has better transferability against  \texttt{MixNet}$_\texttt{large}$ than \texttt{MixNet}$_\texttt{medium}$, even though they have similar architectures.
\texttt{ConViT} transfers exceptionally well to \texttt{Twins} although their architectures are very different. Similar observations hold for the other attacks but with lower transferability scores (see Appendix~\ref{app:model_dependence_matrix}).

From now on, we exclude models whose architecture is similar to the target.

\textbf{Impact on the attack.}
Selecting the right source among several available models is critical for achieving high transferability.
\autoref{fig:model_dependence_n_sources} displays the transferability score~\eqref{eq:TScoreEmpirical} for the best and the worst choices (shown in dotted lines) as the number of source candidates increases. In this evaluation, a single model is selected as the source for building all adversarial examples. On average, the transferability is lower than zero, regardless of the attack method. This means that transferable attacks perform worst than black box attacks on average.
If the attacker knows how to select the best source model, \measure quickly converges to a maximum which is positive but below 0.5. This means that transferability at its best performs closer to a black box attack rather than a white box attack. These remarks highly mitigate the threat of transferability and put emphasis on the crucial selection of the best source model. As far as we know, these facts are not reported in the adversarial examples literature.


\subsection{Image dependence}
\label{sec:image_dependence}

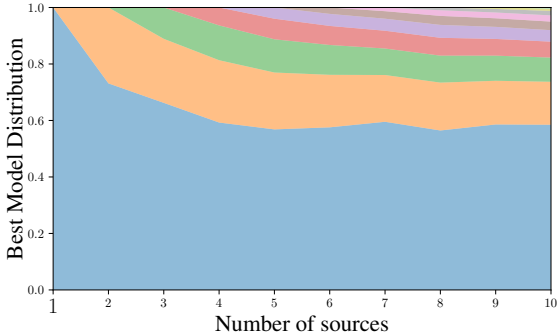
\begin{figure}[bt]
    \centering
    \resizebox{0.9\linewidth}{!}{\input{./images/image_dependences/distribution/Epsilon=8-Attack=DI2FGSM.pgf}}
    \caption{Distribution of best image-model selection function of the number of sources. Adversarial examples obtained with \di.}
    \label{fig:distribution_best_model}
\end{figure}

The difficulty of transferring adversarial examples from a source to a target varies significantly depending on the input $x$. This is illustrated in Fig.~\ref{fig:distribution_best_model}, which shows the distribution of the best-performing model for each image based on the available sources. Even with a large number of sources available, there is typically one source that performs significantly better than the others, but only over 60\% of the images on average. However, this superiority decreases rapidly as the number of available sources increases. It means that a better source exists for 40\% of the images on average.

Supposing that the attacker knows the best source of each input, \autoref{fig:model_dependence_n_sources} shows in solid line that the transferability converges to a value close to 0.5. The performance of the transferable attack lies halfway between the ones of the white box and black box attacks. This highlights the importance of selecting an appropriate source model for a given target and input.


\subsection{Attack dependence}
\label{sec:attack_dependence}

\def\pp{0.45}
\begin{figure}[bt]
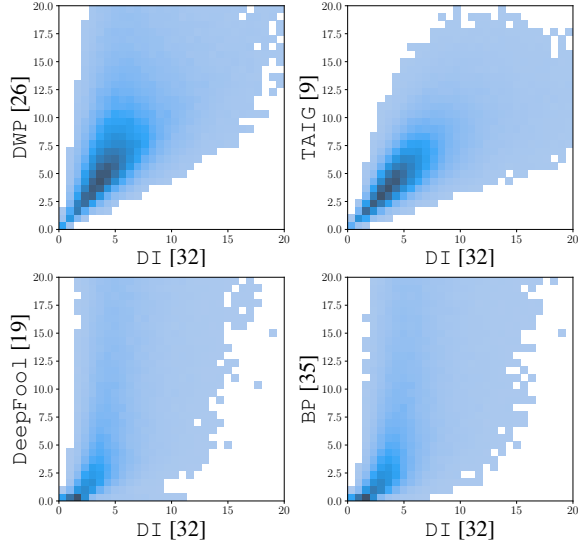

    \centering
    \begin{subfigure}[b]{\pp\linewidth}
        \centering
        \resizebox{\linewidth}{!}{\input{./images/attack_dependences/Attack1=DI2FGSM-Attack2=dwp.pgf}}
    \end{subfigure}
    \begin{subfigure}[b]{\pp\linewidth}
        \centering
        \resizebox{\linewidth}{!}{\input{./images/attack_dependences/Attack1=DI2FGSM-Attack2=taig.pgf}}
    \end{subfigure}
    
    \begin{subfigure}[b]{\pp\linewidth}
        \centering
        \resizebox{\linewidth}{!}{\input{./images/attack_dependences/Attack1=DI2FGSM-Attack2=deepfool.pgf}}
    \end{subfigure}
    \begin{subfigure}[b]{\pp\linewidth}
        \centering
        \resizebox{\linewidth}{!}{\input{./images/attack_dependences/Attack1=DI2FGSM-Attack2=bp.pgf}}
    \end{subfigure}
    \caption{2D Histogram of the minimum distortion for transferable and white box attacks.}
    \label{fig:histogram_model_dependence}
\end{figure}

\textbf{Transferable attack.}
\autoref{fig:histogram_model_dependence} compares transferable attacks with a 2D histogram of the distortion pair for two attacks.
This is computed over all inputs in $\mathcal{X}$ and all pairs of source and target models.
The results show that \dwp exhibits poor transferability compared to \di and \taig, which produce adversarial perturbations with similar distortion. Additionally, regardless of the attack complexity or method, the challenging images remain consistent. If one attack requires a high distortion for a given input / source / target, other attacks are likely to encounter similar difficulties.


\textbf{Traditional white box attack.}
We now compare the methods designed for transferability with a naive approach.
A white box attack (ie. not specific to transferability) is executed on the source, and then the direction found serves as $\dir_{x,s}$ to place the adversarial examples on the target boundary with~\eqref{eq:MinD}. 
The best \measure score is 0.27 for \bp compared to 0.52 for \di.
This confirms the superiority of the recent methods designed for transferability.
However, even though \di is on average better, \autoref{fig:histogram_model_dependence} shows that when the necessary distortion is small, traditional white box attacks like \deepfool and \bp indeed beats \di. On the other hand, \di performs much better for inputs requiring more distortion.


%% file: images/image_dependences/distribution/Epsilon=8-Attack=DI2FGSM.pgf
\begingroup%
\makeatletter%
\begin{pgfpicture}%
\pgfpathrectangle{\pgfpointorigin}{\pgfqpoint{6.148149in}{3.727916in}}%
\pgfusepath{use as bounding box}%
\begin{pgfscope}%
\pgfsetbuttcap%
\pgfsetmiterjoin%
\definecolor{currentfill}{rgb}{1.000000,1.000000,1.000000}%
\pgfsetfillcolor{currentfill}%
\pgfsetlinewidth{0.000000pt}%
\definecolor{currentstroke}{rgb}{1.000000,1.000000,1.000000}%
\pgfsetstrokecolor{currentstroke}%
\pgfsetdash{}{0pt}%
\pgfpathmoveto{\pgfqpoint{0.000000in}{0.000000in}}%
\pgfpathlineto{\pgfqpoint{6.148149in}{0.000000in}}%
\pgfpathlineto{\pgfqpoint{6.148149in}{3.727916in}}%
\pgfpathlineto{\pgfqpoint{0.000000in}{3.727916in}}%
\pgfpathlineto{\pgfqpoint{0.000000in}{0.000000in}}%
\pgfpathclose%
\pgfusepath{fill}%
\end{pgfscope}%
\begin{pgfscope}%
\pgfsetbuttcap%
\pgfsetmiterjoin%
\definecolor{currentfill}{rgb}{1.000000,1.000000,1.000000}%
\pgfsetfillcolor{currentfill}%
\pgfsetlinewidth{0.000000pt}%
\definecolor{currentstroke}{rgb}{0.000000,0.000000,0.000000}%
\pgfsetstrokecolor{currentstroke}%
\pgfsetstrokeopacity{0.000000}%
\pgfsetdash{}{0pt}%
\pgfpathmoveto{\pgfqpoint{0.553704in}{0.499691in}}%
\pgfpathlineto{\pgfqpoint{5.978704in}{0.499691in}}%
\pgfpathlineto{\pgfqpoint{5.978704in}{3.579691in}}%
\pgfpathlineto{\pgfqpoint{0.553704in}{3.579691in}}%
\pgfpathlineto{\pgfqpoint{0.553704in}{0.499691in}}%
\pgfpathclose%
\pgfusepath{fill}%
\end{pgfscope}%
\begin{pgfscope}%
\pgfpathrectangle{\pgfqpoint{0.553704in}{0.499691in}}{\pgfqpoint{5.425000in}{3.080000in}}%
\pgfusepath{clip}%
\pgfsetbuttcap%
\pgfsetroundjoin%
\definecolor{currentfill}{rgb}{0.121569,0.466667,0.705882}%
\pgfsetfillcolor{currentfill}%
\pgfsetfillopacity{0.500000}%
\pgfsetlinewidth{0.000000pt}%
\definecolor{currentstroke}{rgb}{0.000000,0.000000,0.000000}%
\pgfsetstrokecolor{currentstroke}%
\pgfsetdash{}{0pt}%
\pgfpathmoveto{\pgfqpoint{0.553704in}{3.579691in}}%
\pgfpathlineto{\pgfqpoint{0.553704in}{0.499691in}}%
\pgfpathlineto{\pgfqpoint{1.156482in}{0.499691in}}%
\pgfpathlineto{\pgfqpoint{1.759260in}{0.499691in}}%
\pgfpathlineto{\pgfqpoint{2.362038in}{0.499691in}}%
\pgfpathlineto{\pgfqpoint{2.964815in}{0.499691in}}%
\pgfpathlineto{\pgfqpoint{3.567593in}{0.499691in}}%
\pgfpathlineto{\pgfqpoint{4.170371in}{0.499691in}}%
\pgfpathlineto{\pgfqpoint{4.773149in}{0.499691in}}%
\pgfpathlineto{\pgfqpoint{5.375926in}{0.499691in}}%
\pgfpathlineto{\pgfqpoint{5.978704in}{0.499691in}}%
\pgfpathlineto{\pgfqpoint{5.978704in}{2.301183in}}%
\pgfpathlineto{\pgfqpoint{5.978704in}{2.301183in}}%
\pgfpathlineto{\pgfqpoint{5.375926in}{2.304202in}}%
\pgfpathlineto{\pgfqpoint{4.773149in}{2.238598in}}%
\pgfpathlineto{\pgfqpoint{4.170371in}{2.333338in}}%
\pgfpathlineto{\pgfqpoint{3.567593in}{2.272909in}}%
\pgfpathlineto{\pgfqpoint{2.964815in}{2.250486in}}%
\pgfpathlineto{\pgfqpoint{2.362038in}{2.325207in}}%
\pgfpathlineto{\pgfqpoint{1.759260in}{2.539698in}}%
\pgfpathlineto{\pgfqpoint{1.156482in}{2.751849in}}%
\pgfpathlineto{\pgfqpoint{0.553704in}{3.579691in}}%
\pgfpathlineto{\pgfqpoint{0.553704in}{3.579691in}}%
\pgfpathclose%
\pgfusepath{fill}%
\end{pgfscope}%
\begin{pgfscope}%
\pgfpathrectangle{\pgfqpoint{0.553704in}{0.499691in}}{\pgfqpoint{5.425000in}{3.080000in}}%
\pgfusepath{clip}%
\pgfsetbuttcap%
\pgfsetroundjoin%
\definecolor{currentfill}{rgb}{1.000000,0.498039,0.054902}%
\pgfsetfillcolor{currentfill}%
\pgfsetfillopacity{0.500000}%
\pgfsetlinewidth{0.000000pt}%
\definecolor{currentstroke}{rgb}{0.000000,0.000000,0.000000}%
\pgfsetstrokecolor{currentstroke}%
\pgfsetdash{}{0pt}%
\pgfpathmoveto{\pgfqpoint{0.553704in}{3.579691in}}%
\pgfpathlineto{\pgfqpoint{0.553704in}{3.579691in}}%
\pgfpathlineto{\pgfqpoint{1.156482in}{2.751849in}}%
\pgfpathlineto{\pgfqpoint{1.759260in}{2.539698in}}%
\pgfpathlineto{\pgfqpoint{2.362038in}{2.325207in}}%
\pgfpathlineto{\pgfqpoint{2.964815in}{2.250486in}}%
\pgfpathlineto{\pgfqpoint{3.567593in}{2.272909in}}%
\pgfpathlineto{\pgfqpoint{4.170371in}{2.333338in}}%
\pgfpathlineto{\pgfqpoint{4.773149in}{2.238598in}}%
\pgfpathlineto{\pgfqpoint{5.375926in}{2.304202in}}%
\pgfpathlineto{\pgfqpoint{5.978704in}{2.301183in}}%
\pgfpathlineto{\pgfqpoint{5.978704in}{2.769405in}}%
\pgfpathlineto{\pgfqpoint{5.978704in}{2.769405in}}%
\pgfpathlineto{\pgfqpoint{5.375926in}{2.780493in}}%
\pgfpathlineto{\pgfqpoint{4.773149in}{2.760411in}}%
\pgfpathlineto{\pgfqpoint{4.170371in}{2.843263in}}%
\pgfpathlineto{\pgfqpoint{3.567593in}{2.845419in}}%
\pgfpathlineto{\pgfqpoint{2.964815in}{2.870429in}}%
\pgfpathlineto{\pgfqpoint{2.362038in}{3.004717in}}%
\pgfpathlineto{\pgfqpoint{1.759260in}{3.237934in}}%
\pgfpathlineto{\pgfqpoint{1.156482in}{3.579691in}}%
\pgfpathlineto{\pgfqpoint{0.553704in}{3.579691in}}%
\pgfpathlineto{\pgfqpoint{0.553704in}{3.579691in}}%
\pgfpathclose%
\pgfusepath{fill}%
\end{pgfscope}%
\begin{pgfscope}%
\pgfpathrectangle{\pgfqpoint{0.553704in}{0.499691in}}{\pgfqpoint{5.425000in}{3.080000in}}%
\pgfusepath{clip}%
\pgfsetbuttcap%
\pgfsetroundjoin%
\definecolor{currentfill}{rgb}{0.172549,0.627451,0.172549}%
\pgfsetfillcolor{currentfill}%
\pgfsetfillopacity{0.500000}%
\pgfsetlinewidth{0.000000pt}%
\definecolor{currentstroke}{rgb}{0.000000,0.000000,0.000000}%
\pgfsetstrokecolor{currentstroke}%
\pgfsetdash{}{0pt}%
\pgfpathmoveto{\pgfqpoint{0.553704in}{3.579691in}}%
\pgfpathlineto{\pgfqpoint{0.553704in}{3.579691in}}%
\pgfpathlineto{\pgfqpoint{1.156482in}{3.579691in}}%
\pgfpathlineto{\pgfqpoint{1.759260in}{3.237934in}}%
\pgfpathlineto{\pgfqpoint{2.362038in}{3.004717in}}%
\pgfpathlineto{\pgfqpoint{2.964815in}{2.870429in}}%
\pgfpathlineto{\pgfqpoint{3.567593in}{2.845419in}}%
\pgfpathlineto{\pgfqpoint{4.170371in}{2.843263in}}%
\pgfpathlineto{\pgfqpoint{4.773149in}{2.760411in}}%
\pgfpathlineto{\pgfqpoint{5.375926in}{2.780493in}}%
\pgfpathlineto{\pgfqpoint{5.978704in}{2.769405in}}%
\pgfpathlineto{\pgfqpoint{5.978704in}{3.033422in}}%
\pgfpathlineto{\pgfqpoint{5.978704in}{3.033422in}}%
\pgfpathlineto{\pgfqpoint{5.375926in}{3.054551in}}%
\pgfpathlineto{\pgfqpoint{4.773149in}{3.054736in}}%
\pgfpathlineto{\pgfqpoint{4.170371in}{3.132352in}}%
\pgfpathlineto{\pgfqpoint{3.567593in}{3.170975in}}%
\pgfpathlineto{\pgfqpoint{2.964815in}{3.232883in}}%
\pgfpathlineto{\pgfqpoint{2.362038in}{3.384358in}}%
\pgfpathlineto{\pgfqpoint{1.759260in}{3.579691in}}%
\pgfpathlineto{\pgfqpoint{1.156482in}{3.579691in}}%
\pgfpathlineto{\pgfqpoint{0.553704in}{3.579691in}}%
\pgfpathlineto{\pgfqpoint{0.553704in}{3.579691in}}%
\pgfpathclose%
\pgfusepath{fill}%
\end{pgfscope}%
\begin{pgfscope}%
\pgfpathrectangle{\pgfqpoint{0.553704in}{0.499691in}}{\pgfqpoint{5.425000in}{3.080000in}}%
\pgfusepath{clip}%
\pgfsetbuttcap%
\pgfsetroundjoin%
\definecolor{currentfill}{rgb}{0.839216,0.152941,0.156863}%
\pgfsetfillcolor{currentfill}%
\pgfsetfillopacity{0.500000}%
\pgfsetlinewidth{0.000000pt}%
\definecolor{currentstroke}{rgb}{0.000000,0.000000,0.000000}%
\pgfsetstrokecolor{currentstroke}%
\pgfsetdash{}{0pt}%
\pgfpathmoveto{\pgfqpoint{0.553704in}{3.579691in}}%
\pgfpathlineto{\pgfqpoint{0.553704in}{3.579691in}}%
\pgfpathlineto{\pgfqpoint{1.156482in}{3.579691in}}%
\pgfpathlineto{\pgfqpoint{1.759260in}{3.579691in}}%
\pgfpathlineto{\pgfqpoint{2.362038in}{3.384358in}}%
\pgfpathlineto{\pgfqpoint{2.964815in}{3.232883in}}%
\pgfpathlineto{\pgfqpoint{3.567593in}{3.170975in}}%
\pgfpathlineto{\pgfqpoint{4.170371in}{3.132352in}}%
\pgfpathlineto{\pgfqpoint{4.773149in}{3.054736in}}%
\pgfpathlineto{\pgfqpoint{5.375926in}{3.054551in}}%
\pgfpathlineto{\pgfqpoint{5.978704in}{3.033422in}}%
\pgfpathlineto{\pgfqpoint{5.978704in}{3.207073in}}%
\pgfpathlineto{\pgfqpoint{5.978704in}{3.207073in}}%
\pgfpathlineto{\pgfqpoint{5.375926in}{3.236887in}}%
\pgfpathlineto{\pgfqpoint{4.773149in}{3.249269in}}%
\pgfpathlineto{\pgfqpoint{4.170371in}{3.326330in}}%
\pgfpathlineto{\pgfqpoint{3.567593in}{3.378814in}}%
\pgfpathlineto{\pgfqpoint{2.964815in}{3.457600in}}%
\pgfpathlineto{\pgfqpoint{2.362038in}{3.579691in}}%
\pgfpathlineto{\pgfqpoint{1.759260in}{3.579691in}}%
\pgfpathlineto{\pgfqpoint{1.156482in}{3.579691in}}%
\pgfpathlineto{\pgfqpoint{0.553704in}{3.579691in}}%
\pgfpathlineto{\pgfqpoint{0.553704in}{3.579691in}}%
\pgfpathclose%
\pgfusepath{fill}%
\end{pgfscope}%
\begin{pgfscope}%
\pgfpathrectangle{\pgfqpoint{0.553704in}{0.499691in}}{\pgfqpoint{5.425000in}{3.080000in}}%
\pgfusepath{clip}%
\pgfsetbuttcap%
\pgfsetroundjoin%
\definecolor{currentfill}{rgb}{0.580392,0.403922,0.741176}%
\pgfsetfillcolor{currentfill}%
\pgfsetfillopacity{0.500000}%
\pgfsetlinewidth{0.000000pt}%
\definecolor{currentstroke}{rgb}{0.000000,0.000000,0.000000}%
\pgfsetstrokecolor{currentstroke}%
\pgfsetdash{}{0pt}%
\pgfpathmoveto{\pgfqpoint{0.553704in}{3.579691in}}%
\pgfpathlineto{\pgfqpoint{0.553704in}{3.579691in}}%
\pgfpathlineto{\pgfqpoint{1.156482in}{3.579691in}}%
\pgfpathlineto{\pgfqpoint{1.759260in}{3.579691in}}%
\pgfpathlineto{\pgfqpoint{2.362038in}{3.579691in}}%
\pgfpathlineto{\pgfqpoint{2.964815in}{3.457600in}}%
\pgfpathlineto{\pgfqpoint{3.567593in}{3.378814in}}%
\pgfpathlineto{\pgfqpoint{4.170371in}{3.326330in}}%
\pgfpathlineto{\pgfqpoint{4.773149in}{3.249269in}}%
\pgfpathlineto{\pgfqpoint{5.375926in}{3.236887in}}%
\pgfpathlineto{\pgfqpoint{5.978704in}{3.207073in}}%
\pgfpathlineto{\pgfqpoint{5.978704in}{3.332306in}}%
\pgfpathlineto{\pgfqpoint{5.978704in}{3.332306in}}%
\pgfpathlineto{\pgfqpoint{5.375926in}{3.368280in}}%
\pgfpathlineto{\pgfqpoint{4.773149in}{3.389778in}}%
\pgfpathlineto{\pgfqpoint{4.170371in}{3.457415in}}%
\pgfpathlineto{\pgfqpoint{3.567593in}{3.510022in}}%
\pgfpathlineto{\pgfqpoint{2.964815in}{3.579691in}}%
\pgfpathlineto{\pgfqpoint{2.362038in}{3.579691in}}%
\pgfpathlineto{\pgfqpoint{1.759260in}{3.579691in}}%
\pgfpathlineto{\pgfqpoint{1.156482in}{3.579691in}}%
\pgfpathlineto{\pgfqpoint{0.553704in}{3.579691in}}%
\pgfpathlineto{\pgfqpoint{0.553704in}{3.579691in}}%
\pgfpathclose%
\pgfusepath{fill}%
\end{pgfscope}%
\begin{pgfscope}%
\pgfpathrectangle{\pgfqpoint{0.553704in}{0.499691in}}{\pgfqpoint{5.425000in}{3.080000in}}%
\pgfusepath{clip}%
\pgfsetbuttcap%
\pgfsetroundjoin%
\definecolor{currentfill}{rgb}{0.549020,0.337255,0.294118}%
\pgfsetfillcolor{currentfill}%
\pgfsetfillopacity{0.500000}%
\pgfsetlinewidth{0.000000pt}%
\definecolor{currentstroke}{rgb}{0.000000,0.000000,0.000000}%
\pgfsetstrokecolor{currentstroke}%
\pgfsetdash{}{0pt}%
\pgfpathmoveto{\pgfqpoint{0.553704in}{3.579691in}}%
\pgfpathlineto{\pgfqpoint{0.553704in}{3.579691in}}%
\pgfpathlineto{\pgfqpoint{1.156482in}{3.579691in}}%
\pgfpathlineto{\pgfqpoint{1.759260in}{3.579691in}}%
\pgfpathlineto{\pgfqpoint{2.362038in}{3.579691in}}%
\pgfpathlineto{\pgfqpoint{2.964815in}{3.579691in}}%
\pgfpathlineto{\pgfqpoint{3.567593in}{3.510022in}}%
\pgfpathlineto{\pgfqpoint{4.170371in}{3.457415in}}%
\pgfpathlineto{\pgfqpoint{4.773149in}{3.389778in}}%
\pgfpathlineto{\pgfqpoint{5.375926in}{3.368280in}}%
\pgfpathlineto{\pgfqpoint{5.978704in}{3.332306in}}%
\pgfpathlineto{\pgfqpoint{5.978704in}{3.425753in}}%
\pgfpathlineto{\pgfqpoint{5.978704in}{3.425753in}}%
\pgfpathlineto{\pgfqpoint{5.375926in}{3.462158in}}%
\pgfpathlineto{\pgfqpoint{4.773149in}{3.488154in}}%
\pgfpathlineto{\pgfqpoint{4.170371in}{3.539405in}}%
\pgfpathlineto{\pgfqpoint{3.567593in}{3.579691in}}%
\pgfpathlineto{\pgfqpoint{2.964815in}{3.579691in}}%
\pgfpathlineto{\pgfqpoint{2.362038in}{3.579691in}}%
\pgfpathlineto{\pgfqpoint{1.759260in}{3.579691in}}%
\pgfpathlineto{\pgfqpoint{1.156482in}{3.579691in}}%
\pgfpathlineto{\pgfqpoint{0.553704in}{3.579691in}}%
\pgfpathlineto{\pgfqpoint{0.553704in}{3.579691in}}%
\pgfpathclose%
\pgfusepath{fill}%
\end{pgfscope}%
\begin{pgfscope}%
\pgfpathrectangle{\pgfqpoint{0.553704in}{0.499691in}}{\pgfqpoint{5.425000in}{3.080000in}}%
\pgfusepath{clip}%
\pgfsetbuttcap%
\pgfsetroundjoin%
\definecolor{currentfill}{rgb}{0.890196,0.466667,0.760784}%
\pgfsetfillcolor{currentfill}%
\pgfsetfillopacity{0.500000}%
\pgfsetlinewidth{0.000000pt}%
\definecolor{currentstroke}{rgb}{0.000000,0.000000,0.000000}%
\pgfsetstrokecolor{currentstroke}%
\pgfsetdash{}{0pt}%
\pgfpathmoveto{\pgfqpoint{0.553704in}{3.579691in}}%
\pgfpathlineto{\pgfqpoint{0.553704in}{3.579691in}}%
\pgfpathlineto{\pgfqpoint{1.156482in}{3.579691in}}%
\pgfpathlineto{\pgfqpoint{1.759260in}{3.579691in}}%
\pgfpathlineto{\pgfqpoint{2.362038in}{3.579691in}}%
\pgfpathlineto{\pgfqpoint{2.964815in}{3.579691in}}%
\pgfpathlineto{\pgfqpoint{3.567593in}{3.579691in}}%
\pgfpathlineto{\pgfqpoint{4.170371in}{3.539405in}}%
\pgfpathlineto{\pgfqpoint{4.773149in}{3.488154in}}%
\pgfpathlineto{\pgfqpoint{5.375926in}{3.462158in}}%
\pgfpathlineto{\pgfqpoint{5.978704in}{3.425753in}}%
\pgfpathlineto{\pgfqpoint{5.978704in}{3.493451in}}%
\pgfpathlineto{\pgfqpoint{5.978704in}{3.493451in}}%
\pgfpathlineto{\pgfqpoint{5.375926in}{3.524990in}}%
\pgfpathlineto{\pgfqpoint{4.773149in}{3.549569in}}%
\pgfpathlineto{\pgfqpoint{4.170371in}{3.579691in}}%
\pgfpathlineto{\pgfqpoint{3.567593in}{3.579691in}}%
\pgfpathlineto{\pgfqpoint{2.964815in}{3.579691in}}%
\pgfpathlineto{\pgfqpoint{2.362038in}{3.579691in}}%
\pgfpathlineto{\pgfqpoint{1.759260in}{3.579691in}}%
\pgfpathlineto{\pgfqpoint{1.156482in}{3.579691in}}%
\pgfpathlineto{\pgfqpoint{0.553704in}{3.579691in}}%
\pgfpathlineto{\pgfqpoint{0.553704in}{3.579691in}}%
\pgfpathclose%
\pgfusepath{fill}%
\end{pgfscope}%
\begin{pgfscope}%
\pgfpathrectangle{\pgfqpoint{0.553704in}{0.499691in}}{\pgfqpoint{5.425000in}{3.080000in}}%
\pgfusepath{clip}%
\pgfsetbuttcap%
\pgfsetroundjoin%
\definecolor{currentfill}{rgb}{0.498039,0.498039,0.498039}%
\pgfsetfillcolor{currentfill}%
\pgfsetfillopacity{0.500000}%
\pgfsetlinewidth{0.000000pt}%
\definecolor{currentstroke}{rgb}{0.000000,0.000000,0.000000}%
\pgfsetstrokecolor{currentstroke}%
\pgfsetdash{}{0pt}%
\pgfpathmoveto{\pgfqpoint{0.553704in}{3.579691in}}%
\pgfpathlineto{\pgfqpoint{0.553704in}{3.579691in}}%
\pgfpathlineto{\pgfqpoint{1.156482in}{3.579691in}}%
\pgfpathlineto{\pgfqpoint{1.759260in}{3.579691in}}%
\pgfpathlineto{\pgfqpoint{2.362038in}{3.579691in}}%
\pgfpathlineto{\pgfqpoint{2.964815in}{3.579691in}}%
\pgfpathlineto{\pgfqpoint{3.567593in}{3.579691in}}%
\pgfpathlineto{\pgfqpoint{4.170371in}{3.579691in}}%
\pgfpathlineto{\pgfqpoint{4.773149in}{3.549569in}}%
\pgfpathlineto{\pgfqpoint{5.375926in}{3.524990in}}%
\pgfpathlineto{\pgfqpoint{5.978704in}{3.493451in}}%
\pgfpathlineto{\pgfqpoint{5.978704in}{3.541869in}}%
\pgfpathlineto{\pgfqpoint{5.978704in}{3.541869in}}%
\pgfpathlineto{\pgfqpoint{5.375926in}{3.563675in}}%
\pgfpathlineto{\pgfqpoint{4.773149in}{3.579691in}}%
\pgfpathlineto{\pgfqpoint{4.170371in}{3.579691in}}%
\pgfpathlineto{\pgfqpoint{3.567593in}{3.579691in}}%
\pgfpathlineto{\pgfqpoint{2.964815in}{3.579691in}}%
\pgfpathlineto{\pgfqpoint{2.362038in}{3.579691in}}%
\pgfpathlineto{\pgfqpoint{1.759260in}{3.579691in}}%
\pgfpathlineto{\pgfqpoint{1.156482in}{3.579691in}}%
\pgfpathlineto{\pgfqpoint{0.553704in}{3.579691in}}%
\pgfpathlineto{\pgfqpoint{0.553704in}{3.579691in}}%
\pgfpathclose%
\pgfusepath{fill}%
\end{pgfscope}%
\begin{pgfscope}%
\pgfpathrectangle{\pgfqpoint{0.553704in}{0.499691in}}{\pgfqpoint{5.425000in}{3.080000in}}%
\pgfusepath{clip}%
\pgfsetbuttcap%
\pgfsetroundjoin%
\definecolor{currentfill}{rgb}{0.737255,0.741176,0.133333}%
\pgfsetfillcolor{currentfill}%
\pgfsetfillopacity{0.500000}%
\pgfsetlinewidth{0.000000pt}%
\definecolor{currentstroke}{rgb}{0.000000,0.000000,0.000000}%
\pgfsetstrokecolor{currentstroke}%
\pgfsetdash{}{0pt}%
\pgfpathmoveto{\pgfqpoint{0.553704in}{3.579691in}}%
\pgfpathlineto{\pgfqpoint{0.553704in}{3.579691in}}%
\pgfpathlineto{\pgfqpoint{1.156482in}{3.579691in}}%
\pgfpathlineto{\pgfqpoint{1.759260in}{3.579691in}}%
\pgfpathlineto{\pgfqpoint{2.362038in}{3.579691in}}%
\pgfpathlineto{\pgfqpoint{2.964815in}{3.579691in}}%
\pgfpathlineto{\pgfqpoint{3.567593in}{3.579691in}}%
\pgfpathlineto{\pgfqpoint{4.170371in}{3.579691in}}%
\pgfpathlineto{\pgfqpoint{4.773149in}{3.579691in}}%
\pgfpathlineto{\pgfqpoint{5.375926in}{3.563675in}}%
\pgfpathlineto{\pgfqpoint{5.978704in}{3.541869in}}%
\pgfpathlineto{\pgfqpoint{5.978704in}{3.569404in}}%
\pgfpathlineto{\pgfqpoint{5.978704in}{3.569404in}}%
\pgfpathlineto{\pgfqpoint{5.375926in}{3.579691in}}%
\pgfpathlineto{\pgfqpoint{4.773149in}{3.579691in}}%
\pgfpathlineto{\pgfqpoint{4.170371in}{3.579691in}}%
\pgfpathlineto{\pgfqpoint{3.567593in}{3.579691in}}%
\pgfpathlineto{\pgfqpoint{2.964815in}{3.579691in}}%
\pgfpathlineto{\pgfqpoint{2.362038in}{3.579691in}}%
\pgfpathlineto{\pgfqpoint{1.759260in}{3.579691in}}%
\pgfpathlineto{\pgfqpoint{1.156482in}{3.579691in}}%
\pgfpathlineto{\pgfqpoint{0.553704in}{3.579691in}}%
\pgfpathlineto{\pgfqpoint{0.553704in}{3.579691in}}%
\pgfpathclose%
\pgfusepath{fill}%
\end{pgfscope}%
\begin{pgfscope}%
\pgfpathrectangle{\pgfqpoint{0.553704in}{0.499691in}}{\pgfqpoint{5.425000in}{3.080000in}}%
\pgfusepath{clip}%
\pgfsetbuttcap%
\pgfsetroundjoin%
\definecolor{currentfill}{rgb}{0.090196,0.745098,0.811765}%
\pgfsetfillcolor{currentfill}%
\pgfsetfillopacity{0.500000}%
\pgfsetlinewidth{0.000000pt}%
\definecolor{currentstroke}{rgb}{0.000000,0.000000,0.000000}%
\pgfsetstrokecolor{currentstroke}%
\pgfsetdash{}{0pt}%
\pgfpathmoveto{\pgfqpoint{0.553704in}{3.579691in}}%
\pgfpathlineto{\pgfqpoint{0.553704in}{3.579691in}}%
\pgfpathlineto{\pgfqpoint{1.156482in}{3.579691in}}%
\pgfpathlineto{\pgfqpoint{1.759260in}{3.579691in}}%
\pgfpathlineto{\pgfqpoint{2.362038in}{3.579691in}}%
\pgfpathlineto{\pgfqpoint{2.964815in}{3.579691in}}%
\pgfpathlineto{\pgfqpoint{3.567593in}{3.579691in}}%
\pgfpathlineto{\pgfqpoint{4.170371in}{3.579691in}}%
\pgfpathlineto{\pgfqpoint{4.773149in}{3.579691in}}%
\pgfpathlineto{\pgfqpoint{5.375926in}{3.579691in}}%
\pgfpathlineto{\pgfqpoint{5.978704in}{3.569404in}}%
\pgfpathlineto{\pgfqpoint{5.978704in}{3.579691in}}%
\pgfpathlineto{\pgfqpoint{5.978704in}{3.579691in}}%
\pgfpathlineto{\pgfqpoint{5.375926in}{3.579691in}}%
\pgfpathlineto{\pgfqpoint{4.773149in}{3.579691in}}%
\pgfpathlineto{\pgfqpoint{4.170371in}{3.579691in}}%
\pgfpathlineto{\pgfqpoint{3.567593in}{3.579691in}}%
\pgfpathlineto{\pgfqpoint{2.964815in}{3.579691in}}%
\pgfpathlineto{\pgfqpoint{2.362038in}{3.579691in}}%
\pgfpathlineto{\pgfqpoint{1.759260in}{3.579691in}}%
\pgfpathlineto{\pgfqpoint{1.156482in}{3.579691in}}%
\pgfpathlineto{\pgfqpoint{0.553704in}{3.579691in}}%
\pgfpathlineto{\pgfqpoint{0.553704in}{3.579691in}}%
\pgfpathclose%
\pgfusepath{fill}%
\end{pgfscope}%
\begin{pgfscope}%
\pgfsetbuttcap%
\pgfsetroundjoin%
\definecolor{currentfill}{rgb}{0.000000,0.000000,0.000000}%
\pgfsetfillcolor{currentfill}%
\pgfsetlinewidth{0.803000pt}%
\definecolor{currentstroke}{rgb}{0.000000,0.000000,0.000000}%
\pgfsetstrokecolor{currentstroke}%
\pgfsetdash{}{0pt}%
\pgfsys@defobject{currentmarker}{\pgfqpoint{0.000000in}{-0.048611in}}{\pgfqpoint{0.000000in}{0.000000in}}{%
\pgfpathmoveto{\pgfqpoint{0.000000in}{0.000000in}}%
\pgfpathlineto{\pgfqpoint{0.000000in}{-0.048611in}}%
\pgfusepath{stroke,fill}%
}%
\begin{pgfscope}%
\pgfsys@transformshift{0.553704in}{0.499691in}%
\pgfsys@useobject{currentmarker}{}%
\end{pgfscope}%
\end{pgfscope}%
\begin{pgfscope}%
\definecolor{textcolor}{rgb}{0.000000,0.000000,0.000000}%
\pgfsetstrokecolor{textcolor}%
\pgfsetfillcolor{textcolor}%
\pgftext[x=0.553704in,y=0.402469in,,top]{\color{textcolor}\rmfamily\fontsize{18.000000}{12.000000}\selectfont \(\displaystyle {1}\)}%
\end{pgfscope}%
\begin{pgfscope}%
\pgfsetbuttcap%
\pgfsetroundjoin%
\definecolor{currentfill}{rgb}{0.000000,0.000000,0.000000}%
\pgfsetfillcolor{currentfill}%
\pgfsetlinewidth{0.803000pt}%
\definecolor{currentstroke}{rgb}{0.000000,0.000000,0.000000}%
\pgfsetstrokecolor{currentstroke}%
\pgfsetdash{}{0pt}%
\pgfsys@defobject{currentmarker}{\pgfqpoint{0.000000in}{-0.048611in}}{\pgfqpoint{0.000000in}{0.000000in}}{%
\pgfpathmoveto{\pgfqpoint{0.000000in}{0.000000in}}%
\pgfpathlineto{\pgfqpoint{0.000000in}{-0.048611in}}%
\pgfusepath{stroke,fill}%
}%
\begin{pgfscope}%
\pgfsys@transformshift{1.156482in}{0.499691in}%
\pgfsys@useobject{currentmarker}{}%
\end{pgfscope}%
\end{pgfscope}%
\begin{pgfscope}%
\definecolor{textcolor}{rgb}{0.000000,0.000000,0.000000}%
\pgfsetstrokecolor{textcolor}%
\pgfsetfillcolor{textcolor}%
\pgftext[x=1.156482in,y=0.402469in,,top]{\color{textcolor}\rmfamily\fontsize{10.000000}{12.000000}\selectfont \(\displaystyle {2}\)}%
\end{pgfscope}%
\begin{pgfscope}%
\pgfsetbuttcap%
\pgfsetroundjoin%
\definecolor{currentfill}{rgb}{0.000000,0.000000,0.000000}%
\pgfsetfillcolor{currentfill}%
\pgfsetlinewidth{0.803000pt}%
\definecolor{currentstroke}{rgb}{0.000000,0.000000,0.000000}%
\pgfsetstrokecolor{currentstroke}%
\pgfsetdash{}{0pt}%
\pgfsys@defobject{currentmarker}{\pgfqpoint{0.000000in}{-0.048611in}}{\pgfqpoint{0.000000in}{0.000000in}}{%
\pgfpathmoveto{\pgfqpoint{0.000000in}{0.000000in}}%
\pgfpathlineto{\pgfqpoint{0.000000in}{-0.048611in}}%
\pgfusepath{stroke,fill}%
}%
\begin{pgfscope}%
\pgfsys@transformshift{1.759260in}{0.499691in}%
\pgfsys@useobject{currentmarker}{}%
\end{pgfscope}%
\end{pgfscope}%
\begin{pgfscope}%
\definecolor{textcolor}{rgb}{0.000000,0.000000,0.000000}%
\pgfsetstrokecolor{textcolor}%
\pgfsetfillcolor{textcolor}%
\pgftext[x=1.759260in,y=0.402469in,,top]{\color{textcolor}\rmfamily\fontsize{10.000000}{12.000000}\selectfont \(\displaystyle {3}\)}%
\end{pgfscope}%
\begin{pgfscope}%
\pgfsetbuttcap%
\pgfsetroundjoin%
\definecolor{currentfill}{rgb}{0.000000,0.000000,0.000000}%
\pgfsetfillcolor{currentfill}%
\pgfsetlinewidth{0.803000pt}%
\definecolor{currentstroke}{rgb}{0.000000,0.000000,0.000000}%
\pgfsetstrokecolor{currentstroke}%
\pgfsetdash{}{0pt}%
\pgfsys@defobject{currentmarker}{\pgfqpoint{0.000000in}{-0.048611in}}{\pgfqpoint{0.000000in}{0.000000in}}{%
\pgfpathmoveto{\pgfqpoint{0.000000in}{0.000000in}}%
\pgfpathlineto{\pgfqpoint{0.000000in}{-0.048611in}}%
\pgfusepath{stroke,fill}%
}%
\begin{pgfscope}%
\pgfsys@transformshift{2.362038in}{0.499691in}%
\pgfsys@useobject{currentmarker}{}%
\end{pgfscope}%
\end{pgfscope}%
\begin{pgfscope}%
\definecolor{textcolor}{rgb}{0.000000,0.000000,0.000000}%
\pgfsetstrokecolor{textcolor}%
\pgfsetfillcolor{textcolor}%
\pgftext[x=2.362038in,y=0.402469in,,top]{\color{textcolor}\rmfamily\fontsize{10.000000}{12.000000}\selectfont \(\displaystyle {4}\)}%
\end{pgfscope}%
\begin{pgfscope}%
\pgfsetbuttcap%
\pgfsetroundjoin%
\definecolor{currentfill}{rgb}{0.000000,0.000000,0.000000}%
\pgfsetfillcolor{currentfill}%
\pgfsetlinewidth{0.803000pt}%
\definecolor{currentstroke}{rgb}{0.000000,0.000000,0.000000}%
\pgfsetstrokecolor{currentstroke}%
\pgfsetdash{}{0pt}%
\pgfsys@defobject{currentmarker}{\pgfqpoint{0.000000in}{-0.048611in}}{\pgfqpoint{0.000000in}{0.000000in}}{%
\pgfpathmoveto{\pgfqpoint{0.000000in}{0.000000in}}%
\pgfpathlineto{\pgfqpoint{0.000000in}{-0.048611in}}%
\pgfusepath{stroke,fill}%
}%
\begin{pgfscope}%
\pgfsys@transformshift{2.964815in}{0.499691in}%
\pgfsys@useobject{currentmarker}{}%
\end{pgfscope}%
\end{pgfscope}%
\begin{pgfscope}%
\definecolor{textcolor}{rgb}{0.000000,0.000000,0.000000}%
\pgfsetstrokecolor{textcolor}%
\pgfsetfillcolor{textcolor}%
\pgftext[x=2.964815in,y=0.402469in,,top]{\color{textcolor}\rmfamily\fontsize{10.000000}{12.000000}\selectfont \(\displaystyle {5}\)}%
\end{pgfscope}%
\begin{pgfscope}%
\pgfsetbuttcap%
\pgfsetroundjoin%
\definecolor{currentfill}{rgb}{0.000000,0.000000,0.000000}%
\pgfsetfillcolor{currentfill}%
\pgfsetlinewidth{0.803000pt}%
\definecolor{currentstroke}{rgb}{0.000000,0.000000,0.000000}%
\pgfsetstrokecolor{currentstroke}%
\pgfsetdash{}{0pt}%
\pgfsys@defobject{currentmarker}{\pgfqpoint{0.000000in}{-0.048611in}}{\pgfqpoint{0.000000in}{0.000000in}}{%
\pgfpathmoveto{\pgfqpoint{0.000000in}{0.000000in}}%
\pgfpathlineto{\pgfqpoint{0.000000in}{-0.048611in}}%
\pgfusepath{stroke,fill}%
}%
\begin{pgfscope}%
\pgfsys@transformshift{3.567593in}{0.499691in}%
\pgfsys@useobject{currentmarker}{}%
\end{pgfscope}%
\end{pgfscope}%
\begin{pgfscope}%
\definecolor{textcolor}{rgb}{0.000000,0.000000,0.000000}%
\pgfsetstrokecolor{textcolor}%
\pgfsetfillcolor{textcolor}%
\pgftext[x=3.567593in,y=0.402469in,,top]{\color{textcolor}\rmfamily\fontsize{10.000000}{12.000000}\selectfont \(\displaystyle {6}\)}%
\end{pgfscope}%
\begin{pgfscope}%
\pgfsetbuttcap%
\pgfsetroundjoin%
\definecolor{currentfill}{rgb}{0.000000,0.000000,0.000000}%
\pgfsetfillcolor{currentfill}%
\pgfsetlinewidth{0.803000pt}%
\definecolor{currentstroke}{rgb}{0.000000,0.000000,0.000000}%
\pgfsetstrokecolor{currentstroke}%
\pgfsetdash{}{0pt}%
\pgfsys@defobject{currentmarker}{\pgfqpoint{0.000000in}{-0.048611in}}{\pgfqpoint{0.000000in}{0.000000in}}{%
\pgfpathmoveto{\pgfqpoint{0.000000in}{0.000000in}}%
\pgfpathlineto{\pgfqpoint{0.000000in}{-0.048611in}}%
\pgfusepath{stroke,fill}%
}%
\begin{pgfscope}%
\pgfsys@transformshift{4.170371in}{0.499691in}%
\pgfsys@useobject{currentmarker}{}%
\end{pgfscope}%
\end{pgfscope}%
\begin{pgfscope}%
\definecolor{textcolor}{rgb}{0.000000,0.000000,0.000000}%
\pgfsetstrokecolor{textcolor}%
\pgfsetfillcolor{textcolor}%
\pgftext[x=4.170371in,y=0.402469in,,top]{\color{textcolor}\rmfamily\fontsize{10.000000}{12.000000}\selectfont \(\displaystyle {7}\)}%
\end{pgfscope}%
\begin{pgfscope}%
\pgfsetbuttcap%
\pgfsetroundjoin%
\definecolor{currentfill}{rgb}{0.000000,0.000000,0.000000}%
\pgfsetfillcolor{currentfill}%
\pgfsetlinewidth{0.803000pt}%
\definecolor{currentstroke}{rgb}{0.000000,0.000000,0.000000}%
\pgfsetstrokecolor{currentstroke}%
\pgfsetdash{}{0pt}%
\pgfsys@defobject{currentmarker}{\pgfqpoint{0.000000in}{-0.048611in}}{\pgfqpoint{0.000000in}{0.000000in}}{%
\pgfpathmoveto{\pgfqpoint{0.000000in}{0.000000in}}%
\pgfpathlineto{\pgfqpoint{0.000000in}{-0.048611in}}%
\pgfusepath{stroke,fill}%
}%
\begin{pgfscope}%
\pgfsys@transformshift{4.773149in}{0.499691in}%
\pgfsys@useobject{currentmarker}{}%
\end{pgfscope}%
\end{pgfscope}%
\begin{pgfscope}%
\definecolor{textcolor}{rgb}{0.000000,0.000000,0.000000}%
\pgfsetstrokecolor{textcolor}%
\pgfsetfillcolor{textcolor}%
\pgftext[x=4.773149in,y=0.402469in,,top]{\color{textcolor}\rmfamily\fontsize{10.000000}{12.000000}\selectfont \(\displaystyle {8}\)}%
\end{pgfscope}%
\begin{pgfscope}%
\pgfsetbuttcap%
\pgfsetroundjoin%
\definecolor{currentfill}{rgb}{0.000000,0.000000,0.000000}%
\pgfsetfillcolor{currentfill}%
\pgfsetlinewidth{0.803000pt}%
\definecolor{currentstroke}{rgb}{0.000000,0.000000,0.000000}%
\pgfsetstrokecolor{currentstroke}%
\pgfsetdash{}{0pt}%
\pgfsys@defobject{currentmarker}{\pgfqpoint{0.000000in}{-0.048611in}}{\pgfqpoint{0.000000in}{0.000000in}}{%
\pgfpathmoveto{\pgfqpoint{0.000000in}{0.000000in}}%
\pgfpathlineto{\pgfqpoint{0.000000in}{-0.048611in}}%
\pgfusepath{stroke,fill}%
}%
\begin{pgfscope}%
\pgfsys@transformshift{5.375926in}{0.499691in}%
\pgfsys@useobject{currentmarker}{}%
\end{pgfscope}%
\end{pgfscope}%
\begin{pgfscope}%
\definecolor{textcolor}{rgb}{0.000000,0.000000,0.000000}%
\pgfsetstrokecolor{textcolor}%
\pgfsetfillcolor{textcolor}%
\pgftext[x=5.375926in,y=0.402469in,,top]{\color{textcolor}\rmfamily\fontsize{10.000000}{12.000000}\selectfont \(\displaystyle {9}\)}%
\end{pgfscope}%
\begin{pgfscope}%
\pgfsetbuttcap%
\pgfsetroundjoin%
\definecolor{currentfill}{rgb}{0.000000,0.000000,0.000000}%
\pgfsetfillcolor{currentfill}%
\pgfsetlinewidth{0.803000pt}%
\definecolor{currentstroke}{rgb}{0.000000,0.000000,0.000000}%
\pgfsetstrokecolor{currentstroke}%
\pgfsetdash{}{0pt}%
\pgfsys@defobject{currentmarker}{\pgfqpoint{0.000000in}{-0.048611in}}{\pgfqpoint{0.000000in}{0.000000in}}{%
\pgfpathmoveto{\pgfqpoint{0.000000in}{0.000000in}}%
\pgfpathlineto{\pgfqpoint{0.000000in}{-0.048611in}}%
\pgfusepath{stroke,fill}%
}%
\begin{pgfscope}%
\pgfsys@transformshift{5.978704in}{0.499691in}%
\pgfsys@useobject{currentmarker}{}%
\end{pgfscope}%
\end{pgfscope}%
\begin{pgfscope}%
\definecolor{textcolor}{rgb}{0.000000,0.000000,0.000000}%
\pgfsetstrokecolor{textcolor}%
\pgfsetfillcolor{textcolor}%
\pgftext[x=5.978704in,y=0.402469in,,top]{\color{textcolor}\rmfamily\fontsize{10.000000}{12.000000}\selectfont \(\displaystyle {10}\)}%
\end{pgfscope}%
\begin{pgfscope}%
\definecolor{textcolor}{rgb}{0.000000,0.000000,0.000000}%
\pgfsetstrokecolor{textcolor}%
\pgfsetfillcolor{textcolor}%
\pgftext[x=3.266204in,y=0.223457in,,top]{\color{textcolor}\rmfamily\fontsize{18.000000}{12.000000}\selectfont Number of sources}%
\end{pgfscope}%
\begin{pgfscope}%
\pgfsetbuttcap%
\pgfsetroundjoin%
\definecolor{currentfill}{rgb}{0.000000,0.000000,0.000000}%
\pgfsetfillcolor{currentfill}%
\pgfsetlinewidth{0.803000pt}%
\definecolor{currentstroke}{rgb}{0.000000,0.000000,0.000000}%
\pgfsetstrokecolor{currentstroke}%
\pgfsetdash{}{0pt}%
\pgfsys@defobject{currentmarker}{\pgfqpoint{-0.048611in}{0.000000in}}{\pgfqpoint{-0.000000in}{0.000000in}}{%
\pgfpathmoveto{\pgfqpoint{-0.000000in}{0.000000in}}%
\pgfpathlineto{\pgfqpoint{-0.048611in}{0.000000in}}%
\pgfusepath{stroke,fill}%
}%
\begin{pgfscope}%
\pgfsys@transformshift{0.553704in}{0.499691in}%
\pgfsys@useobject{currentmarker}{}%
\end{pgfscope}%
\end{pgfscope}%
\begin{pgfscope}%
\definecolor{textcolor}{rgb}{0.000000,0.000000,0.000000}%
\pgfsetstrokecolor{textcolor}%
\pgfsetfillcolor{textcolor}%
\pgftext[x=0.279012in, y=0.451466in, left, base]{\color{textcolor}\rmfamily\fontsize{10.000000}{12.000000}\selectfont \(\displaystyle {0.0}\)}%
\end{pgfscope}%
\begin{pgfscope}%
\pgfsetbuttcap%
\pgfsetroundjoin%
\definecolor{currentfill}{rgb}{0.000000,0.000000,0.000000}%
\pgfsetfillcolor{currentfill}%
\pgfsetlinewidth{0.803000pt}%
\definecolor{currentstroke}{rgb}{0.000000,0.000000,0.000000}%
\pgfsetstrokecolor{currentstroke}%
\pgfsetdash{}{0pt}%
\pgfsys@defobject{currentmarker}{\pgfqpoint{-0.048611in}{0.000000in}}{\pgfqpoint{-0.000000in}{0.000000in}}{%
\pgfpathmoveto{\pgfqpoint{-0.000000in}{0.000000in}}%
\pgfpathlineto{\pgfqpoint{-0.048611in}{0.000000in}}%
\pgfusepath{stroke,fill}%
}%
\begin{pgfscope}%
\pgfsys@transformshift{0.553704in}{1.115691in}%
\pgfsys@useobject{currentmarker}{}%
\end{pgfscope}%
\end{pgfscope}%
\begin{pgfscope}%
\definecolor{textcolor}{rgb}{0.000000,0.000000,0.000000}%
\pgfsetstrokecolor{textcolor}%
\pgfsetfillcolor{textcolor}%
\pgftext[x=0.279012in, y=1.067466in, left, base]{\color{textcolor}\rmfamily\fontsize{10.000000}{12.000000}\selectfont \(\displaystyle {0.2}\)}%
\end{pgfscope}%
\begin{pgfscope}%
\pgfsetbuttcap%
\pgfsetroundjoin%
\definecolor{currentfill}{rgb}{0.000000,0.000000,0.000000}%
\pgfsetfillcolor{currentfill}%
\pgfsetlinewidth{0.803000pt}%
\definecolor{currentstroke}{rgb}{0.000000,0.000000,0.000000}%
\pgfsetstrokecolor{currentstroke}%
\pgfsetdash{}{0pt}%
\pgfsys@defobject{currentmarker}{\pgfqpoint{-0.048611in}{0.000000in}}{\pgfqpoint{-0.000000in}{0.000000in}}{%
\pgfpathmoveto{\pgfqpoint{-0.000000in}{0.000000in}}%
\pgfpathlineto{\pgfqpoint{-0.048611in}{0.000000in}}%
\pgfusepath{stroke,fill}%
}%
\begin{pgfscope}%
\pgfsys@transformshift{0.553704in}{1.731691in}%
\pgfsys@useobject{currentmarker}{}%
\end{pgfscope}%
\end{pgfscope}%
\begin{pgfscope}%
\definecolor{textcolor}{rgb}{0.000000,0.000000,0.000000}%
\pgfsetstrokecolor{textcolor}%
\pgfsetfillcolor{textcolor}%
\pgftext[x=0.279012in, y=1.683466in, left, base]{\color{textcolor}\rmfamily\fontsize{10.000000}{12.000000}\selectfont \(\displaystyle {0.4}\)}%
\end{pgfscope}%
\begin{pgfscope}%
\pgfsetbuttcap%
\pgfsetroundjoin%
\definecolor{currentfill}{rgb}{0.000000,0.000000,0.000000}%
\pgfsetfillcolor{currentfill}%
\pgfsetlinewidth{0.803000pt}%
\definecolor{currentstroke}{rgb}{0.000000,0.000000,0.000000}%
\pgfsetstrokecolor{currentstroke}%
\pgfsetdash{}{0pt}%
\pgfsys@defobject{currentmarker}{\pgfqpoint{-0.048611in}{0.000000in}}{\pgfqpoint{-0.000000in}{0.000000in}}{%
\pgfpathmoveto{\pgfqpoint{-0.000000in}{0.000000in}}%
\pgfpathlineto{\pgfqpoint{-0.048611in}{0.000000in}}%
\pgfusepath{stroke,fill}%
}%
\begin{pgfscope}%
\pgfsys@transformshift{0.553704in}{2.347691in}%
\pgfsys@useobject{currentmarker}{}%
\end{pgfscope}%
\end{pgfscope}%
\begin{pgfscope}%
\definecolor{textcolor}{rgb}{0.000000,0.000000,0.000000}%
\pgfsetstrokecolor{textcolor}%
\pgfsetfillcolor{textcolor}%
\pgftext[x=0.279012in, y=2.299466in, left, base]{\color{textcolor}\rmfamily\fontsize{10.000000}{12.000000}\selectfont \(\displaystyle {0.6}\)}%
\end{pgfscope}%
\begin{pgfscope}%
\pgfsetbuttcap%
\pgfsetroundjoin%
\definecolor{currentfill}{rgb}{0.000000,0.000000,0.000000}%
\pgfsetfillcolor{currentfill}%
\pgfsetlinewidth{0.803000pt}%
\definecolor{currentstroke}{rgb}{0.000000,0.000000,0.000000}%
\pgfsetstrokecolor{currentstroke}%
\pgfsetdash{}{0pt}%
\pgfsys@defobject{currentmarker}{\pgfqpoint{-0.048611in}{0.000000in}}{\pgfqpoint{-0.000000in}{0.000000in}}{%
\pgfpathmoveto{\pgfqpoint{-0.000000in}{0.000000in}}%
\pgfpathlineto{\pgfqpoint{-0.048611in}{0.000000in}}%
\pgfusepath{stroke,fill}%
}%
\begin{pgfscope}%
\pgfsys@transformshift{0.553704in}{2.963691in}%
\pgfsys@useobject{currentmarker}{}%
\end{pgfscope}%
\end{pgfscope}%
\begin{pgfscope}%
\definecolor{textcolor}{rgb}{0.000000,0.000000,0.000000}%
\pgfsetstrokecolor{textcolor}%
\pgfsetfillcolor{textcolor}%
\pgftext[x=0.279012in, y=2.915466in, left, base]{\color{textcolor}\rmfamily\fontsize{10.000000}{12.000000}\selectfont \(\displaystyle {0.8}\)}%
\end{pgfscope}%
\begin{pgfscope}%
\pgfsetbuttcap%
\pgfsetroundjoin%
\definecolor{currentfill}{rgb}{0.000000,0.000000,0.000000}%
\pgfsetfillcolor{currentfill}%
\pgfsetlinewidth{0.803000pt}%
\definecolor{currentstroke}{rgb}{0.000000,0.000000,0.000000}%
\pgfsetstrokecolor{currentstroke}%
\pgfsetdash{}{0pt}%
\pgfsys@defobject{currentmarker}{\pgfqpoint{-0.048611in}{0.000000in}}{\pgfqpoint{-0.000000in}{0.000000in}}{%
\pgfpathmoveto{\pgfqpoint{-0.000000in}{0.000000in}}%
\pgfpathlineto{\pgfqpoint{-0.048611in}{0.000000in}}%
\pgfusepath{stroke,fill}%
}%
\begin{pgfscope}%
\pgfsys@transformshift{0.553704in}{3.579691in}%
\pgfsys@useobject{currentmarker}{}%
\end{pgfscope}%
\end{pgfscope}%
\begin{pgfscope}%
\definecolor{textcolor}{rgb}{0.000000,0.000000,0.000000}%
\pgfsetstrokecolor{textcolor}%
\pgfsetfillcolor{textcolor}%
\pgftext[x=0.279012in, y=3.531466in, left, base]{\color{textcolor}\rmfamily\fontsize{10.000000}{12.000000}\selectfont \(\displaystyle {1.0}\)}%
\end{pgfscope}%
\begin{pgfscope}%
\definecolor{textcolor}{rgb}{0.000000,0.000000,0.000000}%
\pgfsetstrokecolor{textcolor}%
\pgfsetfillcolor{textcolor}%
\pgftext[x=0.223457in,y=2.039691in,,bottom,rotate=90.000000]{\color{textcolor}\rmfamily\fontsize{18.000000}{12.000000}\selectfont Best Model Distribution}%
\end{pgfscope}%
\begin{pgfscope}%
\pgfsetrectcap%
\pgfsetmiterjoin%
\pgfsetlinewidth{0.803000pt}%
\definecolor{currentstroke}{rgb}{0.000000,0.000000,0.000000}%
\pgfsetstrokecolor{currentstroke}%
\pgfsetdash{}{0pt}%
\pgfpathmoveto{\pgfqpoint{0.553704in}{0.499691in}}%
\pgfpathlineto{\pgfqpoint{0.553704in}{3.579691in}}%
\pgfusepath{stroke}%
\end{pgfscope}%
\begin{pgfscope}%
\pgfsetrectcap%
\pgfsetmiterjoin%
\pgfsetlinewidth{0.803000pt}%
\definecolor{currentstroke}{rgb}{0.000000,0.000000,0.000000}%
\pgfsetstrokecolor{currentstroke}%
\pgfsetdash{}{0pt}%
\pgfpathmoveto{\pgfqpoint{5.978704in}{0.499691in}}%
\pgfpathlineto{\pgfqpoint{5.978704in}{3.579691in}}%
\pgfusepath{stroke}%
\end{pgfscope}%
\begin{pgfscope}%
\pgfsetrectcap%
\pgfsetmiterjoin%
\pgfsetlinewidth{0.803000pt}%
\definecolor{currentstroke}{rgb}{0.000000,0.000000,0.000000}%
\pgfsetstrokecolor{currentstroke}%
\pgfsetdash{}{0pt}%
\pgfpathmoveto{\pgfqpoint{0.553704in}{0.499691in}}%
\pgfpathlineto{\pgfqpoint{5.978704in}{0.499691in}}%
\pgfusepath{stroke}%
\end{pgfscope}%
\begin{pgfscope}%
\pgfsetrectcap%
\pgfsetmiterjoin%
\pgfsetlinewidth{0.803000pt}%
\definecolor{currentstroke}{rgb}{0.000000,0.000000,0.000000}%
\pgfsetstrokecolor{currentstroke}%
\pgfsetdash{}{0pt}%
\pgfpathmoveto{\pgfqpoint{0.553704in}{3.579691in}}%
\pgfpathlineto{\pgfqpoint{5.978704in}{3.579691in}}%
\pgfusepath{stroke}%
\end{pgfscope}%
\end{pgfpicture}%
\makeatother%
\endgroup%

%% file: sections/fit.tex
\section{How to choose the best source}
\label{sec:HowTo}
Section~\ref{sec:Transferability} outlines that the choice of the source is of utmost importance thanks to the transferability measure defined in Sect.~\ref{sec:Methodology}. This section investigates whether the attacker can guess which model is the best source.

We now propose a procedure which combines the dependences with respect to the source and target models (Sect.~\ref{sec:model_dependence}), and the input (Sect.~\ref{sec:image_dependence}).
Model dependence is measured by evaluating the similarity between the source and target models, which is denoted as $\ModSim$. On the other hand, the image dependence is measured by the quality of the adversarial example, denoted as $\TransQ$. Both metrics are combined into the following score:
\begin{equation}
\label{eq:FITScore}
    \fit(s, t, x) :=  \ModSim(s, t) \times \TransQ(s, x).
\end{equation}
These indicators should be easy to compute. We especially pay attention to the number of queries to the target.
This score opens the door to a new strategy for the attacker which first selects the best source among the available models
\begin{equation}
    s^\star(t,x) = \arg \max_{\sigma\in\mathcal{F}_s} \fit(\sigma,t,x),
\end{equation}
and then crafts the adversarial direction $\dir_{x,s^\star(t,x)}$.

\subsection{Criterion $\ModSim(s,t)$}
Section~\ref{sec:model_dependence} highlights the correlation between the transferability and the similarity between the source and target models. Gauging model similarity has been previously studied in the context of fingerprinting as a defense to protect intellectual property (see Sect.~\ref{sec:Fingerprinting}). This paper uses fingerprinting methods as an attack that leaks information about the target.

We consider the fingerprinting method~\cite{maho2022fbi} because it works in a decision-based setup since the target is a black box in our application. Querying two models $s$ and $t$ with few natural images, it computes a distance $\texttt{Dist}(s,t)\in[0,1]$ by comparing their outputs.
Since we look for a similarity, we set $\ModSim(s,t) = 1 - \texttt{Dist}(s,t)$.  
However, this provides a symmetrical similarity, ie. $\ModSim(s,t) = \ModSim(t,s)$, while transferability is not (see Fig.~\ref{fig:gain_score_matrix}). This shows that this criterion alone is not sufficient.


\subsection{Criterion $\TransQ(s,x)$}
This criterion evaluates the general transferability of a given adversarial example crafted by a source. Our idea is to leverage the assumption that the attacker has a set of models $\mathcal{F}_s$. 
Consequently, we can evaluate the transferability thanks to the other models of this set, without querying the target.

For a given input $x$, the source $s$ provides the adversarial direction $\dir_{x,s}$ and we compute the distortion $d_{s,\sigma}$ necessary to delude classifier $\sigma\in\mathcal{F}_s$ with~\eqref{eq:MinD}.
We then aggregate these distortions into a single score with two flavours:
\begin{equation}
\label{eq:TransQ1}
    \TransQ^{(1)}(s, x) := \left(\frac{1}{|\mathcal{F}_s|}{\sum_{\sigma \in \mathcal{F}_s} d_{s,\sigma}}\right)^{-1}.
\end{equation}
A good source for input $x$ gives birth to lower distortions, so that $\TransQ^{(1)}(s, x)$ is large.

\begin{equation}
    \TransQ^{(2)}(s, x) := \frac{\sum_{\sigma\in\mathcal{F}_s} d_{s, \sigma} - \dBB_\sigma}{\sum_{\sigma\in\mathcal{F}_s} \dWB_\sigma - \dBB_\sigma}.
    \label{eq:TransQ2}
\end{equation}
This measure is similar to~\eqref{eq:TScoreEmpirical} except that it is computed over the set of models instead of a set of inputs.








%% file: sections/results.tex
\subsection{Results}
\label{sec:results}
The experimental setup is the same as in Sect.~\ref{sec:experimental_setup}. 
We select the fingerprinting method \fbi with 200 benign natural images to compute the criterion $\ModSim(s,t)$.
It implies that the attacker first makes 200 queries to the target in a preliminary step before forging any adversarial example. Appendix~\ref{app:Fingeprinting} shows that more images improve the fingerprinting accuracy hence the model selection, but the results converge after a number of 200 images.

This section is structured as follows: we use \fit to select the best model for single-model attacks, then we employ it to identify the best subset of models for ensemble-model attacks and different combinations of attacks.


\subsubsection{Single-model attacks}

\begin{table*}
\caption{Transferability \measure for \di, \taig and \dwp for single and ensemble-model attacks.}
    \begin{center}
        \begin{tabular}{|c|c|c|c|c|c|}
            \hline
            Category                               & \multicolumn{2}{|c|}{Selection Method} & \di          & \taig         & \dwp                          \\
            \hline
            \multirow{8}{*}{Single-model attack}   & \multicolumn{2}{c|}{Best}              & 0.52         & 0.46          & 0.34                          \\
                                                   & \multicolumn{2}{c|}{Random}            & -0.16        & -0.12          & -0.72                         \\
            \cline{2-6}
            \cline{2-6}
                                                   & \ModSim                               & \fbi         & 0.18         & 0.12          & -0.39         \\
            \cline{2-6}
                                                   & \multirow{3}{*}{$\TransQ$}              & $ASR$        & -0.21         & -0.24         & -1.14         \\
                                                   &                                        & \TransQOne \eqref{eq:TransQ1}       & 0.38          & 0.24          & 0.10          \\
                                                   &                                        & \TransQTwo \eqref{eq:TransQ2}  & 0.37          & 0.23          & 0.08          \\
            \cline{2-6}
                                                   & \multirow{2}{*}{\fit}                  & \TransQOne \eqref{eq:TransQ1}        & \textbf{0.40} & \textbf{0.27} & \textbf{0.12}          \\
                                                   &                                        & \TransQTwo \eqref{eq:TransQ2}  & 0.39 & 0.25 & 0.10 \\
            \hline
            \hline
             & \multicolumn{2}{c|}{Best}              & 0.72         & 0.62          &      0.46                          \\
                                                   & \multicolumn{2}{c|}{Random}            & 0.43         & 0.40          &      0.02                         \\
            \cline{2-6}
            \cline{2-6}
                                                   & \ModSim                               & \fbi         & 0.59          & 0.49          &       0.05       \\
            \cline{2-6}
             Ensemble-model attack                & \multirow{3}{*}{\TransQ}              & $ASR$        & 0.53          & 0.45          &       0.06        \\
                with three sources                                   &                                        & \TransQOne \eqref{eq:TransQ1}        & 0.62          & 0.54          &       0.33        \\
                                                   &                                        &  \TransQTwo \eqref{eq:TransQ2}  & 0.61          & 0.54          &       0.33        \\
            \cline{2-6}
                                                   & \multirow{2}{*}{\fit}                  & \TransQOne \eqref{eq:TransQ1}        & \textbf{0.64} & 0.55          &       0.35        \\
                                                   &                                        & \TransQTwo \eqref{eq:TransQ2}  & \textbf{0.64} & \textbf{0.57} &       \textbf{0.36}        \\
            \hline
        \end{tabular}
        \label{tab:fit_result}
    \end{center}
\end{table*}

\begin{figure}[t]
    \centering
    \resizebox{0.9\linewidth}{!}{\input{./images/results/sources_available/Gain.pgf}}
    \caption{\measure as a function of the number of available sources for several selection methods. Dotted lines refer to the selection of a unique model for all images and solid lines refer to a model selected per image. Attack is \di.}
    \label{fig:result_n_source_impact}
\end{figure}
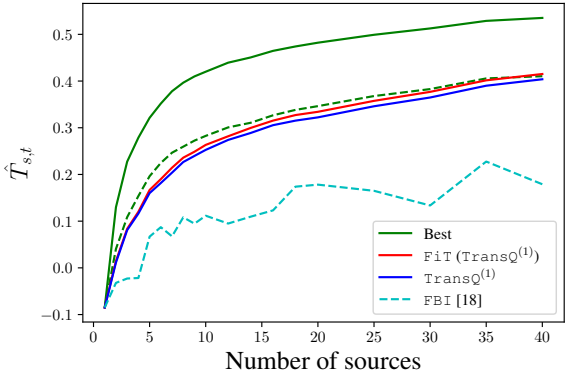

\textbf{Criterion} $\ModSim(s,t)$.
Table~\ref{tab:fit_result} indicates that architectural similarity is a reliable measure of transferability between two models. It can drive the selection of a good source giving birth to a transferable attack outperforming the black box attack since \measure is larger than 0 (except for \dwp). Yet, the results remain low compared to the best results obtained.
As discussed in Sec~\ref{sec:model_dependence}, similarity may not suffice because it implies the selection of one unique for a given target. Better results are achieved when adapting the source to the input.



\textbf{Criterion} $\TransQ(s,x)$.
Improving transferability without querying the target model in a preliminary step is possible thanks to $\TransQ(s,x)$. Adversarial examples that exhibit good transferability on multiple models are more likely to also deceive the unknown targeted model.
Figure~\ref{fig:result_n_source_impact} shows that a significant improvement in transferability is achieved even with only a few available models. Table~\ref{tab:fit_result} confirms this observation for the two other attack methods.
This strategy is indeed better than the selection based on model similarity.

\textbf{Score} \fit(s,t,x).
Combining both criteria together as in~\eqref{eq:FITScore} leads to a slight improvement in transferability compared to $\TransQ(s,x)$ alone. Fig.~\ref{fig:result_n_source_impact} confirms this holds over a wide range of numbers of available sources. For single-model attacks, \TransQOne gives slightly better results than \TransQTwo.

\input{sections/images_figure}

\textbf{Visual results} Figure~\ref{fig:visual} visually demonstrates the impact of different model selection methods on the quality of adversarial examples. Even when \fit is not accurate, the resulting adversarial examples are still close to the best ones obtained from the source models, and the perturbation remains imperceptible. However, random selection generates noisy perturbations, and the worst-case scenario destroys the image entirely.

\subsubsection{Ensemble-model attacks}
\textbf{Importance of model selection in ensemble-model attack.}
Ensemble attacks remain an under-studied area due to the significant computational resources required for evaluating attacks. Consequently, these attacks have only been evaluated with a limited number of sources.
Appendix~\ref{app:EnsModAtt} shows that increasing the number of sources is not necessarily beneficial. 
The \fit score provides a scalable solution: we select a small subset of sources based on their \fit scores (top-3) from the bigger set of available sources. In a way, it is better to put quality above quantity. 
For example, when running an ensemble-model attack against $\texttt{xCiT}_{\texttt{nano}}$, selecting only the three best models using the \fit measure can lead to a significant improvement in transferability compared to using a larger set of models. In our experiments, an ensemble-model of 20 random models achieved a \measure of 0.47, while an ensemble-model of only three models selected with \fit was able to achieve a \measure of 0.56 against the same target.

\textbf{Performance of ensemble-model attacks.}
Table.~\ref{tab:fit_result} shows that ensemble-model transferability surpasses that of single-model attacks. While the average results over a random selection of 3 sources increase, they remain closer to the black box results than the white box ones (transferability lower than 0.5). Choosing the top-3 sources returned by our scores for leading the ensemble-model attack yields better performance, comparable to the best possible results obtainable with ensemble-model attacks. Notably, for the \di and \taig method, it approaches the performance of white box results (transferability greater than 0.5).

\subsubsection{White box vs. transferable attacks}
\begin{table}[b]
\caption{Transferability \measure for white box and transferable attacks (single-model). The \fit score uses \TransQOne.}
    \begin{center}
        \begin{tabular}{|c|c|c|c|}
            \hline

            \multicolumn{2}{|c|}{Attack}           & Best      & \fit         \\
            \hline
            \multirow{4}{*}{White Box}             & \bp       & 0.27          & -0.06 \\
                                                   & \deepfool & 0.03          & -0.34 \\
                                                   & \pgd      & 0.29          & -0.01 \\
                                                   & \ifgsm    & 0.29          & -0.03 \\
            \hline
            \multirow{3}{*}{Transferable}          & \di       & 0.52          & 0.40     \\
                                                   & \taig     & 0.46          & 0.27  \\
                                                   & \dwp      & 0.34          & 0.12  \\
            \hline
            \multicolumn{2}{|c|}{All Attacks}      & 0.65      & \textbf{0.48}         \\
            \hline
        \end{tabular}
        \label{tab:multiple_attacks}
    \end{center}
\end{table}

Our last result is that the selection of a single source with the \fit score does not make traditional white box attacks transferable. Table~\ref{tab:multiple_attacks} shows that \bp, \deepfool, \pgd, and \ifgsm yield negative transferability values.
More precisely, \bp, \ifgsm, and \pgd perform better with \fit than the transferable attacks without any selection mechanism (ie. on average with a random source).
However, section~\ref{sec:attack_dependence} highlights that the traditional white box attacks may be competitive for some inputs demanding low adversarial perturbation distortion.
It is valuable to add them to the options of the attacker and let the \fit score decide the preferable option (source and attack method). This provides our best transferability score for a single model attack, close to 0.5.  



%% file: images/results/sources_available/Gain.pgf
\begingroup%
\makeatletter%
\begin{pgfpicture}%
\pgfpathrectangle{\pgfpointorigin}{\pgfqpoint{5.450579in}{3.703148in}}%
\pgfusepath{use as bounding box}%
\begin{pgfscope}%
\pgfsetbuttcap%
\pgfsetmiterjoin%
\definecolor{currentfill}{rgb}{1.000000,1.000000,1.000000}%
\pgfsetfillcolor{currentfill}%
\pgfsetlinewidth{0.000000pt}%
\definecolor{currentstroke}{rgb}{1.000000,1.000000,1.000000}%
\pgfsetstrokecolor{currentstroke}%
\pgfsetdash{}{0pt}%
\pgfpathmoveto{\pgfqpoint{0.000000in}{0.000000in}}%
\pgfpathlineto{\pgfqpoint{5.450579in}{0.000000in}}%
\pgfpathlineto{\pgfqpoint{5.450579in}{3.703148in}}%
\pgfpathlineto{\pgfqpoint{0.000000in}{3.703148in}}%
\pgfpathlineto{\pgfqpoint{0.000000in}{0.000000in}}%
\pgfpathclose%
\pgfusepath{fill}%
\end{pgfscope}%
\begin{pgfscope}%
\pgfsetbuttcap%
\pgfsetmiterjoin%
\definecolor{currentfill}{rgb}{1.000000,1.000000,1.000000}%
\pgfsetfillcolor{currentfill}%
\pgfsetlinewidth{0.000000pt}%
\definecolor{currentstroke}{rgb}{0.000000,0.000000,0.000000}%
\pgfsetstrokecolor{currentstroke}%
\pgfsetstrokeopacity{0.000000}%
\pgfsetdash{}{0pt}%
\pgfpathmoveto{\pgfqpoint{0.700579in}{0.523148in}}%
\pgfpathlineto{\pgfqpoint{5.350579in}{0.523148in}}%
\pgfpathlineto{\pgfqpoint{5.350579in}{3.603148in}}%
\pgfpathlineto{\pgfqpoint{0.700579in}{3.603148in}}%
\pgfpathlineto{\pgfqpoint{0.700579in}{0.523148in}}%
\pgfpathclose%
\pgfusepath{fill}%
\end{pgfscope}%
\begin{pgfscope}%
\pgfsetbuttcap%
\pgfsetroundjoin%
\definecolor{currentfill}{rgb}{0.000000,0.000000,0.000000}%
\pgfsetfillcolor{currentfill}%
\pgfsetlinewidth{0.803000pt}%
\definecolor{currentstroke}{rgb}{0.000000,0.000000,0.000000}%
\pgfsetstrokecolor{currentstroke}%
\pgfsetdash{}{0pt}%
\pgfsys@defobject{currentmarker}{\pgfqpoint{0.000000in}{-0.048611in}}{\pgfqpoint{0.000000in}{0.000000in}}{%
\pgfpathmoveto{\pgfqpoint{0.000000in}{0.000000in}}%
\pgfpathlineto{\pgfqpoint{0.000000in}{-0.048611in}}%
\pgfusepath{stroke,fill}%
}%
\begin{pgfscope}%
\pgfsys@transformshift{0.803551in}{0.523148in}%
\pgfsys@useobject{currentmarker}{}%
\end{pgfscope}%
\end{pgfscope}%
\begin{pgfscope}%
\definecolor{textcolor}{rgb}{0.000000,0.000000,0.000000}%
\pgfsetstrokecolor{textcolor}%
\pgfsetfillcolor{textcolor}%
\pgftext[x=0.803551in,y=0.425926in,,top]{\color{textcolor}\rmfamily\fontsize{11.000000}{13.200000}\selectfont \(\displaystyle {0}\)}%
\end{pgfscope}%
\begin{pgfscope}%
\pgfsetbuttcap%
\pgfsetroundjoin%
\definecolor{currentfill}{rgb}{0.000000,0.000000,0.000000}%
\pgfsetfillcolor{currentfill}%
\pgfsetlinewidth{0.803000pt}%
\definecolor{currentstroke}{rgb}{0.000000,0.000000,0.000000}%
\pgfsetstrokecolor{currentstroke}%
\pgfsetdash{}{0pt}%
\pgfsys@defobject{currentmarker}{\pgfqpoint{0.000000in}{-0.048611in}}{\pgfqpoint{0.000000in}{0.000000in}}{%
\pgfpathmoveto{\pgfqpoint{0.000000in}{0.000000in}}%
\pgfpathlineto{\pgfqpoint{0.000000in}{-0.048611in}}%
\pgfusepath{stroke,fill}%
}%
\begin{pgfscope}%
\pgfsys@transformshift{1.345509in}{0.523148in}%
\pgfsys@useobject{currentmarker}{}%
\end{pgfscope}%
\end{pgfscope}%
\begin{pgfscope}%
\definecolor{textcolor}{rgb}{0.000000,0.000000,0.000000}%
\pgfsetstrokecolor{textcolor}%
\pgfsetfillcolor{textcolor}%
\pgftext[x=1.345509in,y=0.425926in,,top]{\color{textcolor}\rmfamily\fontsize{11.000000}{13.200000}\selectfont \(\displaystyle {5}\)}%
\end{pgfscope}%
\begin{pgfscope}%
\pgfsetbuttcap%
\pgfsetroundjoin%
\definecolor{currentfill}{rgb}{0.000000,0.000000,0.000000}%
\pgfsetfillcolor{currentfill}%
\pgfsetlinewidth{0.803000pt}%
\definecolor{currentstroke}{rgb}{0.000000,0.000000,0.000000}%
\pgfsetstrokecolor{currentstroke}%
\pgfsetdash{}{0pt}%
\pgfsys@defobject{currentmarker}{\pgfqpoint{0.000000in}{-0.048611in}}{\pgfqpoint{0.000000in}{0.000000in}}{%
\pgfpathmoveto{\pgfqpoint{0.000000in}{0.000000in}}%
\pgfpathlineto{\pgfqpoint{0.000000in}{-0.048611in}}%
\pgfusepath{stroke,fill}%
}%
\begin{pgfscope}%
\pgfsys@transformshift{1.887467in}{0.523148in}%
\pgfsys@useobject{currentmarker}{}%
\end{pgfscope}%
\end{pgfscope}%
\begin{pgfscope}%
\definecolor{textcolor}{rgb}{0.000000,0.000000,0.000000}%
\pgfsetstrokecolor{textcolor}%
\pgfsetfillcolor{textcolor}%
\pgftext[x=1.887467in,y=0.425926in,,top]{\color{textcolor}\rmfamily\fontsize{11.000000}{13.200000}\selectfont \(\displaystyle {10}\)}%
\end{pgfscope}%
\begin{pgfscope}%
\pgfsetbuttcap%
\pgfsetroundjoin%
\definecolor{currentfill}{rgb}{0.000000,0.000000,0.000000}%
\pgfsetfillcolor{currentfill}%
\pgfsetlinewidth{0.803000pt}%
\definecolor{currentstroke}{rgb}{0.000000,0.000000,0.000000}%
\pgfsetstrokecolor{currentstroke}%
\pgfsetdash{}{0pt}%
\pgfsys@defobject{currentmarker}{\pgfqpoint{0.000000in}{-0.048611in}}{\pgfqpoint{0.000000in}{0.000000in}}{%
\pgfpathmoveto{\pgfqpoint{0.000000in}{0.000000in}}%
\pgfpathlineto{\pgfqpoint{0.000000in}{-0.048611in}}%
\pgfusepath{stroke,fill}%
}%
\begin{pgfscope}%
\pgfsys@transformshift{2.429425in}{0.523148in}%
\pgfsys@useobject{currentmarker}{}%
\end{pgfscope}%
\end{pgfscope}%
\begin{pgfscope}%
\definecolor{textcolor}{rgb}{0.000000,0.000000,0.000000}%
\pgfsetstrokecolor{textcolor}%
\pgfsetfillcolor{textcolor}%
\pgftext[x=2.429425in,y=0.425926in,,top]{\color{textcolor}\rmfamily\fontsize{11.000000}{13.200000}\selectfont \(\displaystyle {15}\)}%
\end{pgfscope}%
\begin{pgfscope}%
\pgfsetbuttcap%
\pgfsetroundjoin%
\definecolor{currentfill}{rgb}{0.000000,0.000000,0.000000}%
\pgfsetfillcolor{currentfill}%
\pgfsetlinewidth{0.803000pt}%
\definecolor{currentstroke}{rgb}{0.000000,0.000000,0.000000}%
\pgfsetstrokecolor{currentstroke}%
\pgfsetdash{}{0pt}%
\pgfsys@defobject{currentmarker}{\pgfqpoint{0.000000in}{-0.048611in}}{\pgfqpoint{0.000000in}{0.000000in}}{%
\pgfpathmoveto{\pgfqpoint{0.000000in}{0.000000in}}%
\pgfpathlineto{\pgfqpoint{0.000000in}{-0.048611in}}%
\pgfusepath{stroke,fill}%
}%
\begin{pgfscope}%
\pgfsys@transformshift{2.971384in}{0.523148in}%
\pgfsys@useobject{currentmarker}{}%
\end{pgfscope}%
\end{pgfscope}%
\begin{pgfscope}%
\definecolor{textcolor}{rgb}{0.000000,0.000000,0.000000}%
\pgfsetstrokecolor{textcolor}%
\pgfsetfillcolor{textcolor}%
\pgftext[x=2.971384in,y=0.425926in,,top]{\color{textcolor}\rmfamily\fontsize{11.000000}{13.200000}\selectfont \(\displaystyle {20}\)}%
\end{pgfscope}%
\begin{pgfscope}%
\pgfsetbuttcap%
\pgfsetroundjoin%
\definecolor{currentfill}{rgb}{0.000000,0.000000,0.000000}%
\pgfsetfillcolor{currentfill}%
\pgfsetlinewidth{0.803000pt}%
\definecolor{currentstroke}{rgb}{0.000000,0.000000,0.000000}%
\pgfsetstrokecolor{currentstroke}%
\pgfsetdash{}{0pt}%
\pgfsys@defobject{currentmarker}{\pgfqpoint{0.000000in}{-0.048611in}}{\pgfqpoint{0.000000in}{0.000000in}}{%
\pgfpathmoveto{\pgfqpoint{0.000000in}{0.000000in}}%
\pgfpathlineto{\pgfqpoint{0.000000in}{-0.048611in}}%
\pgfusepath{stroke,fill}%
}%
\begin{pgfscope}%
\pgfsys@transformshift{3.513342in}{0.523148in}%
\pgfsys@useobject{currentmarker}{}%
\end{pgfscope}%
\end{pgfscope}%
\begin{pgfscope}%
\definecolor{textcolor}{rgb}{0.000000,0.000000,0.000000}%
\pgfsetstrokecolor{textcolor}%
\pgfsetfillcolor{textcolor}%
\pgftext[x=3.513342in,y=0.425926in,,top]{\color{textcolor}\rmfamily\fontsize{11.000000}{13.200000}\selectfont \(\displaystyle {25}\)}%
\end{pgfscope}%
\begin{pgfscope}%
\pgfsetbuttcap%
\pgfsetroundjoin%
\definecolor{currentfill}{rgb}{0.000000,0.000000,0.000000}%
\pgfsetfillcolor{currentfill}%
\pgfsetlinewidth{0.803000pt}%
\definecolor{currentstroke}{rgb}{0.000000,0.000000,0.000000}%
\pgfsetstrokecolor{currentstroke}%
\pgfsetdash{}{0pt}%
\pgfsys@defobject{currentmarker}{\pgfqpoint{0.000000in}{-0.048611in}}{\pgfqpoint{0.000000in}{0.000000in}}{%
\pgfpathmoveto{\pgfqpoint{0.000000in}{0.000000in}}%
\pgfpathlineto{\pgfqpoint{0.000000in}{-0.048611in}}%
\pgfusepath{stroke,fill}%
}%
\begin{pgfscope}%
\pgfsys@transformshift{4.055300in}{0.523148in}%
\pgfsys@useobject{currentmarker}{}%
\end{pgfscope}%
\end{pgfscope}%
\begin{pgfscope}%
\definecolor{textcolor}{rgb}{0.000000,0.000000,0.000000}%
\pgfsetstrokecolor{textcolor}%
\pgfsetfillcolor{textcolor}%
\pgftext[x=4.055300in,y=0.425926in,,top]{\color{textcolor}\rmfamily\fontsize{11.000000}{13.200000}\selectfont \(\displaystyle {30}\)}%
\end{pgfscope}%
\begin{pgfscope}%
\pgfsetbuttcap%
\pgfsetroundjoin%
\definecolor{currentfill}{rgb}{0.000000,0.000000,0.000000}%
\pgfsetfillcolor{currentfill}%
\pgfsetlinewidth{0.803000pt}%
\definecolor{currentstroke}{rgb}{0.000000,0.000000,0.000000}%
\pgfsetstrokecolor{currentstroke}%
\pgfsetdash{}{0pt}%
\pgfsys@defobject{currentmarker}{\pgfqpoint{0.000000in}{-0.048611in}}{\pgfqpoint{0.000000in}{0.000000in}}{%
\pgfpathmoveto{\pgfqpoint{0.000000in}{0.000000in}}%
\pgfpathlineto{\pgfqpoint{0.000000in}{-0.048611in}}%
\pgfusepath{stroke,fill}%
}%
\begin{pgfscope}%
\pgfsys@transformshift{4.597258in}{0.523148in}%
\pgfsys@useobject{currentmarker}{}%
\end{pgfscope}%
\end{pgfscope}%
\begin{pgfscope}%
\definecolor{textcolor}{rgb}{0.000000,0.000000,0.000000}%
\pgfsetstrokecolor{textcolor}%
\pgfsetfillcolor{textcolor}%
\pgftext[x=4.597258in,y=0.425926in,,top]{\color{textcolor}\rmfamily\fontsize{11.000000}{13.200000}\selectfont \(\displaystyle {35}\)}%
\end{pgfscope}%
\begin{pgfscope}%
\pgfsetbuttcap%
\pgfsetroundjoin%
\definecolor{currentfill}{rgb}{0.000000,0.000000,0.000000}%
\pgfsetfillcolor{currentfill}%
\pgfsetlinewidth{0.803000pt}%
\definecolor{currentstroke}{rgb}{0.000000,0.000000,0.000000}%
\pgfsetstrokecolor{currentstroke}%
\pgfsetdash{}{0pt}%
\pgfsys@defobject{currentmarker}{\pgfqpoint{0.000000in}{-0.048611in}}{\pgfqpoint{0.000000in}{0.000000in}}{%
\pgfpathmoveto{\pgfqpoint{0.000000in}{0.000000in}}%
\pgfpathlineto{\pgfqpoint{0.000000in}{-0.048611in}}%
\pgfusepath{stroke,fill}%
}%
\begin{pgfscope}%
\pgfsys@transformshift{5.139216in}{0.523148in}%
\pgfsys@useobject{currentmarker}{}%
\end{pgfscope}%
\end{pgfscope}%
\begin{pgfscope}%
\definecolor{textcolor}{rgb}{0.000000,0.000000,0.000000}%
\pgfsetstrokecolor{textcolor}%
\pgfsetfillcolor{textcolor}%
\pgftext[x=5.139216in,y=0.425926in,,top]{\color{textcolor}\rmfamily\fontsize{11.000000}{13.200000}\selectfont \(\displaystyle {40}\)}%
\end{pgfscope}%
\begin{pgfscope}%
\definecolor{textcolor}{rgb}{0.000000,0.000000,0.000000}%
\pgfsetstrokecolor{textcolor}%
\pgfsetfillcolor{textcolor}%
\pgftext[x=3.025579in,y=0.235185in,,top]{\color{textcolor}\rmfamily\fontsize{18.000000}{13.200000}\selectfont Number of sources}%
\end{pgfscope}%
\begin{pgfscope}%
\pgfsetbuttcap%
\pgfsetroundjoin%
\definecolor{currentfill}{rgb}{0.000000,0.000000,0.000000}%
\pgfsetfillcolor{currentfill}%
\pgfsetlinewidth{0.803000pt}%
\definecolor{currentstroke}{rgb}{0.000000,0.000000,0.000000}%
\pgfsetstrokecolor{currentstroke}%
\pgfsetdash{}{0pt}%
\pgfsys@defobject{currentmarker}{\pgfqpoint{-0.048611in}{0.000000in}}{\pgfqpoint{-0.000000in}{0.000000in}}{%
\pgfpathmoveto{\pgfqpoint{-0.000000in}{0.000000in}}%
\pgfpathlineto{\pgfqpoint{-0.048611in}{0.000000in}}%
\pgfusepath{stroke,fill}%
}%
\begin{pgfscope}%
\pgfsys@transformshift{0.700579in}{0.595197in}%
\pgfsys@useobject{currentmarker}{}%
\end{pgfscope}%
\end{pgfscope}%
\begin{pgfscope}%
\definecolor{textcolor}{rgb}{0.000000,0.000000,0.000000}%
\pgfsetstrokecolor{textcolor}%
\pgfsetfillcolor{textcolor}%
\pgftext[x=0.290741in, y=0.542391in, left, base]{\color{textcolor}\rmfamily\fontsize{11.000000}{13.200000}\selectfont \(\displaystyle {\ensuremath{-}0.1}\)}%
\end{pgfscope}%
\begin{pgfscope}%
\pgfsetbuttcap%
\pgfsetroundjoin%
\definecolor{currentfill}{rgb}{0.000000,0.000000,0.000000}%
\pgfsetfillcolor{currentfill}%
\pgfsetlinewidth{0.803000pt}%
\definecolor{currentstroke}{rgb}{0.000000,0.000000,0.000000}%
\pgfsetstrokecolor{currentstroke}%
\pgfsetdash{}{0pt}%
\pgfsys@defobject{currentmarker}{\pgfqpoint{-0.048611in}{0.000000in}}{\pgfqpoint{-0.000000in}{0.000000in}}{%
\pgfpathmoveto{\pgfqpoint{-0.000000in}{0.000000in}}%
\pgfpathlineto{\pgfqpoint{-0.048611in}{0.000000in}}%
\pgfusepath{stroke,fill}%
}%
\begin{pgfscope}%
\pgfsys@transformshift{0.700579in}{1.046702in}%
\pgfsys@useobject{currentmarker}{}%
\end{pgfscope}%
\end{pgfscope}%
\begin{pgfscope}%
\definecolor{textcolor}{rgb}{0.000000,0.000000,0.000000}%
\pgfsetstrokecolor{textcolor}%
\pgfsetfillcolor{textcolor}%
\pgftext[x=0.409028in, y=0.993895in, left, base]{\color{textcolor}\rmfamily\fontsize{11.000000}{13.200000}\selectfont \(\displaystyle {0.0}\)}%
\end{pgfscope}%
\begin{pgfscope}%
\pgfsetbuttcap%
\pgfsetroundjoin%
\definecolor{currentfill}{rgb}{0.000000,0.000000,0.000000}%
\pgfsetfillcolor{currentfill}%
\pgfsetlinewidth{0.803000pt}%
\definecolor{currentstroke}{rgb}{0.000000,0.000000,0.000000}%
\pgfsetstrokecolor{currentstroke}%
\pgfsetdash{}{0pt}%
\pgfsys@defobject{currentmarker}{\pgfqpoint{-0.048611in}{0.000000in}}{\pgfqpoint{-0.000000in}{0.000000in}}{%
\pgfpathmoveto{\pgfqpoint{-0.000000in}{0.000000in}}%
\pgfpathlineto{\pgfqpoint{-0.048611in}{0.000000in}}%
\pgfusepath{stroke,fill}%
}%
\begin{pgfscope}%
\pgfsys@transformshift{0.700579in}{1.498207in}%
\pgfsys@useobject{currentmarker}{}%
\end{pgfscope}%
\end{pgfscope}%
\begin{pgfscope}%
\definecolor{textcolor}{rgb}{0.000000,0.000000,0.000000}%
\pgfsetstrokecolor{textcolor}%
\pgfsetfillcolor{textcolor}%
\pgftext[x=0.409028in, y=1.445400in, left, base]{\color{textcolor}\rmfamily\fontsize{11.000000}{13.200000}\selectfont \(\displaystyle {0.1}\)}%
\end{pgfscope}%
\begin{pgfscope}%
\pgfsetbuttcap%
\pgfsetroundjoin%
\definecolor{currentfill}{rgb}{0.000000,0.000000,0.000000}%
\pgfsetfillcolor{currentfill}%
\pgfsetlinewidth{0.803000pt}%
\definecolor{currentstroke}{rgb}{0.000000,0.000000,0.000000}%
\pgfsetstrokecolor{currentstroke}%
\pgfsetdash{}{0pt}%
\pgfsys@defobject{currentmarker}{\pgfqpoint{-0.048611in}{0.000000in}}{\pgfqpoint{-0.000000in}{0.000000in}}{%
\pgfpathmoveto{\pgfqpoint{-0.000000in}{0.000000in}}%
\pgfpathlineto{\pgfqpoint{-0.048611in}{0.000000in}}%
\pgfusepath{stroke,fill}%
}%
\begin{pgfscope}%
\pgfsys@transformshift{0.700579in}{1.949711in}%
\pgfsys@useobject{currentmarker}{}%
\end{pgfscope}%
\end{pgfscope}%
\begin{pgfscope}%
\definecolor{textcolor}{rgb}{0.000000,0.000000,0.000000}%
\pgfsetstrokecolor{textcolor}%
\pgfsetfillcolor{textcolor}%
\pgftext[x=0.409028in, y=1.896905in, left, base]{\color{textcolor}\rmfamily\fontsize{11.000000}{13.200000}\selectfont \(\displaystyle {0.2}\)}%
\end{pgfscope}%
\begin{pgfscope}%
\pgfsetbuttcap%
\pgfsetroundjoin%
\definecolor{currentfill}{rgb}{0.000000,0.000000,0.000000}%
\pgfsetfillcolor{currentfill}%
\pgfsetlinewidth{0.803000pt}%
\definecolor{currentstroke}{rgb}{0.000000,0.000000,0.000000}%
\pgfsetstrokecolor{currentstroke}%
\pgfsetdash{}{0pt}%
\pgfsys@defobject{currentmarker}{\pgfqpoint{-0.048611in}{0.000000in}}{\pgfqpoint{-0.000000in}{0.000000in}}{%
\pgfpathmoveto{\pgfqpoint{-0.000000in}{0.000000in}}%
\pgfpathlineto{\pgfqpoint{-0.048611in}{0.000000in}}%
\pgfusepath{stroke,fill}%
}%
\begin{pgfscope}%
\pgfsys@transformshift{0.700579in}{2.401216in}%
\pgfsys@useobject{currentmarker}{}%
\end{pgfscope}%
\end{pgfscope}%
\begin{pgfscope}%
\definecolor{textcolor}{rgb}{0.000000,0.000000,0.000000}%
\pgfsetstrokecolor{textcolor}%
\pgfsetfillcolor{textcolor}%
\pgftext[x=0.409028in, y=2.348409in, left, base]{\color{textcolor}\rmfamily\fontsize{11.000000}{13.200000}\selectfont \(\displaystyle {0.3}\)}%
\end{pgfscope}%
\begin{pgfscope}%
\pgfsetbuttcap%
\pgfsetroundjoin%
\definecolor{currentfill}{rgb}{0.000000,0.000000,0.000000}%
\pgfsetfillcolor{currentfill}%
\pgfsetlinewidth{0.803000pt}%
\definecolor{currentstroke}{rgb}{0.000000,0.000000,0.000000}%
\pgfsetstrokecolor{currentstroke}%
\pgfsetdash{}{0pt}%
\pgfsys@defobject{currentmarker}{\pgfqpoint{-0.048611in}{0.000000in}}{\pgfqpoint{-0.000000in}{0.000000in}}{%
\pgfpathmoveto{\pgfqpoint{-0.000000in}{0.000000in}}%
\pgfpathlineto{\pgfqpoint{-0.048611in}{0.000000in}}%
\pgfusepath{stroke,fill}%
}%
\begin{pgfscope}%
\pgfsys@transformshift{0.700579in}{2.852721in}%
\pgfsys@useobject{currentmarker}{}%
\end{pgfscope}%
\end{pgfscope}%
\begin{pgfscope}%
\definecolor{textcolor}{rgb}{0.000000,0.000000,0.000000}%
\pgfsetstrokecolor{textcolor}%
\pgfsetfillcolor{textcolor}%
\pgftext[x=0.409028in, y=2.799914in, left, base]{\color{textcolor}\rmfamily\fontsize{11.000000}{13.200000}\selectfont \(\displaystyle {0.4}\)}%
\end{pgfscope}%
\begin{pgfscope}%
\pgfsetbuttcap%
\pgfsetroundjoin%
\definecolor{currentfill}{rgb}{0.000000,0.000000,0.000000}%
\pgfsetfillcolor{currentfill}%
\pgfsetlinewidth{0.803000pt}%
\definecolor{currentstroke}{rgb}{0.000000,0.000000,0.000000}%
\pgfsetstrokecolor{currentstroke}%
\pgfsetdash{}{0pt}%
\pgfsys@defobject{currentmarker}{\pgfqpoint{-0.048611in}{0.000000in}}{\pgfqpoint{-0.000000in}{0.000000in}}{%
\pgfpathmoveto{\pgfqpoint{-0.000000in}{0.000000in}}%
\pgfpathlineto{\pgfqpoint{-0.048611in}{0.000000in}}%
\pgfusepath{stroke,fill}%
}%
\begin{pgfscope}%
\pgfsys@transformshift{0.700579in}{3.304225in}%
\pgfsys@useobject{currentmarker}{}%
\end{pgfscope}%
\end{pgfscope}%
\begin{pgfscope}%
\definecolor{textcolor}{rgb}{0.000000,0.000000,0.000000}%
\pgfsetstrokecolor{textcolor}%
\pgfsetfillcolor{textcolor}%
\pgftext[x=0.409028in, y=3.251419in, left, base]{\color{textcolor}\rmfamily\fontsize{11.000000}{13.200000}\selectfont \(\displaystyle {0.5}\)}%
\end{pgfscope}%
\begin{pgfscope}%
\definecolor{textcolor}{rgb}{0.000000,0.000000,0.000000}%
\pgfsetstrokecolor{textcolor}%
\pgfsetfillcolor{textcolor}%
\pgftext[x=0.235185in,y=2.063148in,,bottom,rotate=90.000000]{\color{textcolor}\rmfamily\fontsize{18.000000}{13.200000}\selectfont \measure}%
\end{pgfscope}%
\begin{pgfscope}%
\pgfpathrectangle{\pgfqpoint{0.700579in}{0.523148in}}{\pgfqpoint{4.650000in}{3.080000in}}%
\pgfusepath{clip}%
\pgfsetrectcap%
\pgfsetroundjoin%
\pgfsetlinewidth{1.505625pt}%
\definecolor{currentstroke}{rgb}{0.000000,0.500000,0.000000}%
\pgfsetstrokecolor{currentstroke}%
\pgfsetdash{}{0pt}%
\pgfpathmoveto{\pgfqpoint{0.911943in}{0.663148in}}%
\pgfpathlineto{\pgfqpoint{1.020335in}{1.632949in}}%
\pgfpathlineto{\pgfqpoint{1.128726in}{2.073525in}}%
\pgfpathlineto{\pgfqpoint{1.237118in}{2.305412in}}%
\pgfpathlineto{\pgfqpoint{1.345509in}{2.497467in}}%
\pgfpathlineto{\pgfqpoint{1.453901in}{2.635156in}}%
\pgfpathlineto{\pgfqpoint{1.562293in}{2.754454in}}%
\pgfpathlineto{\pgfqpoint{1.670684in}{2.837108in}}%
\pgfpathlineto{\pgfqpoint{1.779076in}{2.897744in}}%
\pgfpathlineto{\pgfqpoint{1.887467in}{2.942995in}}%
\pgfpathlineto{\pgfqpoint{2.104251in}{3.030383in}}%
\pgfpathlineto{\pgfqpoint{2.321034in}{3.080797in}}%
\pgfpathlineto{\pgfqpoint{2.537817in}{3.144597in}}%
\pgfpathlineto{\pgfqpoint{2.754600in}{3.187170in}}%
\pgfpathlineto{\pgfqpoint{2.971384in}{3.223581in}}%
\pgfpathlineto{\pgfqpoint{3.513342in}{3.299962in}}%
\pgfpathlineto{\pgfqpoint{4.055300in}{3.361389in}}%
\pgfpathlineto{\pgfqpoint{4.597258in}{3.434467in}}%
\pgfpathlineto{\pgfqpoint{5.139216in}{3.463148in}}%
\pgfusepath{stroke}%
\end{pgfscope}%
\begin{pgfscope}%
\pgfpathrectangle{\pgfqpoint{0.700579in}{0.523148in}}{\pgfqpoint{4.650000in}{3.080000in}}%
\pgfusepath{clip}%
\pgfsetbuttcap%
\pgfsetroundjoin%
\pgfsetlinewidth{1.505625pt}%
\definecolor{currentstroke}{rgb}{0.000000,0.500000,0.000000}%
\pgfsetstrokecolor{currentstroke}%
\pgfsetdash{{5.550000pt}{2.400000pt}}{0.000000pt}%
\pgfpathmoveto{\pgfqpoint{0.911943in}{0.663148in}}%
\pgfpathlineto{\pgfqpoint{1.020335in}{1.229786in}}%
\pgfpathlineto{\pgfqpoint{1.128726in}{1.536963in}}%
\pgfpathlineto{\pgfqpoint{1.237118in}{1.741325in}}%
\pgfpathlineto{\pgfqpoint{1.345509in}{1.932891in}}%
\pgfpathlineto{\pgfqpoint{1.453901in}{2.061289in}}%
\pgfpathlineto{\pgfqpoint{1.562293in}{2.160241in}}%
\pgfpathlineto{\pgfqpoint{1.670684in}{2.218570in}}%
\pgfpathlineto{\pgfqpoint{1.779076in}{2.277063in}}%
\pgfpathlineto{\pgfqpoint{1.887467in}{2.322636in}}%
\pgfpathlineto{\pgfqpoint{2.104251in}{2.404681in}}%
\pgfpathlineto{\pgfqpoint{2.321034in}{2.449240in}}%
\pgfpathlineto{\pgfqpoint{2.537817in}{2.521637in}}%
\pgfpathlineto{\pgfqpoint{2.754600in}{2.574020in}}%
\pgfpathlineto{\pgfqpoint{2.971384in}{2.610282in}}%
\pgfpathlineto{\pgfqpoint{3.513342in}{2.707743in}}%
\pgfpathlineto{\pgfqpoint{4.055300in}{2.775063in}}%
\pgfpathlineto{\pgfqpoint{4.597258in}{2.878101in}}%
\pgfpathlineto{\pgfqpoint{5.139216in}{2.899830in}}%
\pgfusepath{stroke}%
\end{pgfscope}%
\begin{pgfscope}%
\pgfpathrectangle{\pgfqpoint{0.700579in}{0.523148in}}{\pgfqpoint{4.650000in}{3.080000in}}%
\pgfusepath{clip}%
\pgfsetrectcap%
\pgfsetroundjoin%
\pgfsetlinewidth{1.505625pt}%
\definecolor{currentstroke}{rgb}{1.000000,0.000000,0.000000}%
\pgfsetstrokecolor{currentstroke}%
\pgfsetdash{}{0pt}%
\pgfpathmoveto{\pgfqpoint{0.911943in}{0.663148in}}%
\pgfpathlineto{\pgfqpoint{1.020335in}{1.112611in}}%
\pgfpathlineto{\pgfqpoint{1.128726in}{1.424495in}}%
\pgfpathlineto{\pgfqpoint{1.237118in}{1.589077in}}%
\pgfpathlineto{\pgfqpoint{1.345509in}{1.799656in}}%
\pgfpathlineto{\pgfqpoint{1.453901in}{1.904609in}}%
\pgfpathlineto{\pgfqpoint{1.562293in}{2.015073in}}%
\pgfpathlineto{\pgfqpoint{1.670684in}{2.112061in}}%
\pgfpathlineto{\pgfqpoint{1.779076in}{2.169678in}}%
\pgfpathlineto{\pgfqpoint{1.887467in}{2.235501in}}%
\pgfpathlineto{\pgfqpoint{2.104251in}{2.318884in}}%
\pgfpathlineto{\pgfqpoint{2.321034in}{2.399262in}}%
\pgfpathlineto{\pgfqpoint{2.537817in}{2.470300in}}%
\pgfpathlineto{\pgfqpoint{2.754600in}{2.523933in}}%
\pgfpathlineto{\pgfqpoint{2.971384in}{2.555404in}}%
\pgfpathlineto{\pgfqpoint{3.513342in}{2.661188in}}%
\pgfpathlineto{\pgfqpoint{4.055300in}{2.748203in}}%
\pgfpathlineto{\pgfqpoint{4.597258in}{2.859645in}}%
\pgfpathlineto{\pgfqpoint{5.139216in}{2.920815in}}%
\pgfusepath{stroke}%
\end{pgfscope}%
\begin{pgfscope}%
\pgfpathrectangle{\pgfqpoint{0.700579in}{0.523148in}}{\pgfqpoint{4.650000in}{3.080000in}}%
\pgfusepath{clip}%
\pgfsetrectcap%
\pgfsetroundjoin%
\pgfsetlinewidth{1.505625pt}%
\definecolor{currentstroke}{rgb}{0.000000,0.000000,1.000000}%
\pgfsetstrokecolor{currentstroke}%
\pgfsetdash{}{0pt}%
\pgfpathmoveto{\pgfqpoint{0.911943in}{0.663148in}}%
\pgfpathlineto{\pgfqpoint{1.020335in}{1.097995in}}%
\pgfpathlineto{\pgfqpoint{1.128726in}{1.411652in}}%
\pgfpathlineto{\pgfqpoint{1.237118in}{1.571797in}}%
\pgfpathlineto{\pgfqpoint{1.345509in}{1.769416in}}%
\pgfpathlineto{\pgfqpoint{1.453901in}{1.866990in}}%
\pgfpathlineto{\pgfqpoint{1.562293in}{1.967435in}}%
\pgfpathlineto{\pgfqpoint{1.670684in}{2.069990in}}%
\pgfpathlineto{\pgfqpoint{1.779076in}{2.129061in}}%
\pgfpathlineto{\pgfqpoint{1.887467in}{2.187289in}}%
\pgfpathlineto{\pgfqpoint{2.104251in}{2.283608in}}%
\pgfpathlineto{\pgfqpoint{2.321034in}{2.350485in}}%
\pgfpathlineto{\pgfqpoint{2.537817in}{2.426161in}}%
\pgfpathlineto{\pgfqpoint{2.754600in}{2.470064in}}%
\pgfpathlineto{\pgfqpoint{2.971384in}{2.501671in}}%
\pgfpathlineto{\pgfqpoint{3.513342in}{2.608932in}}%
\pgfpathlineto{\pgfqpoint{4.055300in}{2.693626in}}%
\pgfpathlineto{\pgfqpoint{4.597258in}{2.807678in}}%
\pgfpathlineto{\pgfqpoint{5.139216in}{2.869899in}}%
\pgfusepath{stroke}%
\end{pgfscope}%
\begin{pgfscope}%
\pgfpathrectangle{\pgfqpoint{0.700579in}{0.523148in}}{\pgfqpoint{4.650000in}{3.080000in}}%
\pgfusepath{clip}%
\pgfsetbuttcap%
\pgfsetroundjoin%
\pgfsetlinewidth{1.505625pt}%
\definecolor{currentstroke}{rgb}{0.000000,0.750000,0.750000}%
\pgfsetstrokecolor{currentstroke}%
\pgfsetdash{{5.550000pt}{2.400000pt}}{0.000000pt}%
\pgfpathmoveto{\pgfqpoint{0.911943in}{0.663148in}}%
\pgfpathlineto{\pgfqpoint{1.020335in}{0.901514in}}%
\pgfpathlineto{\pgfqpoint{1.128726in}{0.943666in}}%
\pgfpathlineto{\pgfqpoint{1.237118in}{0.948553in}}%
\pgfpathlineto{\pgfqpoint{1.345509in}{1.348608in}}%
\pgfpathlineto{\pgfqpoint{1.453901in}{1.440154in}}%
\pgfpathlineto{\pgfqpoint{1.562293in}{1.351371in}}%
\pgfpathlineto{\pgfqpoint{1.670684in}{1.535236in}}%
\pgfpathlineto{\pgfqpoint{1.779076in}{1.474035in}}%
\pgfpathlineto{\pgfqpoint{1.887467in}{1.550891in}}%
\pgfpathlineto{\pgfqpoint{2.104251in}{1.475150in}}%
\pgfpathlineto{\pgfqpoint{2.321034in}{1.541227in}}%
\pgfpathlineto{\pgfqpoint{2.537817in}{1.602436in}}%
\pgfpathlineto{\pgfqpoint{2.754600in}{1.831654in}}%
\pgfpathlineto{\pgfqpoint{2.971384in}{1.851201in}}%
\pgfpathlineto{\pgfqpoint{3.513342in}{1.791553in}}%
\pgfpathlineto{\pgfqpoint{4.055300in}{1.649811in}}%
\pgfpathlineto{\pgfqpoint{4.597258in}{2.074454in}}%
\pgfpathlineto{\pgfqpoint{5.139216in}{1.855128in}}%
\pgfusepath{stroke}%
\end{pgfscope}%
\begin{pgfscope}%
\pgfsetrectcap%
\pgfsetmiterjoin%
\pgfsetlinewidth{0.803000pt}%
\definecolor{currentstroke}{rgb}{0.000000,0.000000,0.000000}%
\pgfsetstrokecolor{currentstroke}%
\pgfsetdash{}{0pt}%
\pgfpathmoveto{\pgfqpoint{0.700579in}{0.523148in}}%
\pgfpathlineto{\pgfqpoint{0.700579in}{3.603148in}}%
\pgfusepath{stroke}%
\end{pgfscope}%
\begin{pgfscope}%
\pgfsetrectcap%
\pgfsetmiterjoin%
\pgfsetlinewidth{0.803000pt}%
\definecolor{currentstroke}{rgb}{0.000000,0.000000,0.000000}%
\pgfsetstrokecolor{currentstroke}%
\pgfsetdash{}{0pt}%
\pgfpathmoveto{\pgfqpoint{5.350579in}{0.523148in}}%
\pgfpathlineto{\pgfqpoint{5.350579in}{3.603148in}}%
\pgfusepath{stroke}%
\end{pgfscope}%
\begin{pgfscope}%
\pgfsetrectcap%
\pgfsetmiterjoin%
\pgfsetlinewidth{0.803000pt}%
\definecolor{currentstroke}{rgb}{0.000000,0.000000,0.000000}%
\pgfsetstrokecolor{currentstroke}%
\pgfsetdash{}{0pt}%
\pgfpathmoveto{\pgfqpoint{0.700579in}{0.523148in}}%
\pgfpathlineto{\pgfqpoint{5.350579in}{0.523148in}}%
\pgfusepath{stroke}%
\end{pgfscope}%
\begin{pgfscope}%
\pgfsetrectcap%
\pgfsetmiterjoin%
\pgfsetlinewidth{0.803000pt}%
\definecolor{currentstroke}{rgb}{0.000000,0.000000,0.000000}%
\pgfsetstrokecolor{currentstroke}%
\pgfsetdash{}{0pt}%
\pgfpathmoveto{\pgfqpoint{0.700579in}{3.603148in}}%
\pgfpathlineto{\pgfqpoint{5.350579in}{3.603148in}}%
\pgfusepath{stroke}%
\end{pgfscope}%
\begin{pgfscope}%
\pgfsetbuttcap%
\pgfsetmiterjoin%
\definecolor{currentfill}{rgb}{1.000000,1.000000,1.000000}%
\pgfsetfillcolor{currentfill}%
\pgfsetfillopacity{0.800000}%
\pgfsetlinewidth{1.003750pt}%
\definecolor{currentstroke}{rgb}{0.800000,0.800000,0.800000}%
\pgfsetstrokecolor{currentstroke}%
\pgfsetstrokeopacity{0.800000}%
\pgfsetdash{}{0pt}%
\pgfpathmoveto{\pgfqpoint{3.535883in}{0.599537in}}%
\pgfpathlineto{\pgfqpoint{5.243635in}{0.599537in}}%
\pgfpathquadraticcurveto{\pgfqpoint{5.274190in}{0.599537in}}{\pgfqpoint{5.274190in}{0.630092in}}%
\pgfpathlineto{\pgfqpoint{5.274190in}{1.466435in}}%
\pgfpathquadraticcurveto{\pgfqpoint{5.274190in}{1.496991in}}{\pgfqpoint{5.243635in}{1.496991in}}%
\pgfpathlineto{\pgfqpoint{3.535883in}{1.496991in}}%
\pgfpathquadraticcurveto{\pgfqpoint{3.505327in}{1.496991in}}{\pgfqpoint{3.505327in}{1.466435in}}%
\pgfpathlineto{\pgfqpoint{3.505327in}{0.630092in}}%
\pgfpathquadraticcurveto{\pgfqpoint{3.505327in}{0.599537in}}{\pgfqpoint{3.535883in}{0.599537in}}%
\pgfpathlineto{\pgfqpoint{3.535883in}{0.599537in}}%
\pgfpathclose%
\pgfusepath{stroke,fill}%
\end{pgfscope}%
\begin{pgfscope}%
\pgfsetrectcap%
\pgfsetroundjoin%
\pgfsetlinewidth{1.505625pt}%
\definecolor{currentstroke}{rgb}{0.000000,0.500000,0.000000}%
\pgfsetstrokecolor{currentstroke}%
\pgfsetdash{}{0pt}%
\pgfpathmoveto{\pgfqpoint{3.566438in}{1.382408in}}%
\pgfpathlineto{\pgfqpoint{3.719216in}{1.382408in}}%
\pgfpathlineto{\pgfqpoint{3.871994in}{1.382408in}}%
\pgfusepath{stroke}%
\end{pgfscope}%
\begin{pgfscope}%
\definecolor{textcolor}{rgb}{0.000000,0.000000,0.000000}%
\pgfsetstrokecolor{textcolor}%
\pgfsetfillcolor{textcolor}%
\pgftext[x=3.994216in,y=1.328935in,left,base]{\color{textcolor}\rmfamily\fontsize{11.000000}{13.200000}\selectfont Best}%
\end{pgfscope}%
\begin{pgfscope}%
\pgfsetrectcap%
\pgfsetroundjoin%
\pgfsetlinewidth{1.505625pt}%
\definecolor{currentstroke}{rgb}{1.000000,0.000000,0.000000}%
\pgfsetstrokecolor{currentstroke}%
\pgfsetdash{}{0pt}%
\pgfpathmoveto{\pgfqpoint{3.566438in}{1.169502in}}%
\pgfpathlineto{\pgfqpoint{3.719216in}{1.169502in}}%
\pgfpathlineto{\pgfqpoint{3.871994in}{1.169502in}}%
\pgfusepath{stroke}%
\end{pgfscope}%
\begin{pgfscope}%
\definecolor{textcolor}{rgb}{0.000000,0.000000,0.000000}%
\pgfsetstrokecolor{textcolor}%
\pgfsetfillcolor{textcolor}%
\pgftext[x=3.994216in,y=1.116030in,left,base]{\color{textcolor}\rmfamily\fontsize{11.000000}{13.200000}\selectfont \fit (\TransQOne)}%
\end{pgfscope}%
\begin{pgfscope}%
\pgfsetrectcap%
\pgfsetroundjoin%
\pgfsetlinewidth{1.505625pt}%
\definecolor{currentstroke}{rgb}{0.000000,0.000000,1.000000}%
\pgfsetstrokecolor{currentstroke}%
\pgfsetdash{}{0pt}%
\pgfpathmoveto{\pgfqpoint{3.566438in}{0.956597in}}%
\pgfpathlineto{\pgfqpoint{3.719216in}{0.956597in}}%
\pgfpathlineto{\pgfqpoint{3.871994in}{0.956597in}}%
\pgfusepath{stroke}%
\end{pgfscope}%
\begin{pgfscope}%
\definecolor{textcolor}{rgb}{0.000000,0.000000,0.000000}%
\pgfsetstrokecolor{textcolor}%
\pgfsetfillcolor{textcolor}%
\pgftext[x=3.994216in,y=0.903125in,left,base]{\color{textcolor}\rmfamily\fontsize{11.000000}{13.200000}\selectfont \TransQOne}%
\end{pgfscope}%
\begin{pgfscope}%
\pgfsetbuttcap%
\pgfsetroundjoin%
\pgfsetlinewidth{1.505625pt}%
\definecolor{currentstroke}{rgb}{0.000000,0.750000,0.750000}%
\pgfsetstrokecolor{currentstroke}%
\pgfsetdash{{5.550000pt}{2.400000pt}}{0.000000pt}%
\pgfpathmoveto{\pgfqpoint{3.566438in}{0.743692in}}%
\pgfpathlineto{\pgfqpoint{3.719216in}{0.743692in}}%
\pgfpathlineto{\pgfqpoint{3.871994in}{0.743692in}}%
\pgfusepath{stroke}%
\end{pgfscope}%
\begin{pgfscope}%
\definecolor{textcolor}{rgb}{0.000000,0.000000,0.000000}%
\pgfsetstrokecolor{textcolor}%
\pgfsetfillcolor{textcolor}%
\pgftext[x=3.994216in,y=0.690220in,left,base]{\color{textcolor}\rmfamily\fontsize{11.000000}{13.200000}\selectfont \fbi}%
\end{pgfscope}%
\end{pgfpicture}%
\makeatother%
\endgroup%

%% file: sections/images_figure.tex
\def\ppi{0.12}
\newcommand{\textsize}[1]{\footnotesize{#1}}
\newcommand{\textsizetitle}[1]{\small{#1}}
\begin{table*}[t]
\caption{Visual impact of the source selection with \di attacking \texttt{ConViT}$_\texttt{base}$ (first row) and \texttt{DPN}$_\texttt{92}$ (second row).}
\centering
\resizebox{0.80\linewidth}{!}{%
{
\setlength\tabcolsep{1pt}
\begin{tabular}{cccccc}
\hline
  & \textsizetitle{Original} & \textsizetitle{Best} & \textsizetitle{\fit (\TransQOne)} & \textsizetitle{Random} & \textsizetitle{Worst} \\
\hline

 & \includegraphics[width=\ppi\textwidth]{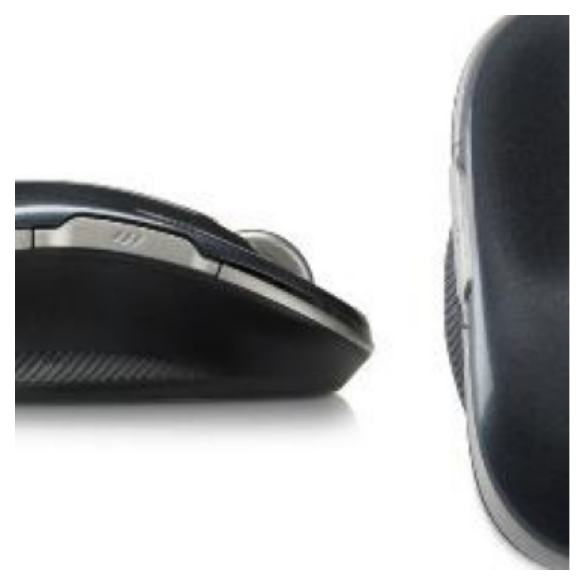} 
 & \includegraphics[width=\ppi\textwidth]{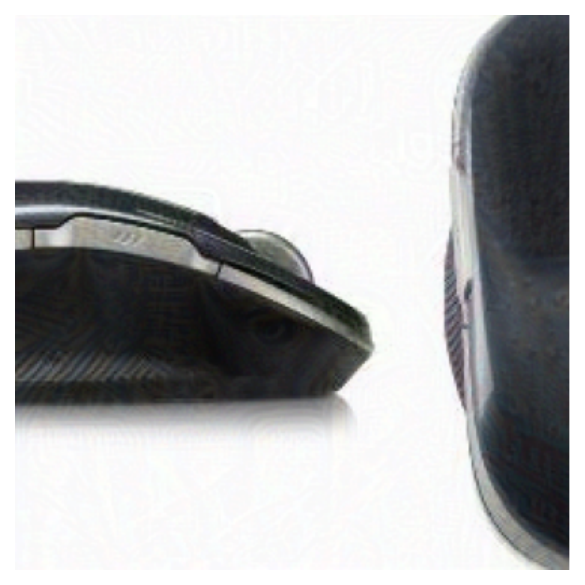} 
 & \includegraphics[width=\ppi\textwidth]{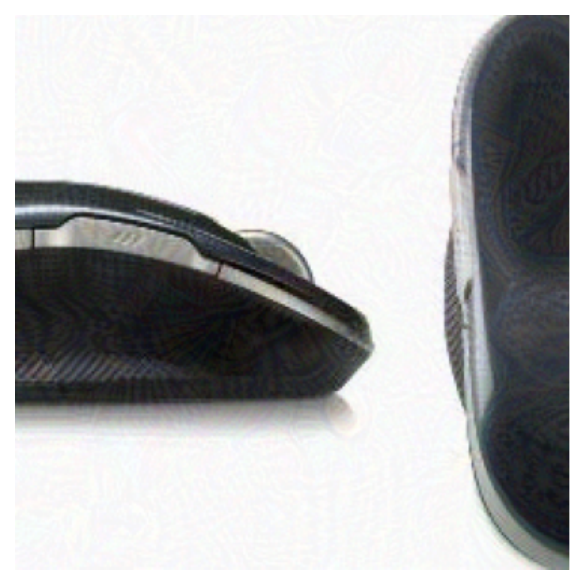} 
 & \includegraphics[width=\ppi\textwidth]{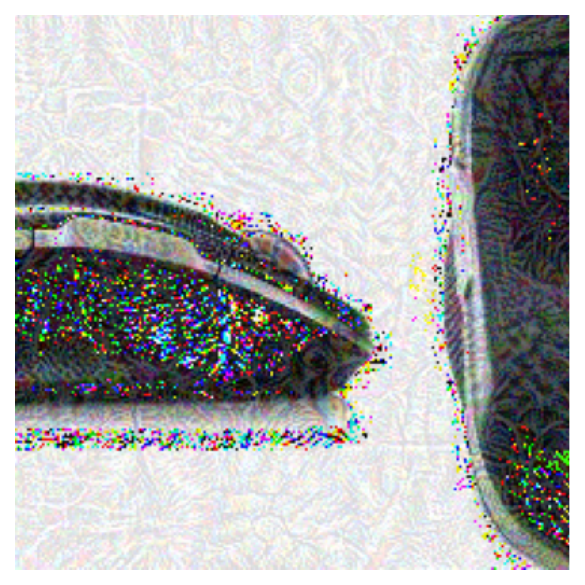}
 & \includegraphics[width=\ppi\textwidth]{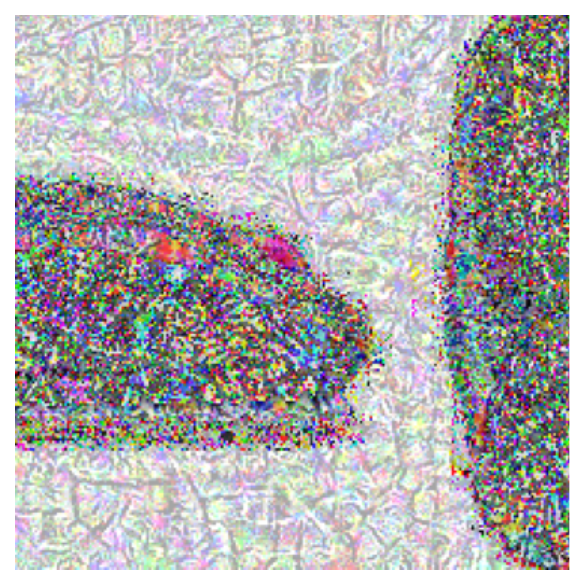} \\
 \textsize{label} &  \textsize{mouse} &  \textsize{oil filter} &  \textsize{purse} &  \textsize{purse} &  \textsize{jigsaw puzzle} \\
  \textsize{source} & & \textsize{\texttt{PiT}$_\texttt{small-dist}$} &  \textsize{\texttt{ConViT}$_\texttt{base}$} &  \textsize{\texttt{MobileNetV2}$_\texttt{110d}$} &  \textsize{\texttt{DenseNet}$_\texttt{121}$} \\
  \textsize{distortion}  &  &  \textsize{5.62} &  \textsize{8.13} &  \textsize{35.9} &  \textsize{87.8} \\
\hline
 & \includegraphics[width=\ppi\textwidth]{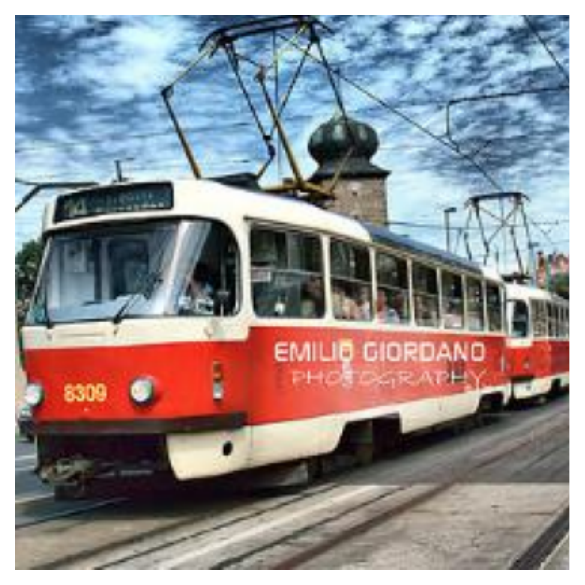} & \includegraphics[width=\ppi\textwidth]{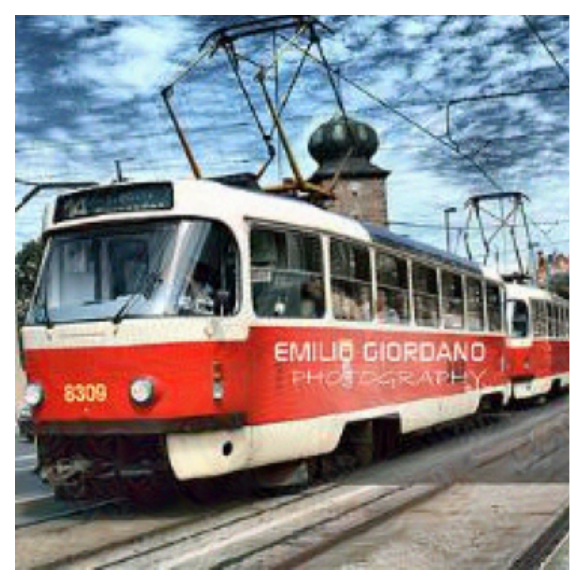} & \includegraphics[width=\ppi\textwidth]{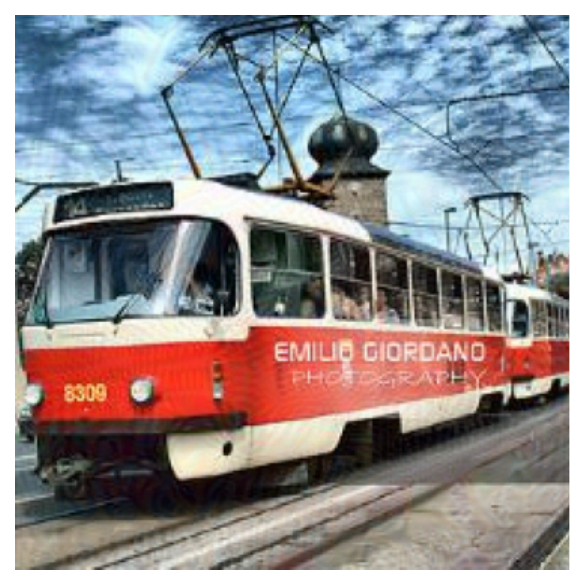}& \includegraphics[width=\ppi\textwidth]{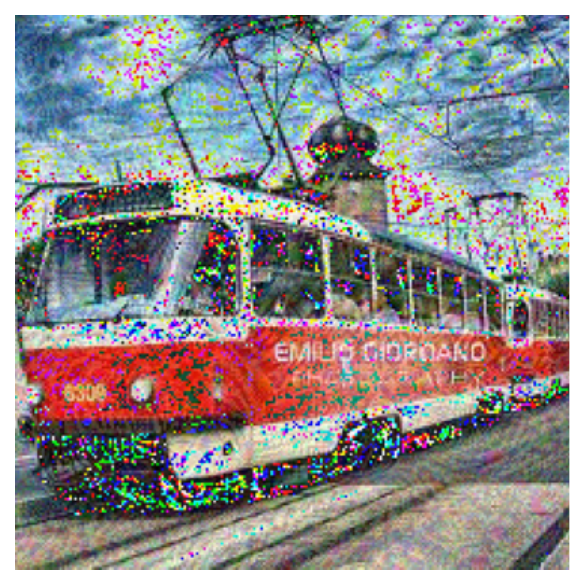} & \includegraphics[width=\ppi\textwidth]{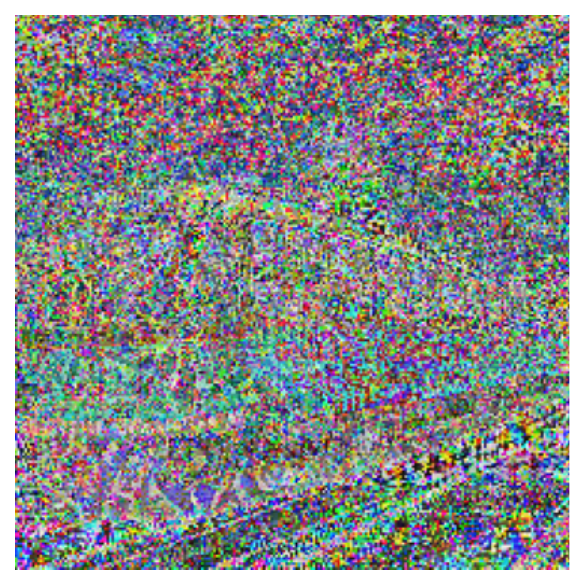}\\
 \textsize{label} &  \textsize{tram} &  \textsize{elec. locomotive}  &  \textsize{elec. locomotive}  &  \textsize{racing car} &  \textsize{jigsaw puzzle} \\
 \textsize{source} & &  \textsize{\texttt{ResNetV2}$_\texttt{50x1-dist}$}
&  \textsize{\texttt{PiT}$_\texttt{small-dist}$}  &  \textsize{\texttt{DLA}$_\texttt{60}$} &  \textsize{\texttt{MixNet}$_\texttt{medium}$} \\
 \textsize{distortion} & &  \textsize{6.94} &  \textsize{7.50} &  \textsize{33.3} &  \textsize{128.7}\\
\hline

\end{tabular}
}}
\label{fig:visual}
\end{table*} 

%% file: sections/conclusion.tex
\section{Conclusion}

Transferability is a crucial feature of adversarial examples as it allows a single perturbation to deceive multiple models. However, solely relying on Attack Success Rate (ASR) to measure transferability overlooks the degree of distortion needed to fool a model. This paper introduces a novel approach to assess transferability by comparing it to the distortion of two reference attacks: white box and black box attacks. We show that transferable attacks can perform worse than black box attacks without an appropriate selection of the source model, highlighting the need to choose the best source model to target a specific model.

The proposed solution, named \fit, allows the attacker to choose one of the best source models with minimal queries to the target. Our experiments demonstrate that the proposed solution performs well in multiple attack scenarios.

This study has highlighted the differences in transferability between images for the same source model and their particularity for this specific network. Further research could focus on addressing this issue and investigating its underlying causes, with the hope of designing an even better selection mechanism able to spot the best source  model.

%% file: sections/appendices/setup.tex
\section{Experimental Setup}
\label{app:setup}

In this study, we evaluated the transferability of adversarial attacks on a diverse set of 48 models trained for image classification on the ImageNet dataset with over one million annotated $224 \times 224$ images.
The models were obtained from the Timm library~\cite{rw2019timm}, with the exception of \texttt{ResNet50}$_\texttt{AdvTrain}$, which was obtained from the GitHub repository of the original paper~\footnote{https://github.com/MadryLab/robustness}.
To ensure adequate representation, we randomly selected models from each architecture, with a minimum of three models per architecture. The only exception was the \texttt{ReXNet} architecture, which had two distinct models.
The 48 selected models are:

\begin{itemize}
    \item \texttt{ConViT} architecture: \texttt{ConViT}$_\texttt{base}$, \texttt{ConViT}$_\texttt{small}$, \texttt{ConViT}$_\texttt{tiny}$
    \item \texttt{LeViT} architecture: \texttt{LeViT}$_\texttt{192}$, \texttt{LeViT}$_\texttt{256}$, \texttt{LeViT}$_\texttt{128}$
    \item \texttt{DenseNet} architecture: \texttt{DenseNet}$_\texttt{169}$, \texttt{DenseNet}$_\texttt{121}$, \texttt{DenseNet}$_\texttt{161}$
    \item \texttt{PiT} architecture: \texttt{PiT}$_\texttt{small}$, \texttt{PiT}$_\texttt{tight}$, \texttt{PiT}$_\texttt{tight-dist}$, \texttt{PiT}$_\texttt{small-dist}$
    \item \texttt{MobileNet} architecture (V2): \texttt{MobileNetV2}$_\texttt{110d}$, \texttt{MobileNetV2}$_\texttt{100}$, \texttt{MobileNetV2}$_\texttt{120d}$
    \item \texttt{CoaT} architecture: \texttt{CoatLite}$_\texttt{tiny}$, \texttt{CoatLite}$_\texttt{mini}$, \texttt{CoatLite}$_\texttt{small}$
    \item \texttt{xCiT} architecture: \texttt{xCiT}$_\texttt{medium}$, \texttt{xCiT}$_\texttt{nano}$, \texttt{xCiT}$_\texttt{small}$
    \item \texttt{Twins} architecture: \texttt{Twins}$_\texttt{small}$, \texttt{Twins}$_\texttt{large}$, \texttt{Twins}$_\texttt{base}$, 
    \item \texttt{MixNet} architecture: \texttt{MixNet}$_\texttt{large}$, \texttt{MixNet}$_\texttt{small}$, \texttt{MixNet}$_\texttt{medium}$, \texttt{MixNet}$_\texttt{small-TensorFlow}$, \texttt{MixNet}$_\texttt{large-TensorFlow}$, \texttt{MixNet}$_\texttt{medium-TensorFlow}$
    \item \texttt{EfficientNet} architecture: \texttt{EfficientNetB0}, \texttt{EfficientNetB0}$_\texttt{AdvProp}$, \texttt{EfficientNetB0}$_\texttt{NS}$
    \item \texttt{ResNet} architecture: \texttt{ResNet}$_\texttt{50}$, \texttt{ResNet}$_\texttt{50d}$, \texttt{ResNet50}$_\texttt{AdvTrain}$
    \item \texttt{ResNetV2} architecture: \texttt{ResNetV2}$_\texttt{50x1-dist}$, \texttt{ResNetV2}$_\texttt{101}$, \texttt{ResNetV2}$_\texttt{50}$
    \item \texttt{ReXNet} architecture: \texttt{RexNet}$_\texttt{150}$, \texttt{RexNet}$_\texttt{130}$
    \item \texttt{DPN} architecture: \texttt{DPN}$_\texttt{92}$, \texttt{DPN}$_\texttt{107}$, \texttt{DPN}$_\texttt{68b}$
    \item \texttt{DLA} architecture: \texttt{DLA}$_\texttt{60}$, \texttt{DLA}$_\texttt{102}$, \texttt{DLA}$_\texttt{169}$
\end{itemize}

%% file: sections/appendices/preliminaries.tex
\section{Preliminaries}


\subsection{Epsilon Parameter}
\label{app:epsilon_parameter}

All transferable attacks share a common parameter $\epsilon$. 
It controls the maximum perturbation norm added on a single pixel for the adversarial example built.
Fig.~\ref{fig:score_different_epsilon} demonstrates the ASR obtained for various values of $\epsilon$ as a function of the perturbation norm. 
It shows that even if more freedom is given to the perturbation, in the sense that a larger maximum perturbation norm is allowed, the transferable directions remain consistent.
Irrespective of the value of $\epsilon$ for a given attack, all scores for a given norm of the perturbation are similar.

\def\pp{0.3}
\begin{figure}[bt]
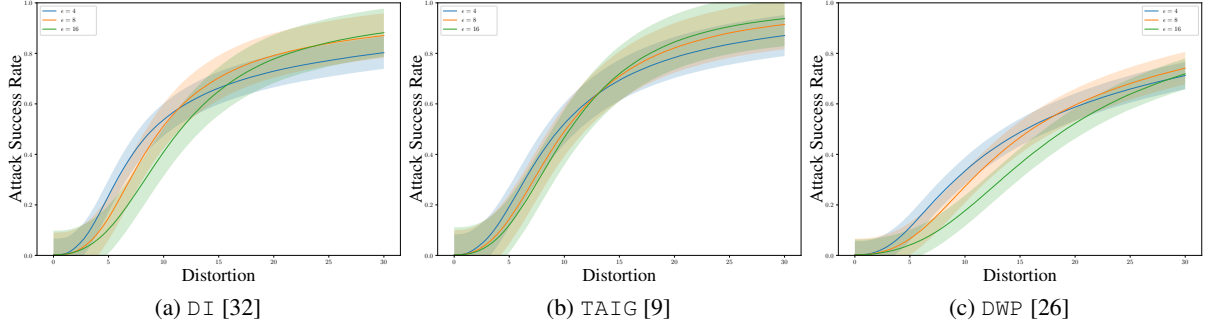

    \centering
    \begin{subfigure}[b]{\pp\linewidth}
        \centering
        \resizebox{\linewidth}{!}{\input{./images/preliminaries/asr_with_different_epsilon/DI2FGSM.pgf}}
        \caption{\di}
    \end{subfigure}
    \begin{subfigure}[b]{\pp\linewidth}
        \centering
        \resizebox{\linewidth}{!}{\input{./images/preliminaries/asr_with_different_epsilon/TAIG.pgf}}
        \caption{\taig}
    \end{subfigure}
    \begin{subfigure}[b]{\pp\linewidth}
        \centering
        \resizebox{\linewidth}{!}{\input{./images/preliminaries/asr_with_different_epsilon/DWP.pgf}}
        \caption{\dwp}
    \end{subfigure}
    \caption{Attack Success Rate function of the perturbation norm for different values of $\epsilon$.}
    \label{fig:score_different_epsilon}
\end{figure}

%% file: sections/appendices/matrix.tex
\section{Transferability Dependences}
\label{app:model_dependence_matrix}

The 48 models considered in \ref{app:setup} are evaluated as both sources and targets in this study.
For each possible pair of models, each source model is evaluated for its ability to transfer to each target model.
This results in a total of $48^2 = 2304$ evaluations.
The transferability is evaluated using the score defined in \ref{sec:measure} and their matrices for the attacks \di, \taig, and \dwp are presented in \ref{fig:matrix_attacks}.
Each matrix exhibits a similar structure, with models that have high transferability values appearing in each matrix.
However, the values achieved are different for each attack.
The \di and \taig attacks achieve higher values than \dwp, indicating that these attacks create better quality transferable examples.

\def\pp{0.3}
\begin{figure*}[ht]
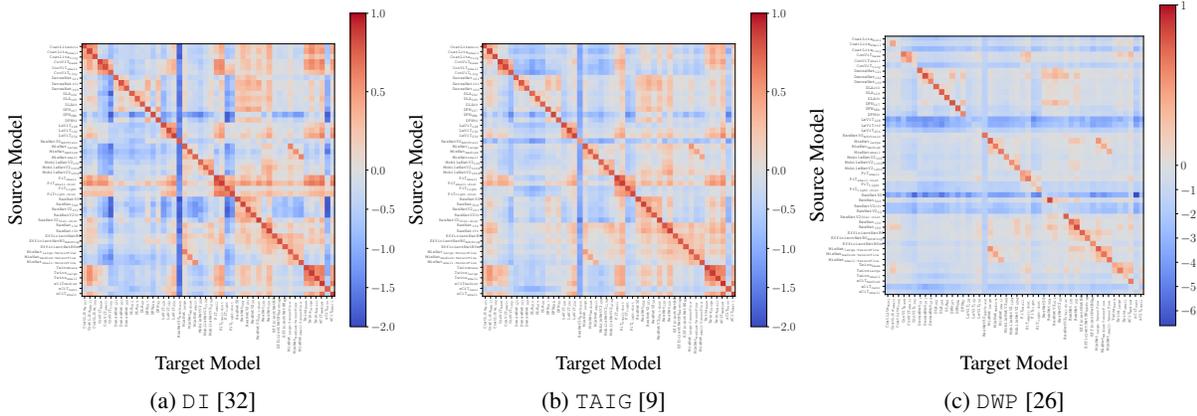

\centering
    \begin{subfigure}[b]{\pp\linewidth}
        \centering
        \resizebox{\linewidth}{!}{\input{./images/matrix/measure=gain-attack=DI2FGSM-epsilon=8.pgf}}
        \caption{\di}
    \end{subfigure}
    \begin{subfigure}[b]{\pp\linewidth}
        \centering
        \resizebox{\linewidth}{!}{\input{./images/matrix/measure=gain-attack=TAIG-epsilon=8.pgf}}
        \caption{\taig}
    \end{subfigure}
    \begin{subfigure}[b]{\pp\linewidth}
        \centering
        \resizebox{\linewidth}{!}{\input{./images/matrix/measure=gain-attack=DWP-epsilon=8.pgf}}
        \caption{\dwp}
    \end{subfigure}
    \caption{Transferability score \measure matrix of 48 sources and 48 targets listed in \ref{app:setup} for \di, \taig and \dwp. }
    \label{fig:matrix_attacks}
\end{figure*}

%% file: sections/appendices/results.tex
\section{Results}

\subsection{Fingerprinting}
\label{app:Fingeprinting}

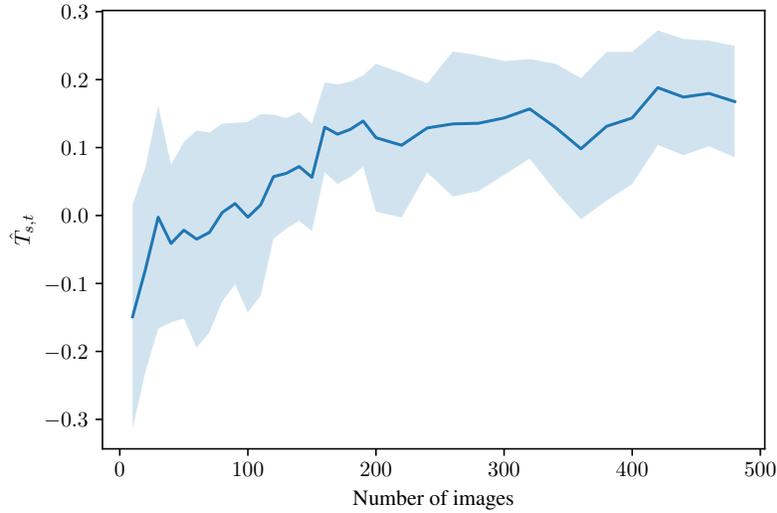
\begin{figure}[bt]
    \centering
    \resizebox{0.6\linewidth}{!}{\input{./images/results/fbi_number_of_images/Gain.pgf}}
    \caption{\measure function of the number of images used for \fbi to estimate the transferability between 45 sources and one target. Adversarial obtained with \di and $\epsilon=8$.}
    \label{fig:fbi_number_of_images}
\end{figure}

Transferability can be divided into three components: the attack, the model, and the attacked image.
To estimate transferability, the \fit measure defined in \ref{sec:measure} first estimate the similarity between the source and the target models.
In a defensive scenario, fingerprinting methods have been proposed to estimate model similarity without accessing one of the models.
These methods do not modify the model during training but instead take an already trained model and find images that are its signatures.
They usually generate adversarial examples specially designed for this model~\cite{ipguard, lukasdeep, https://doi.org/10.48550/arxiv.2202.08602}.
\fbi is the only method using benign images to assess the similarity of two models by measuring the independence between the two models using mutual information.
All fingerprinting methods are sensitive to the number of images used for fingerprinting.
More images lead to more accurate similarity scores, but they also have a cost.
In the scenario considered here, the number of images submitted must be minimized.
Figure~\ref{fig:fbi_number_of_images} shows the \measure function of the number of images used for \fbi.
Increasing the number of images submitted provides a better estimation of the transferability.
The score reaches a plateau at 200 images submitted.

\subsection{Ensemble model attack}
\label{app:EnsModAtt}

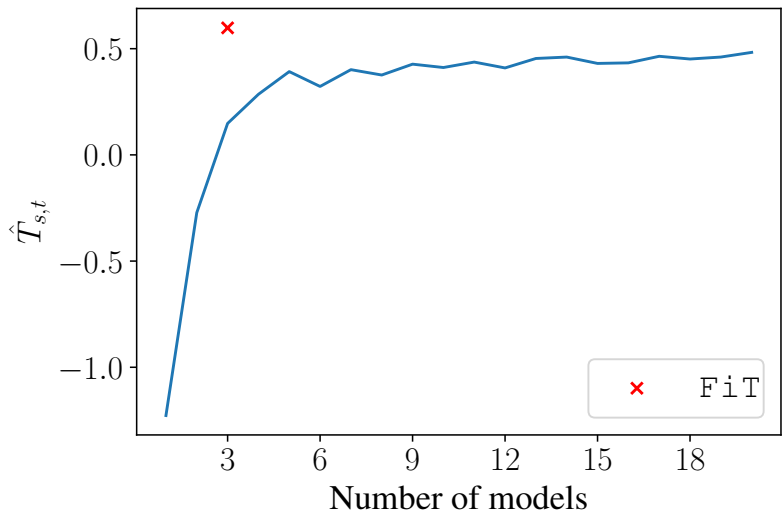
\begin{figure}[t]
    \centering
    \resizebox{0.6\linewidth}{!}{\input{./images/results/ensemble_progressive/target=xcit_nano_12_p16_224-model_seed=20-20_images_done.pgf}}
    \caption{\measure function of the number of models used for ensemble-model to attack $\texttt{xCiT}_{\texttt{nano}}$. The models are randomly selected and added one by one and compared with \fit selecting the three best models for ensemble-model among the 20 models available.}
    \label{fig:ensemble_sources_increasing}
\end{figure}

When attackers have access to multiple models, they can perform an ensemble-model attack to generate transferable adversarial examples. 
This approach has been shown to offer better transferability than the best single-model attack. 
However, existing methods for performing ensemble-model attacks have only been evaluated with a limited number of source models, typically with a maximum of three models. 
In this paper, a high number of models is used to build large ensemble-model attacks in the scenario described in the experimental setup in \ref{sec:experimental_setup}. 
At each step of the attack, a model is randomly selected from the available sources and added to the ensemble-model. 
To build transferable adversarial examples, the logits of the models are averaged together, as proposed in~\cite{wang2021boosting}.
Transferability is computed for ensemble-model attacks of up to 20 models. 
Figure~\ref{fig:ensemble_sources_increasing} shows the \fit score as a function of the ensemble-model size and compares the results with \fit scores obtained by selecting the three best models for the ensemble-model among the 20 models available.
Ensemble-model attacks demonstrate significant improvements when only a few models are considered, but beyond 5 models, the improvements become negligible. 
Additionally, the \fit score for the ensemble-model attack with 20 models was lower than that of the ensemble-model attack with only three models, which were carefully selected using \fit. 
These findings suggest that the quality of the selected models is more crucial than the quantity of models for effective ensemble-model attacks.

%% file: images/results/fbi_number_of_images/Gain.pgf
\begingroup%
\makeatletter%
\begin{pgfpicture}%
\pgfpathrectangle{\pgfpointorigin}{\pgfqpoint{5.533162in}{3.747763in}}%
\pgfusepath{use as bounding box, clip}%
\begin{pgfscope}%
\pgfsetbuttcap%
\pgfsetmiterjoin%
\definecolor{currentfill}{rgb}{1.000000,1.000000,1.000000}%
\pgfsetfillcolor{currentfill}%
\pgfsetlinewidth{0.000000pt}%
\definecolor{currentstroke}{rgb}{1.000000,1.000000,1.000000}%
\pgfsetstrokecolor{currentstroke}%
\pgfsetdash{}{0pt}%
\pgfpathmoveto{\pgfqpoint{0.000000in}{0.000000in}}%
\pgfpathlineto{\pgfqpoint{5.533162in}{0.000000in}}%
\pgfpathlineto{\pgfqpoint{5.533162in}{3.747763in}}%
\pgfpathlineto{\pgfqpoint{0.000000in}{3.747763in}}%
\pgfpathlineto{\pgfqpoint{0.000000in}{0.000000in}}%
\pgfpathclose%
\pgfusepath{fill}%
\end{pgfscope}%
\begin{pgfscope}%
\pgfsetbuttcap%
\pgfsetmiterjoin%
\definecolor{currentfill}{rgb}{1.000000,1.000000,1.000000}%
\pgfsetfillcolor{currentfill}%
\pgfsetlinewidth{0.000000pt}%
\definecolor{currentstroke}{rgb}{0.000000,0.000000,0.000000}%
\pgfsetstrokecolor{currentstroke}%
\pgfsetstrokeopacity{0.000000}%
\pgfsetdash{}{0pt}%
\pgfpathmoveto{\pgfqpoint{0.700579in}{0.523148in}}%
\pgfpathlineto{\pgfqpoint{5.350579in}{0.523148in}}%
\pgfpathlineto{\pgfqpoint{5.350579in}{3.603148in}}%
\pgfpathlineto{\pgfqpoint{0.700579in}{3.603148in}}%
\pgfpathlineto{\pgfqpoint{0.700579in}{0.523148in}}%
\pgfpathclose%
\pgfusepath{fill}%
\end{pgfscope}%
\begin{pgfscope}%
\pgfpathrectangle{\pgfqpoint{0.700579in}{0.523148in}}{\pgfqpoint{4.650000in}{3.080000in}}%
\pgfusepath{clip}%
\pgfsetbuttcap%
\pgfsetroundjoin%
\definecolor{currentfill}{rgb}{0.121569,0.466667,0.705882}%
\pgfsetfillcolor{currentfill}%
\pgfsetfillopacity{0.200000}%
\pgfsetlinewidth{0.000000pt}%
\definecolor{currentstroke}{rgb}{0.000000,0.000000,0.000000}%
\pgfsetstrokecolor{currentstroke}%
\pgfsetdash{}{0pt}%
\pgfpathmoveto{\pgfqpoint{0.911943in}{2.238728in}}%
\pgfpathlineto{\pgfqpoint{0.911943in}{0.663148in}}%
\pgfpathlineto{\pgfqpoint{1.001885in}{1.063338in}}%
\pgfpathlineto{\pgfqpoint{1.091827in}{1.367230in}}%
\pgfpathlineto{\pgfqpoint{1.181769in}{1.411779in}}%
\pgfpathlineto{\pgfqpoint{1.271711in}{1.440144in}}%
\pgfpathlineto{\pgfqpoint{1.361653in}{1.232738in}}%
\pgfpathlineto{\pgfqpoint{1.451595in}{1.341207in}}%
\pgfpathlineto{\pgfqpoint{1.541537in}{1.559081in}}%
\pgfpathlineto{\pgfqpoint{1.631479in}{1.679734in}}%
\pgfpathlineto{\pgfqpoint{1.721421in}{1.481484in}}%
\pgfpathlineto{\pgfqpoint{1.811363in}{1.600209in}}%
\pgfpathlineto{\pgfqpoint{1.901305in}{1.999967in}}%
\pgfpathlineto{\pgfqpoint{1.991247in}{2.070649in}}%
\pgfpathlineto{\pgfqpoint{2.081189in}{2.124472in}}%
\pgfpathlineto{\pgfqpoint{2.171131in}{2.053863in}}%
\pgfpathlineto{\pgfqpoint{2.261073in}{2.467790in}}%
\pgfpathlineto{\pgfqpoint{2.351015in}{2.384628in}}%
\pgfpathlineto{\pgfqpoint{2.440956in}{2.432278in}}%
\pgfpathlineto{\pgfqpoint{2.530898in}{2.505590in}}%
\pgfpathlineto{\pgfqpoint{2.620840in}{2.190121in}}%
\pgfpathlineto{\pgfqpoint{2.800724in}{2.148120in}}%
\pgfpathlineto{\pgfqpoint{2.980608in}{2.464465in}}%
\pgfpathlineto{\pgfqpoint{3.160492in}{2.296685in}}%
\pgfpathlineto{\pgfqpoint{3.340376in}{2.334061in}}%
\pgfpathlineto{\pgfqpoint{3.520260in}{2.447902in}}%
\pgfpathlineto{\pgfqpoint{3.700144in}{2.561383in}}%
\pgfpathlineto{\pgfqpoint{3.880028in}{2.335422in}}%
\pgfpathlineto{\pgfqpoint{4.059912in}{2.135752in}}%
\pgfpathlineto{\pgfqpoint{4.239796in}{2.267042in}}%
\pgfpathlineto{\pgfqpoint{4.419680in}{2.383394in}}%
\pgfpathlineto{\pgfqpoint{4.599564in}{2.658429in}}%
\pgfpathlineto{\pgfqpoint{4.779448in}{2.586577in}}%
\pgfpathlineto{\pgfqpoint{4.959332in}{2.648741in}}%
\pgfpathlineto{\pgfqpoint{5.139216in}{2.571122in}}%
\pgfpathlineto{\pgfqpoint{5.139216in}{3.354939in}}%
\pgfpathlineto{\pgfqpoint{5.139216in}{3.354939in}}%
\pgfpathlineto{\pgfqpoint{4.959332in}{3.391805in}}%
\pgfpathlineto{\pgfqpoint{4.779448in}{3.402581in}}%
\pgfpathlineto{\pgfqpoint{4.599564in}{3.463148in}}%
\pgfpathlineto{\pgfqpoint{4.419680in}{3.312677in}}%
\pgfpathlineto{\pgfqpoint{4.239796in}{3.311843in}}%
\pgfpathlineto{\pgfqpoint{4.059912in}{3.127298in}}%
\pgfpathlineto{\pgfqpoint{3.880028in}{3.229273in}}%
\pgfpathlineto{\pgfqpoint{3.700144in}{3.261876in}}%
\pgfpathlineto{\pgfqpoint{3.520260in}{3.248045in}}%
\pgfpathlineto{\pgfqpoint{3.340376in}{3.287245in}}%
\pgfpathlineto{\pgfqpoint{3.160492in}{3.315800in}}%
\pgfpathlineto{\pgfqpoint{2.980608in}{3.090952in}}%
\pgfpathlineto{\pgfqpoint{2.800724in}{3.164280in}}%
\pgfpathlineto{\pgfqpoint{2.620840in}{3.228819in}}%
\pgfpathlineto{\pgfqpoint{2.530898in}{3.147338in}}%
\pgfpathlineto{\pgfqpoint{2.440956in}{3.103554in}}%
\pgfpathlineto{\pgfqpoint{2.351015in}{3.083210in}}%
\pgfpathlineto{\pgfqpoint{2.261073in}{3.098552in}}%
\pgfpathlineto{\pgfqpoint{2.171131in}{2.807611in}}%
\pgfpathlineto{\pgfqpoint{2.081189in}{2.889408in}}%
\pgfpathlineto{\pgfqpoint{1.991247in}{2.847187in}}%
\pgfpathlineto{\pgfqpoint{1.901305in}{2.871725in}}%
\pgfpathlineto{\pgfqpoint{1.811363in}{2.875109in}}%
\pgfpathlineto{\pgfqpoint{1.721421in}{2.819943in}}%
\pgfpathlineto{\pgfqpoint{1.631479in}{2.813251in}}%
\pgfpathlineto{\pgfqpoint{1.541537in}{2.808230in}}%
\pgfpathlineto{\pgfqpoint{1.451595in}{2.746427in}}%
\pgfpathlineto{\pgfqpoint{1.361653in}{2.760112in}}%
\pgfpathlineto{\pgfqpoint{1.271711in}{2.679195in}}%
\pgfpathlineto{\pgfqpoint{1.181769in}{2.520210in}}%
\pgfpathlineto{\pgfqpoint{1.091827in}{2.934508in}}%
\pgfpathlineto{\pgfqpoint{1.001885in}{2.501121in}}%
\pgfpathlineto{\pgfqpoint{0.911943in}{2.238728in}}%
\pgfpathlineto{\pgfqpoint{0.911943in}{2.238728in}}%
\pgfpathclose%
\pgfusepath{fill}%
\end{pgfscope}%
\begin{pgfscope}%
\pgfsetbuttcap%
\pgfsetroundjoin%
\definecolor{currentfill}{rgb}{0.000000,0.000000,0.000000}%
\pgfsetfillcolor{currentfill}%
\pgfsetlinewidth{0.803000pt}%
\definecolor{currentstroke}{rgb}{0.000000,0.000000,0.000000}%
\pgfsetstrokecolor{currentstroke}%
\pgfsetdash{}{0pt}%
\pgfsys@defobject{currentmarker}{\pgfqpoint{0.000000in}{-0.048611in}}{\pgfqpoint{0.000000in}{0.000000in}}{%
\pgfpathmoveto{\pgfqpoint{0.000000in}{0.000000in}}%
\pgfpathlineto{\pgfqpoint{0.000000in}{-0.048611in}}%
\pgfusepath{stroke,fill}%
}%
\begin{pgfscope}%
\pgfsys@transformshift{0.822001in}{0.523148in}%
\pgfsys@useobject{currentmarker}{}%
\end{pgfscope}%
\end{pgfscope}%
\begin{pgfscope}%
\definecolor{textcolor}{rgb}{0.000000,0.000000,0.000000}%
\pgfsetstrokecolor{textcolor}%
\pgfsetfillcolor{textcolor}%
\pgftext[x=0.822001in,y=0.425926in,,top]{\color{textcolor}\rmfamily\fontsize{11.000000}{13.200000}\selectfont \(\displaystyle {0}\)}%
\end{pgfscope}%
\begin{pgfscope}%
\pgfsetbuttcap%
\pgfsetroundjoin%
\definecolor{currentfill}{rgb}{0.000000,0.000000,0.000000}%
\pgfsetfillcolor{currentfill}%
\pgfsetlinewidth{0.803000pt}%
\definecolor{currentstroke}{rgb}{0.000000,0.000000,0.000000}%
\pgfsetstrokecolor{currentstroke}%
\pgfsetdash{}{0pt}%
\pgfsys@defobject{currentmarker}{\pgfqpoint{0.000000in}{-0.048611in}}{\pgfqpoint{0.000000in}{0.000000in}}{%
\pgfpathmoveto{\pgfqpoint{0.000000in}{0.000000in}}%
\pgfpathlineto{\pgfqpoint{0.000000in}{-0.048611in}}%
\pgfusepath{stroke,fill}%
}%
\begin{pgfscope}%
\pgfsys@transformshift{1.721421in}{0.523148in}%
\pgfsys@useobject{currentmarker}{}%
\end{pgfscope}%
\end{pgfscope}%
\begin{pgfscope}%
\definecolor{textcolor}{rgb}{0.000000,0.000000,0.000000}%
\pgfsetstrokecolor{textcolor}%
\pgfsetfillcolor{textcolor}%
\pgftext[x=1.721421in,y=0.425926in,,top]{\color{textcolor}\rmfamily\fontsize{11.000000}{13.200000}\selectfont \(\displaystyle {100}\)}%
\end{pgfscope}%
\begin{pgfscope}%
\pgfsetbuttcap%
\pgfsetroundjoin%
\definecolor{currentfill}{rgb}{0.000000,0.000000,0.000000}%
\pgfsetfillcolor{currentfill}%
\pgfsetlinewidth{0.803000pt}%
\definecolor{currentstroke}{rgb}{0.000000,0.000000,0.000000}%
\pgfsetstrokecolor{currentstroke}%
\pgfsetdash{}{0pt}%
\pgfsys@defobject{currentmarker}{\pgfqpoint{0.000000in}{-0.048611in}}{\pgfqpoint{0.000000in}{0.000000in}}{%
\pgfpathmoveto{\pgfqpoint{0.000000in}{0.000000in}}%
\pgfpathlineto{\pgfqpoint{0.000000in}{-0.048611in}}%
\pgfusepath{stroke,fill}%
}%
\begin{pgfscope}%
\pgfsys@transformshift{2.620840in}{0.523148in}%
\pgfsys@useobject{currentmarker}{}%
\end{pgfscope}%
\end{pgfscope}%
\begin{pgfscope}%
\definecolor{textcolor}{rgb}{0.000000,0.000000,0.000000}%
\pgfsetstrokecolor{textcolor}%
\pgfsetfillcolor{textcolor}%
\pgftext[x=2.620840in,y=0.425926in,,top]{\color{textcolor}\rmfamily\fontsize{11.000000}{13.200000}\selectfont \(\displaystyle {200}\)}%
\end{pgfscope}%
\begin{pgfscope}%
\pgfsetbuttcap%
\pgfsetroundjoin%
\definecolor{currentfill}{rgb}{0.000000,0.000000,0.000000}%
\pgfsetfillcolor{currentfill}%
\pgfsetlinewidth{0.803000pt}%
\definecolor{currentstroke}{rgb}{0.000000,0.000000,0.000000}%
\pgfsetstrokecolor{currentstroke}%
\pgfsetdash{}{0pt}%
\pgfsys@defobject{currentmarker}{\pgfqpoint{0.000000in}{-0.048611in}}{\pgfqpoint{0.000000in}{0.000000in}}{%
\pgfpathmoveto{\pgfqpoint{0.000000in}{0.000000in}}%
\pgfpathlineto{\pgfqpoint{0.000000in}{-0.048611in}}%
\pgfusepath{stroke,fill}%
}%
\begin{pgfscope}%
\pgfsys@transformshift{3.520260in}{0.523148in}%
\pgfsys@useobject{currentmarker}{}%
\end{pgfscope}%
\end{pgfscope}%
\begin{pgfscope}%
\definecolor{textcolor}{rgb}{0.000000,0.000000,0.000000}%
\pgfsetstrokecolor{textcolor}%
\pgfsetfillcolor{textcolor}%
\pgftext[x=3.520260in,y=0.425926in,,top]{\color{textcolor}\rmfamily\fontsize{11.000000}{13.200000}\selectfont \(\displaystyle {300}\)}%
\end{pgfscope}%
\begin{pgfscope}%
\pgfsetbuttcap%
\pgfsetroundjoin%
\definecolor{currentfill}{rgb}{0.000000,0.000000,0.000000}%
\pgfsetfillcolor{currentfill}%
\pgfsetlinewidth{0.803000pt}%
\definecolor{currentstroke}{rgb}{0.000000,0.000000,0.000000}%
\pgfsetstrokecolor{currentstroke}%
\pgfsetdash{}{0pt}%
\pgfsys@defobject{currentmarker}{\pgfqpoint{0.000000in}{-0.048611in}}{\pgfqpoint{0.000000in}{0.000000in}}{%
\pgfpathmoveto{\pgfqpoint{0.000000in}{0.000000in}}%
\pgfpathlineto{\pgfqpoint{0.000000in}{-0.048611in}}%
\pgfusepath{stroke,fill}%
}%
\begin{pgfscope}%
\pgfsys@transformshift{4.419680in}{0.523148in}%
\pgfsys@useobject{currentmarker}{}%
\end{pgfscope}%
\end{pgfscope}%
\begin{pgfscope}%
\definecolor{textcolor}{rgb}{0.000000,0.000000,0.000000}%
\pgfsetstrokecolor{textcolor}%
\pgfsetfillcolor{textcolor}%
\pgftext[x=4.419680in,y=0.425926in,,top]{\color{textcolor}\rmfamily\fontsize{11.000000}{13.200000}\selectfont \(\displaystyle {400}\)}%
\end{pgfscope}%
\begin{pgfscope}%
\pgfsetbuttcap%
\pgfsetroundjoin%
\definecolor{currentfill}{rgb}{0.000000,0.000000,0.000000}%
\pgfsetfillcolor{currentfill}%
\pgfsetlinewidth{0.803000pt}%
\definecolor{currentstroke}{rgb}{0.000000,0.000000,0.000000}%
\pgfsetstrokecolor{currentstroke}%
\pgfsetdash{}{0pt}%
\pgfsys@defobject{currentmarker}{\pgfqpoint{0.000000in}{-0.048611in}}{\pgfqpoint{0.000000in}{0.000000in}}{%
\pgfpathmoveto{\pgfqpoint{0.000000in}{0.000000in}}%
\pgfpathlineto{\pgfqpoint{0.000000in}{-0.048611in}}%
\pgfusepath{stroke,fill}%
}%
\begin{pgfscope}%
\pgfsys@transformshift{5.319100in}{0.523148in}%
\pgfsys@useobject{currentmarker}{}%
\end{pgfscope}%
\end{pgfscope}%
\begin{pgfscope}%
\definecolor{textcolor}{rgb}{0.000000,0.000000,0.000000}%
\pgfsetstrokecolor{textcolor}%
\pgfsetfillcolor{textcolor}%
\pgftext[x=5.319100in,y=0.425926in,,top]{\color{textcolor}\rmfamily\fontsize{11.000000}{13.200000}\selectfont \(\displaystyle {500}\)}%
\end{pgfscope}%
\begin{pgfscope}%
\definecolor{textcolor}{rgb}{0.000000,0.000000,0.000000}%
\pgfsetstrokecolor{textcolor}%
\pgfsetfillcolor{textcolor}%
\pgftext[x=3.025579in,y=0.235185in,,top]{\color{textcolor}\rmfamily\fontsize{11.000000}{13.200000}\selectfont Number of images}%
\end{pgfscope}%
\begin{pgfscope}%
\pgfsetbuttcap%
\pgfsetroundjoin%
\definecolor{currentfill}{rgb}{0.000000,0.000000,0.000000}%
\pgfsetfillcolor{currentfill}%
\pgfsetlinewidth{0.803000pt}%
\definecolor{currentstroke}{rgb}{0.000000,0.000000,0.000000}%
\pgfsetstrokecolor{currentstroke}%
\pgfsetdash{}{0pt}%
\pgfsys@defobject{currentmarker}{\pgfqpoint{-0.048611in}{0.000000in}}{\pgfqpoint{-0.000000in}{0.000000in}}{%
\pgfpathmoveto{\pgfqpoint{-0.000000in}{0.000000in}}%
\pgfpathlineto{\pgfqpoint{-0.048611in}{0.000000in}}%
\pgfusepath{stroke,fill}%
}%
\begin{pgfscope}%
\pgfsys@transformshift{0.700579in}{0.730189in}%
\pgfsys@useobject{currentmarker}{}%
\end{pgfscope}%
\end{pgfscope}%
\begin{pgfscope}%
\definecolor{textcolor}{rgb}{0.000000,0.000000,0.000000}%
\pgfsetstrokecolor{textcolor}%
\pgfsetfillcolor{textcolor}%
\pgftext[x=0.290741in, y=0.677382in, left, base]{\color{textcolor}\rmfamily\fontsize{11.000000}{13.200000}\selectfont \(\displaystyle {\ensuremath{-}0.3}\)}%
\end{pgfscope}%
\begin{pgfscope}%
\pgfsetbuttcap%
\pgfsetroundjoin%
\definecolor{currentfill}{rgb}{0.000000,0.000000,0.000000}%
\pgfsetfillcolor{currentfill}%
\pgfsetlinewidth{0.803000pt}%
\definecolor{currentstroke}{rgb}{0.000000,0.000000,0.000000}%
\pgfsetstrokecolor{currentstroke}%
\pgfsetdash{}{0pt}%
\pgfsys@defobject{currentmarker}{\pgfqpoint{-0.048611in}{0.000000in}}{\pgfqpoint{-0.000000in}{0.000000in}}{%
\pgfpathmoveto{\pgfqpoint{-0.000000in}{0.000000in}}%
\pgfpathlineto{\pgfqpoint{-0.048611in}{0.000000in}}%
\pgfusepath{stroke,fill}%
}%
\begin{pgfscope}%
\pgfsys@transformshift{0.700579in}{1.207650in}%
\pgfsys@useobject{currentmarker}{}%
\end{pgfscope}%
\end{pgfscope}%
\begin{pgfscope}%
\definecolor{textcolor}{rgb}{0.000000,0.000000,0.000000}%
\pgfsetstrokecolor{textcolor}%
\pgfsetfillcolor{textcolor}%
\pgftext[x=0.290741in, y=1.154843in, left, base]{\color{textcolor}\rmfamily\fontsize{11.000000}{13.200000}\selectfont \(\displaystyle {\ensuremath{-}0.2}\)}%
\end{pgfscope}%
\begin{pgfscope}%
\pgfsetbuttcap%
\pgfsetroundjoin%
\definecolor{currentfill}{rgb}{0.000000,0.000000,0.000000}%
\pgfsetfillcolor{currentfill}%
\pgfsetlinewidth{0.803000pt}%
\definecolor{currentstroke}{rgb}{0.000000,0.000000,0.000000}%
\pgfsetstrokecolor{currentstroke}%
\pgfsetdash{}{0pt}%
\pgfsys@defobject{currentmarker}{\pgfqpoint{-0.048611in}{0.000000in}}{\pgfqpoint{-0.000000in}{0.000000in}}{%
\pgfpathmoveto{\pgfqpoint{-0.000000in}{0.000000in}}%
\pgfpathlineto{\pgfqpoint{-0.048611in}{0.000000in}}%
\pgfusepath{stroke,fill}%
}%
\begin{pgfscope}%
\pgfsys@transformshift{0.700579in}{1.685111in}%
\pgfsys@useobject{currentmarker}{}%
\end{pgfscope}%
\end{pgfscope}%
\begin{pgfscope}%
\definecolor{textcolor}{rgb}{0.000000,0.000000,0.000000}%
\pgfsetstrokecolor{textcolor}%
\pgfsetfillcolor{textcolor}%
\pgftext[x=0.290741in, y=1.632305in, left, base]{\color{textcolor}\rmfamily\fontsize{11.000000}{13.200000}\selectfont \(\displaystyle {\ensuremath{-}0.1}\)}%
\end{pgfscope}%
\begin{pgfscope}%
\pgfsetbuttcap%
\pgfsetroundjoin%
\definecolor{currentfill}{rgb}{0.000000,0.000000,0.000000}%
\pgfsetfillcolor{currentfill}%
\pgfsetlinewidth{0.803000pt}%
\definecolor{currentstroke}{rgb}{0.000000,0.000000,0.000000}%
\pgfsetstrokecolor{currentstroke}%
\pgfsetdash{}{0pt}%
\pgfsys@defobject{currentmarker}{\pgfqpoint{-0.048611in}{0.000000in}}{\pgfqpoint{-0.000000in}{0.000000in}}{%
\pgfpathmoveto{\pgfqpoint{-0.000000in}{0.000000in}}%
\pgfpathlineto{\pgfqpoint{-0.048611in}{0.000000in}}%
\pgfusepath{stroke,fill}%
}%
\begin{pgfscope}%
\pgfsys@transformshift{0.700579in}{2.162573in}%
\pgfsys@useobject{currentmarker}{}%
\end{pgfscope}%
\end{pgfscope}%
\begin{pgfscope}%
\definecolor{textcolor}{rgb}{0.000000,0.000000,0.000000}%
\pgfsetstrokecolor{textcolor}%
\pgfsetfillcolor{textcolor}%
\pgftext[x=0.409028in, y=2.109766in, left, base]{\color{textcolor}\rmfamily\fontsize{11.000000}{13.200000}\selectfont \(\displaystyle {0.0}\)}%
\end{pgfscope}%
\begin{pgfscope}%
\pgfsetbuttcap%
\pgfsetroundjoin%
\definecolor{currentfill}{rgb}{0.000000,0.000000,0.000000}%
\pgfsetfillcolor{currentfill}%
\pgfsetlinewidth{0.803000pt}%
\definecolor{currentstroke}{rgb}{0.000000,0.000000,0.000000}%
\pgfsetstrokecolor{currentstroke}%
\pgfsetdash{}{0pt}%
\pgfsys@defobject{currentmarker}{\pgfqpoint{-0.048611in}{0.000000in}}{\pgfqpoint{-0.000000in}{0.000000in}}{%
\pgfpathmoveto{\pgfqpoint{-0.000000in}{0.000000in}}%
\pgfpathlineto{\pgfqpoint{-0.048611in}{0.000000in}}%
\pgfusepath{stroke,fill}%
}%
\begin{pgfscope}%
\pgfsys@transformshift{0.700579in}{2.640034in}%
\pgfsys@useobject{currentmarker}{}%
\end{pgfscope}%
\end{pgfscope}%
\begin{pgfscope}%
\definecolor{textcolor}{rgb}{0.000000,0.000000,0.000000}%
\pgfsetstrokecolor{textcolor}%
\pgfsetfillcolor{textcolor}%
\pgftext[x=0.409028in, y=2.587227in, left, base]{\color{textcolor}\rmfamily\fontsize{11.000000}{13.200000}\selectfont \(\displaystyle {0.1}\)}%
\end{pgfscope}%
\begin{pgfscope}%
\pgfsetbuttcap%
\pgfsetroundjoin%
\definecolor{currentfill}{rgb}{0.000000,0.000000,0.000000}%
\pgfsetfillcolor{currentfill}%
\pgfsetlinewidth{0.803000pt}%
\definecolor{currentstroke}{rgb}{0.000000,0.000000,0.000000}%
\pgfsetstrokecolor{currentstroke}%
\pgfsetdash{}{0pt}%
\pgfsys@defobject{currentmarker}{\pgfqpoint{-0.048611in}{0.000000in}}{\pgfqpoint{-0.000000in}{0.000000in}}{%
\pgfpathmoveto{\pgfqpoint{-0.000000in}{0.000000in}}%
\pgfpathlineto{\pgfqpoint{-0.048611in}{0.000000in}}%
\pgfusepath{stroke,fill}%
}%
\begin{pgfscope}%
\pgfsys@transformshift{0.700579in}{3.117495in}%
\pgfsys@useobject{currentmarker}{}%
\end{pgfscope}%
\end{pgfscope}%
\begin{pgfscope}%
\definecolor{textcolor}{rgb}{0.000000,0.000000,0.000000}%
\pgfsetstrokecolor{textcolor}%
\pgfsetfillcolor{textcolor}%
\pgftext[x=0.409028in, y=3.064688in, left, base]{\color{textcolor}\rmfamily\fontsize{11.000000}{13.200000}\selectfont \(\displaystyle {0.2}\)}%
\end{pgfscope}%
\begin{pgfscope}%
\pgfsetbuttcap%
\pgfsetroundjoin%
\definecolor{currentfill}{rgb}{0.000000,0.000000,0.000000}%
\pgfsetfillcolor{currentfill}%
\pgfsetlinewidth{0.803000pt}%
\definecolor{currentstroke}{rgb}{0.000000,0.000000,0.000000}%
\pgfsetstrokecolor{currentstroke}%
\pgfsetdash{}{0pt}%
\pgfsys@defobject{currentmarker}{\pgfqpoint{-0.048611in}{0.000000in}}{\pgfqpoint{-0.000000in}{0.000000in}}{%
\pgfpathmoveto{\pgfqpoint{-0.000000in}{0.000000in}}%
\pgfpathlineto{\pgfqpoint{-0.048611in}{0.000000in}}%
\pgfusepath{stroke,fill}%
}%
\begin{pgfscope}%
\pgfsys@transformshift{0.700579in}{3.594956in}%
\pgfsys@useobject{currentmarker}{}%
\end{pgfscope}%
\end{pgfscope}%
\begin{pgfscope}%
\definecolor{textcolor}{rgb}{0.000000,0.000000,0.000000}%
\pgfsetstrokecolor{textcolor}%
\pgfsetfillcolor{textcolor}%
\pgftext[x=0.409028in, y=3.542150in, left, base]{\color{textcolor}\rmfamily\fontsize{11.000000}{13.200000}\selectfont \(\displaystyle {0.3}\)}%
\end{pgfscope}%
\begin{pgfscope}%
\definecolor{textcolor}{rgb}{0.000000,0.000000,0.000000}%
\pgfsetstrokecolor{textcolor}%
\pgfsetfillcolor{textcolor}%
\pgftext[x=0.235185in,y=2.063148in,,bottom,rotate=90.000000]{\color{textcolor}\rmfamily\fontsize{11.000000}{13.200000}\selectfont \measure}%
\end{pgfscope}%
\begin{pgfscope}%
\pgfpathrectangle{\pgfqpoint{0.700579in}{0.523148in}}{\pgfqpoint{4.650000in}{3.080000in}}%
\pgfusepath{clip}%
\pgfsetrectcap%
\pgfsetroundjoin%
\pgfsetlinewidth{1.505625pt}%
\definecolor{currentstroke}{rgb}{0.121569,0.466667,0.705882}%
\pgfsetstrokecolor{currentstroke}%
\pgfsetdash{}{0pt}%
\pgfpathmoveto{\pgfqpoint{0.911943in}{1.450938in}}%
\pgfpathlineto{\pgfqpoint{1.001885in}{1.782229in}}%
\pgfpathlineto{\pgfqpoint{1.091827in}{2.150869in}}%
\pgfpathlineto{\pgfqpoint{1.181769in}{1.965994in}}%
\pgfpathlineto{\pgfqpoint{1.271711in}{2.059670in}}%
\pgfpathlineto{\pgfqpoint{1.361653in}{1.996425in}}%
\pgfpathlineto{\pgfqpoint{1.451595in}{2.043817in}}%
\pgfpathlineto{\pgfqpoint{1.541537in}{2.183656in}}%
\pgfpathlineto{\pgfqpoint{1.631479in}{2.246492in}}%
\pgfpathlineto{\pgfqpoint{1.721421in}{2.150714in}}%
\pgfpathlineto{\pgfqpoint{1.811363in}{2.237659in}}%
\pgfpathlineto{\pgfqpoint{1.901305in}{2.435846in}}%
\pgfpathlineto{\pgfqpoint{1.991247in}{2.458918in}}%
\pgfpathlineto{\pgfqpoint{2.081189in}{2.506940in}}%
\pgfpathlineto{\pgfqpoint{2.171131in}{2.430737in}}%
\pgfpathlineto{\pgfqpoint{2.261073in}{2.783171in}}%
\pgfpathlineto{\pgfqpoint{2.351015in}{2.733919in}}%
\pgfpathlineto{\pgfqpoint{2.440956in}{2.767916in}}%
\pgfpathlineto{\pgfqpoint{2.530898in}{2.826464in}}%
\pgfpathlineto{\pgfqpoint{2.620840in}{2.709470in}}%
\pgfpathlineto{\pgfqpoint{2.800724in}{2.656200in}}%
\pgfpathlineto{\pgfqpoint{2.980608in}{2.777708in}}%
\pgfpathlineto{\pgfqpoint{3.160492in}{2.806243in}}%
\pgfpathlineto{\pgfqpoint{3.340376in}{2.810653in}}%
\pgfpathlineto{\pgfqpoint{3.520260in}{2.847973in}}%
\pgfpathlineto{\pgfqpoint{3.700144in}{2.911630in}}%
\pgfpathlineto{\pgfqpoint{3.880028in}{2.782348in}}%
\pgfpathlineto{\pgfqpoint{4.059912in}{2.631525in}}%
\pgfpathlineto{\pgfqpoint{4.239796in}{2.789442in}}%
\pgfpathlineto{\pgfqpoint{4.419680in}{2.848036in}}%
\pgfpathlineto{\pgfqpoint{4.599564in}{3.060789in}}%
\pgfpathlineto{\pgfqpoint{4.779448in}{2.994579in}}%
\pgfpathlineto{\pgfqpoint{4.959332in}{3.020273in}}%
\pgfpathlineto{\pgfqpoint{5.139216in}{2.963031in}}%
\pgfusepath{stroke}%
\end{pgfscope}%
\begin{pgfscope}%
\pgfsetrectcap%
\pgfsetmiterjoin%
\pgfsetlinewidth{0.803000pt}%
\definecolor{currentstroke}{rgb}{0.000000,0.000000,0.000000}%
\pgfsetstrokecolor{currentstroke}%
\pgfsetdash{}{0pt}%
\pgfpathmoveto{\pgfqpoint{0.700579in}{0.523148in}}%
\pgfpathlineto{\pgfqpoint{0.700579in}{3.603148in}}%
\pgfusepath{stroke}%
\end{pgfscope}%
\begin{pgfscope}%
\pgfsetrectcap%
\pgfsetmiterjoin%
\pgfsetlinewidth{0.803000pt}%
\definecolor{currentstroke}{rgb}{0.000000,0.000000,0.000000}%
\pgfsetstrokecolor{currentstroke}%
\pgfsetdash{}{0pt}%
\pgfpathmoveto{\pgfqpoint{5.350579in}{0.523148in}}%
\pgfpathlineto{\pgfqpoint{5.350579in}{3.603148in}}%
\pgfusepath{stroke}%
\end{pgfscope}%
\begin{pgfscope}%
\pgfsetrectcap%
\pgfsetmiterjoin%
\pgfsetlinewidth{0.803000pt}%
\definecolor{currentstroke}{rgb}{0.000000,0.000000,0.000000}%
\pgfsetstrokecolor{currentstroke}%
\pgfsetdash{}{0pt}%
\pgfpathmoveto{\pgfqpoint{0.700579in}{0.523148in}}%
\pgfpathlineto{\pgfqpoint{5.350579in}{0.523148in}}%
\pgfusepath{stroke}%
\end{pgfscope}%
\begin{pgfscope}%
\pgfsetrectcap%
\pgfsetmiterjoin%
\pgfsetlinewidth{0.803000pt}%
\definecolor{currentstroke}{rgb}{0.000000,0.000000,0.000000}%
\pgfsetstrokecolor{currentstroke}%
\pgfsetdash{}{0pt}%
\pgfpathmoveto{\pgfqpoint{0.700579in}{3.603148in}}%
\pgfpathlineto{\pgfqpoint{5.350579in}{3.603148in}}%
\pgfusepath{stroke}%
\end{pgfscope}%
\end{pgfpicture}%
\makeatother%
\endgroup%

%% file: images/results/ensemble_progressive/target=xcit_nano_12_p16_224-model_seed=20-20_images_done.pgf
\begingroup%
\makeatletter%
\begin{pgfpicture}%
\pgfpathrectangle{\pgfpointorigin}{\pgfqpoint{5.688207in}{3.859475in}}%
\pgfusepath{use as bounding box, clip}%
\begin{pgfscope}%
\pgfsetbuttcap%
\pgfsetmiterjoin%
\definecolor{currentfill}{rgb}{1.000000,1.000000,1.000000}%
\pgfsetfillcolor{currentfill}%
\pgfsetlinewidth{0.000000pt}%
\definecolor{currentstroke}{rgb}{1.000000,1.000000,1.000000}%
\pgfsetstrokecolor{currentstroke}%
\pgfsetdash{}{0pt}%
\pgfpathmoveto{\pgfqpoint{0.000000in}{0.000000in}}%
\pgfpathlineto{\pgfqpoint{5.688207in}{0.000000in}}%
\pgfpathlineto{\pgfqpoint{5.688207in}{3.859475in}}%
\pgfpathlineto{\pgfqpoint{0.000000in}{3.859475in}}%
\pgfpathlineto{\pgfqpoint{0.000000in}{0.000000in}}%
\pgfpathclose%
\pgfusepath{fill}%
\end{pgfscope}%
\begin{pgfscope}%
\pgfsetbuttcap%
\pgfsetmiterjoin%
\definecolor{currentfill}{rgb}{1.000000,1.000000,1.000000}%
\pgfsetfillcolor{currentfill}%
\pgfsetlinewidth{0.000000pt}%
\definecolor{currentstroke}{rgb}{0.000000,0.000000,0.000000}%
\pgfsetstrokecolor{currentstroke}%
\pgfsetstrokeopacity{0.000000}%
\pgfsetdash{}{0pt}%
\pgfpathmoveto{\pgfqpoint{0.938207in}{0.679475in}}%
\pgfpathlineto{\pgfqpoint{5.588207in}{0.679475in}}%
\pgfpathlineto{\pgfqpoint{5.588207in}{3.759475in}}%
\pgfpathlineto{\pgfqpoint{0.938207in}{3.759475in}}%
\pgfpathlineto{\pgfqpoint{0.938207in}{0.679475in}}%
\pgfpathclose%
\pgfusepath{fill}%
\end{pgfscope}%
\begin{pgfscope}%
\pgfpathrectangle{\pgfqpoint{0.938207in}{0.679475in}}{\pgfqpoint{4.650000in}{3.080000in}}%
\pgfusepath{clip}%
\pgfsetbuttcap%
\pgfsetroundjoin%
\definecolor{currentfill}{rgb}{1.000000,0.000000,0.000000}%
\pgfsetfillcolor{currentfill}%
\pgfsetlinewidth{1.505625pt}%
\definecolor{currentstroke}{rgb}{1.000000,0.000000,0.000000}%
\pgfsetstrokecolor{currentstroke}%
\pgfsetdash{}{0pt}%
\pgfsys@defobject{currentmarker}{\pgfqpoint{-0.041667in}{-0.041667in}}{\pgfqpoint{0.041667in}{0.041667in}}{%
\pgfpathmoveto{\pgfqpoint{-0.041667in}{-0.041667in}}%
\pgfpathlineto{\pgfqpoint{0.041667in}{0.041667in}}%
\pgfpathmoveto{\pgfqpoint{-0.041667in}{0.041667in}}%
\pgfpathlineto{\pgfqpoint{0.041667in}{-0.041667in}}%
\pgfusepath{stroke,fill}%
}%
\begin{pgfscope}%
\pgfsys@transformshift{1.594547in}{3.619475in}%
\pgfsys@useobject{currentmarker}{}%
\end{pgfscope}%
\end{pgfscope}%
\begin{pgfscope}%
\pgfsetbuttcap%
\pgfsetroundjoin%
\definecolor{currentfill}{rgb}{0.000000,0.000000,0.000000}%
\pgfsetfillcolor{currentfill}%
\pgfsetlinewidth{0.803000pt}%
\definecolor{currentstroke}{rgb}{0.000000,0.000000,0.000000}%
\pgfsetstrokecolor{currentstroke}%
\pgfsetdash{}{0pt}%
\pgfsys@defobject{currentmarker}{\pgfqpoint{0.000000in}{-0.048611in}}{\pgfqpoint{0.000000in}{0.000000in}}{%
\pgfpathmoveto{\pgfqpoint{0.000000in}{0.000000in}}%
\pgfpathlineto{\pgfqpoint{0.000000in}{-0.048611in}}%
\pgfusepath{stroke,fill}%
}%
\begin{pgfscope}%
\pgfsys@transformshift{1.594547in}{0.679475in}%
\pgfsys@useobject{currentmarker}{}%
\end{pgfscope}%
\end{pgfscope}%
\begin{pgfscope}%
\definecolor{textcolor}{rgb}{0.000000,0.000000,0.000000}%
\pgfsetstrokecolor{textcolor}%
\pgfsetfillcolor{textcolor}%
\pgftext[x=1.594547in,y=0.582253in,,top]{\color{textcolor}\rmfamily\fontsize{18.000000}{21.600000}\selectfont \(\displaystyle {3}\)}%
\end{pgfscope}%
\begin{pgfscope}%
\pgfsetbuttcap%
\pgfsetroundjoin%
\definecolor{currentfill}{rgb}{0.000000,0.000000,0.000000}%
\pgfsetfillcolor{currentfill}%
\pgfsetlinewidth{0.803000pt}%
\definecolor{currentstroke}{rgb}{0.000000,0.000000,0.000000}%
\pgfsetstrokecolor{currentstroke}%
\pgfsetdash{}{0pt}%
\pgfsys@defobject{currentmarker}{\pgfqpoint{0.000000in}{-0.048611in}}{\pgfqpoint{0.000000in}{0.000000in}}{%
\pgfpathmoveto{\pgfqpoint{0.000000in}{0.000000in}}%
\pgfpathlineto{\pgfqpoint{0.000000in}{-0.048611in}}%
\pgfusepath{stroke,fill}%
}%
\begin{pgfscope}%
\pgfsys@transformshift{2.262011in}{0.679475in}%
\pgfsys@useobject{currentmarker}{}%
\end{pgfscope}%
\end{pgfscope}%
\begin{pgfscope}%
\definecolor{textcolor}{rgb}{0.000000,0.000000,0.000000}%
\pgfsetstrokecolor{textcolor}%
\pgfsetfillcolor{textcolor}%
\pgftext[x=2.262011in,y=0.582253in,,top]{\color{textcolor}\rmfamily\fontsize{18.000000}{21.600000}\selectfont \(\displaystyle {6}\)}%
\end{pgfscope}%
\begin{pgfscope}%
\pgfsetbuttcap%
\pgfsetroundjoin%
\definecolor{currentfill}{rgb}{0.000000,0.000000,0.000000}%
\pgfsetfillcolor{currentfill}%
\pgfsetlinewidth{0.803000pt}%
\definecolor{currentstroke}{rgb}{0.000000,0.000000,0.000000}%
\pgfsetstrokecolor{currentstroke}%
\pgfsetdash{}{0pt}%
\pgfsys@defobject{currentmarker}{\pgfqpoint{0.000000in}{-0.048611in}}{\pgfqpoint{0.000000in}{0.000000in}}{%
\pgfpathmoveto{\pgfqpoint{0.000000in}{0.000000in}}%
\pgfpathlineto{\pgfqpoint{0.000000in}{-0.048611in}}%
\pgfusepath{stroke,fill}%
}%
\begin{pgfscope}%
\pgfsys@transformshift{2.929475in}{0.679475in}%
\pgfsys@useobject{currentmarker}{}%
\end{pgfscope}%
\end{pgfscope}%
\begin{pgfscope}%
\definecolor{textcolor}{rgb}{0.000000,0.000000,0.000000}%
\pgfsetstrokecolor{textcolor}%
\pgfsetfillcolor{textcolor}%
\pgftext[x=2.929475in,y=0.582253in,,top]{\color{textcolor}\rmfamily\fontsize{18.000000}{21.600000}\selectfont \(\displaystyle {9}\)}%
\end{pgfscope}%
\begin{pgfscope}%
\pgfsetbuttcap%
\pgfsetroundjoin%
\definecolor{currentfill}{rgb}{0.000000,0.000000,0.000000}%
\pgfsetfillcolor{currentfill}%
\pgfsetlinewidth{0.803000pt}%
\definecolor{currentstroke}{rgb}{0.000000,0.000000,0.000000}%
\pgfsetstrokecolor{currentstroke}%
\pgfsetdash{}{0pt}%
\pgfsys@defobject{currentmarker}{\pgfqpoint{0.000000in}{-0.048611in}}{\pgfqpoint{0.000000in}{0.000000in}}{%
\pgfpathmoveto{\pgfqpoint{0.000000in}{0.000000in}}%
\pgfpathlineto{\pgfqpoint{0.000000in}{-0.048611in}}%
\pgfusepath{stroke,fill}%
}%
\begin{pgfscope}%
\pgfsys@transformshift{3.596939in}{0.679475in}%
\pgfsys@useobject{currentmarker}{}%
\end{pgfscope}%
\end{pgfscope}%
\begin{pgfscope}%
\definecolor{textcolor}{rgb}{0.000000,0.000000,0.000000}%
\pgfsetstrokecolor{textcolor}%
\pgfsetfillcolor{textcolor}%
\pgftext[x=3.596939in,y=0.582253in,,top]{\color{textcolor}\rmfamily\fontsize{18.000000}{21.600000}\selectfont \(\displaystyle {12}\)}%
\end{pgfscope}%
\begin{pgfscope}%
\pgfsetbuttcap%
\pgfsetroundjoin%
\definecolor{currentfill}{rgb}{0.000000,0.000000,0.000000}%
\pgfsetfillcolor{currentfill}%
\pgfsetlinewidth{0.803000pt}%
\definecolor{currentstroke}{rgb}{0.000000,0.000000,0.000000}%
\pgfsetstrokecolor{currentstroke}%
\pgfsetdash{}{0pt}%
\pgfsys@defobject{currentmarker}{\pgfqpoint{0.000000in}{-0.048611in}}{\pgfqpoint{0.000000in}{0.000000in}}{%
\pgfpathmoveto{\pgfqpoint{0.000000in}{0.000000in}}%
\pgfpathlineto{\pgfqpoint{0.000000in}{-0.048611in}}%
\pgfusepath{stroke,fill}%
}%
\begin{pgfscope}%
\pgfsys@transformshift{4.264403in}{0.679475in}%
\pgfsys@useobject{currentmarker}{}%
\end{pgfscope}%
\end{pgfscope}%
\begin{pgfscope}%
\definecolor{textcolor}{rgb}{0.000000,0.000000,0.000000}%
\pgfsetstrokecolor{textcolor}%
\pgfsetfillcolor{textcolor}%
\pgftext[x=4.264403in,y=0.582253in,,top]{\color{textcolor}\rmfamily\fontsize{18.000000}{21.600000}\selectfont \(\displaystyle {15}\)}%
\end{pgfscope}%
\begin{pgfscope}%
\pgfsetbuttcap%
\pgfsetroundjoin%
\definecolor{currentfill}{rgb}{0.000000,0.000000,0.000000}%
\pgfsetfillcolor{currentfill}%
\pgfsetlinewidth{0.803000pt}%
\definecolor{currentstroke}{rgb}{0.000000,0.000000,0.000000}%
\pgfsetstrokecolor{currentstroke}%
\pgfsetdash{}{0pt}%
\pgfsys@defobject{currentmarker}{\pgfqpoint{0.000000in}{-0.048611in}}{\pgfqpoint{0.000000in}{0.000000in}}{%
\pgfpathmoveto{\pgfqpoint{0.000000in}{0.000000in}}%
\pgfpathlineto{\pgfqpoint{0.000000in}{-0.048611in}}%
\pgfusepath{stroke,fill}%
}%
\begin{pgfscope}%
\pgfsys@transformshift{4.931867in}{0.679475in}%
\pgfsys@useobject{currentmarker}{}%
\end{pgfscope}%
\end{pgfscope}%
\begin{pgfscope}%
\definecolor{textcolor}{rgb}{0.000000,0.000000,0.000000}%
\pgfsetstrokecolor{textcolor}%
\pgfsetfillcolor{textcolor}%
\pgftext[x=4.931867in,y=0.582253in,,top]{\color{textcolor}\rmfamily\fontsize{18.000000}{21.600000}\selectfont \(\displaystyle {18}\)}%
\end{pgfscope}%
\begin{pgfscope}%
\definecolor{textcolor}{rgb}{0.000000,0.000000,0.000000}%
\pgfsetstrokecolor{textcolor}%
\pgfsetfillcolor{textcolor}%
\pgftext[x=3.263207in,y=0.313349in,,top]{\color{textcolor}\rmfamily\fontsize{18.000000}{21.600000}\selectfont Number of models}%
\end{pgfscope}%
\begin{pgfscope}%
\pgfsetbuttcap%
\pgfsetroundjoin%
\definecolor{currentfill}{rgb}{0.000000,0.000000,0.000000}%
\pgfsetfillcolor{currentfill}%
\pgfsetlinewidth{0.803000pt}%
\definecolor{currentstroke}{rgb}{0.000000,0.000000,0.000000}%
\pgfsetstrokecolor{currentstroke}%
\pgfsetdash{}{0pt}%
\pgfsys@defobject{currentmarker}{\pgfqpoint{-0.048611in}{0.000000in}}{\pgfqpoint{-0.000000in}{0.000000in}}{%
\pgfpathmoveto{\pgfqpoint{-0.000000in}{0.000000in}}%
\pgfpathlineto{\pgfqpoint{-0.048611in}{0.000000in}}%
\pgfusepath{stroke,fill}%
}%
\begin{pgfscope}%
\pgfsys@transformshift{0.938207in}{1.168114in}%
\pgfsys@useobject{currentmarker}{}%
\end{pgfscope}%
\end{pgfscope}%
\begin{pgfscope}%
\definecolor{textcolor}{rgb}{0.000000,0.000000,0.000000}%
\pgfsetstrokecolor{textcolor}%
\pgfsetfillcolor{textcolor}%
\pgftext[x=0.368904in, y=1.084781in, left, base]{\color{textcolor}\rmfamily\fontsize{18.000000}{21.600000}\selectfont \(\displaystyle {\ensuremath{-}1.0}\)}%
\end{pgfscope}%
\begin{pgfscope}%
\pgfsetbuttcap%
\pgfsetroundjoin%
\definecolor{currentfill}{rgb}{0.000000,0.000000,0.000000}%
\pgfsetfillcolor{currentfill}%
\pgfsetlinewidth{0.803000pt}%
\definecolor{currentstroke}{rgb}{0.000000,0.000000,0.000000}%
\pgfsetstrokecolor{currentstroke}%
\pgfsetdash{}{0pt}%
\pgfsys@defobject{currentmarker}{\pgfqpoint{-0.048611in}{0.000000in}}{\pgfqpoint{-0.000000in}{0.000000in}}{%
\pgfpathmoveto{\pgfqpoint{-0.000000in}{0.000000in}}%
\pgfpathlineto{\pgfqpoint{-0.048611in}{0.000000in}}%
\pgfusepath{stroke,fill}%
}%
\begin{pgfscope}%
\pgfsys@transformshift{0.938207in}{1.935145in}%
\pgfsys@useobject{currentmarker}{}%
\end{pgfscope}%
\end{pgfscope}%
\begin{pgfscope}%
\definecolor{textcolor}{rgb}{0.000000,0.000000,0.000000}%
\pgfsetstrokecolor{textcolor}%
\pgfsetfillcolor{textcolor}%
\pgftext[x=0.368904in, y=1.851812in, left, base]{\color{textcolor}\rmfamily\fontsize{18.000000}{21.600000}\selectfont \(\displaystyle {\ensuremath{-}0.5}\)}%
\end{pgfscope}%
\begin{pgfscope}%
\pgfsetbuttcap%
\pgfsetroundjoin%
\definecolor{currentfill}{rgb}{0.000000,0.000000,0.000000}%
\pgfsetfillcolor{currentfill}%
\pgfsetlinewidth{0.803000pt}%
\definecolor{currentstroke}{rgb}{0.000000,0.000000,0.000000}%
\pgfsetstrokecolor{currentstroke}%
\pgfsetdash{}{0pt}%
\pgfsys@defobject{currentmarker}{\pgfqpoint{-0.048611in}{0.000000in}}{\pgfqpoint{-0.000000in}{0.000000in}}{%
\pgfpathmoveto{\pgfqpoint{-0.000000in}{0.000000in}}%
\pgfpathlineto{\pgfqpoint{-0.048611in}{0.000000in}}%
\pgfusepath{stroke,fill}%
}%
\begin{pgfscope}%
\pgfsys@transformshift{0.938207in}{2.702175in}%
\pgfsys@useobject{currentmarker}{}%
\end{pgfscope}%
\end{pgfscope}%
\begin{pgfscope}%
\definecolor{textcolor}{rgb}{0.000000,0.000000,0.000000}%
\pgfsetstrokecolor{textcolor}%
\pgfsetfillcolor{textcolor}%
\pgftext[x=0.555571in, y=2.618842in, left, base]{\color{textcolor}\rmfamily\fontsize{18.000000}{21.600000}\selectfont \(\displaystyle {0.0}\)}%
\end{pgfscope}%
\begin{pgfscope}%
\pgfsetbuttcap%
\pgfsetroundjoin%
\definecolor{currentfill}{rgb}{0.000000,0.000000,0.000000}%
\pgfsetfillcolor{currentfill}%
\pgfsetlinewidth{0.803000pt}%
\definecolor{currentstroke}{rgb}{0.000000,0.000000,0.000000}%
\pgfsetstrokecolor{currentstroke}%
\pgfsetdash{}{0pt}%
\pgfsys@defobject{currentmarker}{\pgfqpoint{-0.048611in}{0.000000in}}{\pgfqpoint{-0.000000in}{0.000000in}}{%
\pgfpathmoveto{\pgfqpoint{-0.000000in}{0.000000in}}%
\pgfpathlineto{\pgfqpoint{-0.048611in}{0.000000in}}%
\pgfusepath{stroke,fill}%
}%
\begin{pgfscope}%
\pgfsys@transformshift{0.938207in}{3.469206in}%
\pgfsys@useobject{currentmarker}{}%
\end{pgfscope}%
\end{pgfscope}%
\begin{pgfscope}%
\definecolor{textcolor}{rgb}{0.000000,0.000000,0.000000}%
\pgfsetstrokecolor{textcolor}%
\pgfsetfillcolor{textcolor}%
\pgftext[x=0.555571in, y=3.385873in, left, base]{\color{textcolor}\rmfamily\fontsize{18.000000}{21.600000}\selectfont \(\displaystyle {0.5}\)}%
\end{pgfscope}%
\begin{pgfscope}%
\definecolor{textcolor}{rgb}{0.000000,0.000000,0.000000}%
\pgfsetstrokecolor{textcolor}%
\pgfsetfillcolor{textcolor}%
\pgftext[x=0.313349in,y=2.219475in,,bottom,rotate=90.000000]{\color{textcolor}\rmfamily\fontsize{18.000000}{21.600000}\selectfont \measure}%
\end{pgfscope}%
\begin{pgfscope}%
\pgfpathrectangle{\pgfqpoint{0.938207in}{0.679475in}}{\pgfqpoint{4.650000in}{3.080000in}}%
\pgfusepath{clip}%
\pgfsetrectcap%
\pgfsetroundjoin%
\pgfsetlinewidth{1.505625pt}%
\definecolor{currentstroke}{rgb}{0.121569,0.466667,0.705882}%
\pgfsetstrokecolor{currentstroke}%
\pgfsetdash{}{0pt}%
\pgfpathmoveto{\pgfqpoint{1.149570in}{0.819475in}}%
\pgfpathlineto{\pgfqpoint{1.372058in}{2.284756in}}%
\pgfpathlineto{\pgfqpoint{1.594547in}{2.929248in}}%
\pgfpathlineto{\pgfqpoint{1.817035in}{3.139715in}}%
\pgfpathlineto{\pgfqpoint{2.039523in}{3.303680in}}%
\pgfpathlineto{\pgfqpoint{2.262011in}{3.196381in}}%
\pgfpathlineto{\pgfqpoint{2.484499in}{3.317716in}}%
\pgfpathlineto{\pgfqpoint{2.706987in}{3.278980in}}%
\pgfpathlineto{\pgfqpoint{2.929475in}{3.357389in}}%
\pgfpathlineto{\pgfqpoint{3.151963in}{3.333140in}}%
\pgfpathlineto{\pgfqpoint{3.374451in}{3.372473in}}%
\pgfpathlineto{\pgfqpoint{3.596939in}{3.330132in}}%
\pgfpathlineto{\pgfqpoint{3.819427in}{3.398459in}}%
\pgfpathlineto{\pgfqpoint{4.041915in}{3.408636in}}%
\pgfpathlineto{\pgfqpoint{4.264403in}{3.362742in}}%
\pgfpathlineto{\pgfqpoint{4.486891in}{3.366725in}}%
\pgfpathlineto{\pgfqpoint{4.709379in}{3.414262in}}%
\pgfpathlineto{\pgfqpoint{4.931867in}{3.394546in}}%
\pgfpathlineto{\pgfqpoint{5.154355in}{3.408820in}}%
\pgfpathlineto{\pgfqpoint{5.376843in}{3.442483in}}%
\pgfusepath{stroke}%
\end{pgfscope}%
\begin{pgfscope}%
\pgfsetrectcap%
\pgfsetmiterjoin%
\pgfsetlinewidth{0.803000pt}%
\definecolor{currentstroke}{rgb}{0.000000,0.000000,0.000000}%
\pgfsetstrokecolor{currentstroke}%
\pgfsetdash{}{0pt}%
\pgfpathmoveto{\pgfqpoint{0.938207in}{0.679475in}}%
\pgfpathlineto{\pgfqpoint{0.938207in}{3.759475in}}%
\pgfusepath{stroke}%
\end{pgfscope}%
\begin{pgfscope}%
\pgfsetrectcap%
\pgfsetmiterjoin%
\pgfsetlinewidth{0.803000pt}%
\definecolor{currentstroke}{rgb}{0.000000,0.000000,0.000000}%
\pgfsetstrokecolor{currentstroke}%
\pgfsetdash{}{0pt}%
\pgfpathmoveto{\pgfqpoint{5.588207in}{0.679475in}}%
\pgfpathlineto{\pgfqpoint{5.588207in}{3.759475in}}%
\pgfusepath{stroke}%
\end{pgfscope}%
\begin{pgfscope}%
\pgfsetrectcap%
\pgfsetmiterjoin%
\pgfsetlinewidth{0.803000pt}%
\definecolor{currentstroke}{rgb}{0.000000,0.000000,0.000000}%
\pgfsetstrokecolor{currentstroke}%
\pgfsetdash{}{0pt}%
\pgfpathmoveto{\pgfqpoint{0.938207in}{0.679475in}}%
\pgfpathlineto{\pgfqpoint{5.588207in}{0.679475in}}%
\pgfusepath{stroke}%
\end{pgfscope}%
\begin{pgfscope}%
\pgfsetrectcap%
\pgfsetmiterjoin%
\pgfsetlinewidth{0.803000pt}%
\definecolor{currentstroke}{rgb}{0.000000,0.000000,0.000000}%
\pgfsetstrokecolor{currentstroke}%
\pgfsetdash{}{0pt}%
\pgfpathmoveto{\pgfqpoint{0.938207in}{3.759475in}}%
\pgfpathlineto{\pgfqpoint{5.588207in}{3.759475in}}%
\pgfusepath{stroke}%
\end{pgfscope}%
\begin{pgfscope}%
\pgfsetbuttcap%
\pgfsetmiterjoin%
\definecolor{currentfill}{rgb}{1.000000,1.000000,1.000000}%
\pgfsetfillcolor{currentfill}%
\pgfsetfillopacity{0.800000}%
\pgfsetlinewidth{1.003750pt}%
\definecolor{currentstroke}{rgb}{0.800000,0.800000,0.800000}%
\pgfsetstrokecolor{currentstroke}%
\pgfsetstrokeopacity{0.800000}%
\pgfsetdash{}{0pt}%
\pgfpathmoveto{\pgfqpoint{4.248569in}{0.804475in}}%
\pgfpathlineto{\pgfqpoint{5.413207in}{0.804475in}}%
\pgfpathquadraticcurveto{\pgfqpoint{5.463207in}{0.804475in}}{\pgfqpoint{5.463207in}{0.854475in}}%
\pgfpathlineto{\pgfqpoint{5.463207in}{1.176157in}}%
\pgfpathquadraticcurveto{\pgfqpoint{5.463207in}{1.226157in}}{\pgfqpoint{5.413207in}{1.226157in}}%
\pgfpathlineto{\pgfqpoint{4.248569in}{1.226157in}}%
\pgfpathquadraticcurveto{\pgfqpoint{4.198569in}{1.226157in}}{\pgfqpoint{4.198569in}{1.176157in}}%
\pgfpathlineto{\pgfqpoint{4.198569in}{0.854475in}}%
\pgfpathquadraticcurveto{\pgfqpoint{4.198569in}{0.804475in}}{\pgfqpoint{4.248569in}{0.804475in}}%
\pgfpathlineto{\pgfqpoint{4.248569in}{0.804475in}}%
\pgfpathclose%
\pgfusepath{stroke,fill}%
\end{pgfscope}%
\begin{pgfscope}%
\pgfsetbuttcap%
\pgfsetroundjoin%
\definecolor{currentfill}{rgb}{1.000000,0.000000,0.000000}%
\pgfsetfillcolor{currentfill}%
\pgfsetlinewidth{1.505625pt}%
\definecolor{currentstroke}{rgb}{1.000000,0.000000,0.000000}%
\pgfsetstrokecolor{currentstroke}%
\pgfsetdash{}{0pt}%
\pgfsys@defobject{currentmarker}{\pgfqpoint{-0.041667in}{-0.041667in}}{\pgfqpoint{0.041667in}{0.041667in}}{%
\pgfpathmoveto{\pgfqpoint{-0.041667in}{-0.041667in}}%
\pgfpathlineto{\pgfqpoint{0.041667in}{0.041667in}}%
\pgfpathmoveto{\pgfqpoint{-0.041667in}{0.041667in}}%
\pgfpathlineto{\pgfqpoint{0.041667in}{-0.041667in}}%
\pgfusepath{stroke,fill}%
}%
\begin{pgfscope}%
\pgfsys@transformshift{4.548569in}{1.016782in}%
\pgfsys@useobject{currentmarker}{}%
\end{pgfscope}%
\end{pgfscope}%
\begin{pgfscope}%
\definecolor{textcolor}{rgb}{0.000000,0.000000,0.000000}%
\pgfsetstrokecolor{textcolor}%
\pgfsetfillcolor{textcolor}%
\pgftext[x=4.998569in,y=0.951157in,left,base]{\color{textcolor}\rmfamily\fontsize{18.000000}{21.600000}\selectfont \fit}%
\end{pgfscope}%
\end{pgfpicture}%
\makeatother%
\endgroup%